\documentclass[%
 article,twocolumn,
 amsmath,amssymb,
 aps,
pra,
floatfix
]{revtex4-2}

\usepackage{amsmath}%
\usepackage{amsfonts}%
\usepackage{amssymb}%
\usepackage{amstext}
\usepackage{textgreek}
\usepackage{multirow}
\usepackage{esint}
\usepackage{graphicx}
\usepackage[table,xcdraw]{xcolor}
\usepackage[sort&compress]{natbib}
\usepackage{color}
\usepackage{xcolor}
\usepackage{booktabs}
\usepackage{threeparttable}
\usepackage{hyperref}
\hypersetup{colorlinks=true,allcolors=blue}
\usepackage[page,toc,titletoc,title]{appendix}
\usepackage{orcidlink}

\makeatletter
\renewcommand{\p@subsection}{}
\renewcommand{\p@subsubsection}{}
\makeatother

\usepackage{xspace} 

\begin{document}
\title{The science of compressional heating on the LM26 magnetized target fusion experiment}

%
\author{ S.~J. Howard~\orcidlink{0009-0007-3247-6096}}
\affiliation{General Fusion Inc., Richmond, British Columbia, Canada}
\author{ D.~P. Brennan~\orcidlink{0000-0003-4927-8196}}
\affiliation{General Fusion Inc., Richmond, British Columbia, Canada}
\author{ K. Epp }
\affiliation{General Fusion Inc., Richmond, British Columbia, Canada}
\author{ P. Forysinski }
\affiliation{General Fusion Inc., Richmond, British Columbia, Canada}
\author{ D. Plant }
\affiliation{General Fusion Inc., Richmond, British Columbia, Canada}
\author{ M. Reynolds~\orcidlink{0000-0001-5880-2290}}
\affiliation{General Fusion Inc., Richmond, British Columbia, Canada}
\author{A.~Froese~\orcidlink{0000-0002-1669-7397}}
\affiliation{General Fusion Inc., Richmond, British Columbia, Canada}
\author{N. Sirmas~\orcidlink{0000-0002-1672-1633}}
\affiliation{General Fusion Inc., Richmond, British Columbia, Canada}
\author{D.~Krotez~\orcidlink{0009-0000-8302-6879}}
\affiliation{General Fusion Inc., Richmond, British Columbia, Canada}
\author{V. Suponitsky }
\affiliation{General Fusion Inc., Richmond, British Columbia, Canada}
\author{R. Zindler }
\affiliation{General Fusion Inc., Richmond, British Columbia, Canada}
\author{E. Love }
\affiliation{General Fusion Inc., Richmond, British Columbia, Canada}
\author{C. Macdonald }
\affiliation{General Fusion Inc., Richmond, British Columbia, Canada}
\author{N. Kumar }
\affiliation{General Fusion Inc., Richmond, British Columbia, Canada}
\author{Z.~Seifollahi~Moghadam }
\affiliation{General Fusion Inc., Richmond, British Columbia, Canada}
\author{K.~Conquergood}
\affiliation{General Fusion Inc., Richmond, British Columbia, Canada}
\author{A. Wong}
\affiliation{General Fusion Inc., Richmond, British Columbia, Canada}
\author{W. Zawalski}
\affiliation{General Fusion Inc., Richmond, British Columbia, Canada}
\author{B. Rablah}
\affiliation{General Fusion Inc., Richmond, British Columbia, Canada}
\author{W.~Kozicki}
\affiliation{General Fusion Inc., Richmond, British Columbia, Canada}
\author{P.~Carle~\orcidlink{0000-0003-4317-1451}}
\affiliation{General Fusion Inc., Richmond, British Columbia, Canada}
\author{A.~Rohollahi~\orcidlink{0000-0001-6688-8696}}
\affiliation{General Fusion Inc., Richmond, British Columbia, Canada}
\author{C.~Preston~\orcidlink{0000-0003-1372-1598}}
\affiliation{General Fusion Inc., Richmond, British Columbia, Canada}
\author{A. M. D. Lee~\orcidlink{0000-0001-6327-2351}}
\affiliation{General Fusion Inc., Richmond, British Columbia, Canada}
\author{J.~Hobbis}
\affiliation{General Fusion Inc., Richmond, British Columbia, Canada}
\author{L.~Santos}
\affiliation{General Fusion Inc., Richmond, British Columbia, Canada}
\author{X. Feng}
\affiliation{General Fusion Inc., Richmond, British Columbia, Canada}
\author{M. Schellenberg-Beaver}
\affiliation{General Fusion Inc., Richmond, British Columbia, Canada}
\author{R.~Tingley}
\affiliation{General Fusion Inc., Richmond, British Columbia, Canada}
\author{R. Underwood}
\affiliation{General Fusion Inc., Richmond, British Columbia, Canada}
\author{J.~Sanchez Rojo}
\affiliation{General Fusion Inc., Richmond, British Columbia, Canada}
\author{J.~Gorenstein}
\affiliation{General Fusion Inc., Richmond, British Columbia, Canada}
\author{ D. Froese}
\affiliation{General Fusion Inc., Richmond, British Columbia, Canada}
\author{E. Cessford }
\affiliation{General Fusion Inc., Richmond, British Columbia, Canada}
\author{J.~Pratt}
\affiliation{General Fusion Inc., Richmond, British Columbia, Canada}
\author{J.~Crofts}
\affiliation{General Fusion Inc., Richmond, British Columbia, Canada}
\author{J. Sardari}
\affiliation{General Fusion Inc., Richmond, British Columbia, Canada}
\author{G. Faust}
\affiliation{General Fusion Inc., Richmond, British Columbia, Canada}
\author{D. Ross}
\affiliation{General Fusion Inc., Richmond, British Columbia, Canada}
\author{J.~Wilkie}
\affiliation{General Fusion Inc., Richmond, British Columbia, Canada}
\author{S. Bernard}
\affiliation{General Fusion Inc., Richmond, British Columbia, Canada}
\author{S. Edwards}
\affiliation{General Fusion Inc., Richmond, British Columbia, Canada}
\author{R. Oosterom}
\affiliation{General Fusion Inc., Richmond, British Columbia, Canada}
\author{M. Yurkiv}
\affiliation{General Fusion Inc., Richmond, British Columbia, Canada}
\author{J. Y. J. Cheng}
\affiliation{General Fusion Inc., Richmond, British Columbia, Canada}
\author{ C. Connor}
\affiliation{General Fusion Inc., Richmond, British Columbia, Canada}
\author{S.~Bolanos }
\affiliation{General Fusion Inc., Richmond, British Columbia, Canada}
\author{ C. Gutjahr}
\affiliation{General Fusion Inc., Richmond, British Columbia, Canada}
\author{ E. Chan}
\affiliation{General Fusion Inc., Richmond, British Columbia, Canada}
\author{ M. Greenwood}
\affiliation{General Fusion Inc., Richmond, British Columbia, Canada}
\author{ E. Ng}
\affiliation{General Fusion Inc., Richmond, British Columbia, Canada}
\author{ A. Massey}
\affiliation{General Fusion Inc., Richmond, British Columbia, Canada}
\author{ K. Chen}
\affiliation{General Fusion Inc., Richmond, British Columbia, Canada}
\author{ R. Svihra}
\affiliation{General Fusion Inc., Richmond, British Columbia, Canada}
\author{ A. Gromer}
\affiliation{General Fusion Inc., Richmond, British Columbia, Canada}
\author{ S. Lee}
\affiliation{General Fusion Inc., Richmond, British Columbia, Canada}
\author{ X.~Zhu}
\affiliation{General Fusion Inc., Richmond, British Columbia, Canada}
\author{ L. Marshall}
\affiliation{General Fusion Inc., Richmond, British Columbia, Canada}
\author{ C. Eyrich }
\affiliation{General Fusion Inc., Richmond, British Columbia, Canada}
\author{ A. Mahoney }
\affiliation{General Fusion Inc., Richmond, British Columbia, Canada}
\author{M. Davidson}
\affiliation{General Fusion Inc., Richmond, British Columbia, Canada}
\author{H. Feng}
\affiliation{General Fusion Inc., Richmond, British Columbia, Canada} 
\author{A. Rudy}
\affiliation{General Fusion Inc., Richmond, British Columbia, Canada} 
\author{ M. Laberge~\orcidlink{0009-0006-0093-2168}} 
\affiliation{General Fusion Inc., Richmond, British Columbia, Canada}

\begin{abstract}

The Lawson Machine 26 (LM26) operating at General Fusion has demonstrated compressional heating of a spherical tokamak deuterium plasma as it was compressed by an imploding solid lithium liner.  Results from the first 11 compression experiments on LM26 are presented, the highest-performing of which show more than a three-fold increase in electron temperature, a ten-fold increase in density, and a ten-fold increase in poloidal field in the plasma driven by three-fold radial compression.   
The experimental device and its instrumentation are reviewed in detail, followed by direct observations from each of the key diagnostics for liner trajectory and plasma properties, measuring increases in magnetic field, electron density, as well as emission of neutrons, X-rays, and visible radiation. 
Observations from fast-camera images during compression provide detailed context for interpreting the spatial structure of plasma-wall interaction. 
Overviews of the various models developed and used in the analysis are presented.  Diagnostic data are used to reconstruct the experimental equilibrium state in computational modeling as a function of time.  
The results indicate the reconstructions are sufficiently accurate to match the essential observations from the experiment, validating the models and building confidence in the stability and transport analyses that support the key conclusions.  
Trends across the full set of 11 compression shots are presented, and a more detailed examination of the high-performance shots are given individually.  
The key conclusions of the integrated physics model specifically indicate that compressional heating was achieved in this set of experiments, as evidenced by the balance of heating power from compression, Ohmic heating from plasma current, and losses to the boundary necessary to match the experimental data. 
A majority of the increase in temperature is attributable to compressional heating.  
An increase in neutron flux is also observed during compression. The results provide a basis for planned improvements to the LM26 facility that will enable the compression of magnetized plasma to increasingly higher densities and temperatures.

\end{abstract}
\date{\today}

\maketitle


\clearpage

\section{Introduction}

Magnetized target fusion (MTF) is a fusion power concept in which a fuel 
is confined by magnetic fields while being compressed and heated inside an imploding conductive liner. This method is economically attractive because the energy used to bring the fuel up to fusion conditions is primarily supplied by the motion of the walls, which can be produced inexpensively.  Using a solid lithium liner in Lawson Machine 26 (LM26), we have made 
progress at demonstrating the viability of heating a deuterium plasma to fusion-relevant temperatures using rapid compression as the primary means of heating. 

This paper will report on the evidence of compressional heating and observations of magnetically stable confinement of the plasma, which has been gathered during the sequence of plasma compression experiments conducted so far on LM26. 
At the time of this writing 11 plasmas have been successfully compressed by imploding a lithium liner; those discussed below are denoted as LMC-$N$ (Lawson Machine Compression), with $N$ ranging from 1--11. 
The initial proof-of-concept design of the experiment uses a cylindrical domain to hold a spherical tokamak (ST) plasma which is compressed towards a straight central shaft, relying on poloidal flux conservation within  the lithium liner to compress the confining magnetic field of the plasma. Solid lithium is used due to its low mass density, high electrical conductivity, mechanical softness, low atomic number, and its chemical getter ability that enables a low recycling-coefficient boundary. The trajectory of the lithium liner is designed to shape the plasma boundary, control the speed of compression, and determine when the confining chamber is closed by making contact with the axial endplates. 
Future versions of the experiment will include an upgrade to the center conductor geometry, providing a double-napped cone to increase the rate of geometric convergence and better maintain self-similarity during compression. 

A suite of diagnostics is used to directly measure fundamental plasma quantities, such as magnetic field, temperature, and density.  Those measurements are fitted to two separate models to reconstruct the evolving equilibrium state.  The Bayesian Plasma Reconstruction (BPR) model~\cite{Howard_2025} treats the discharge as sequential independent time slices, using time-dependent boundary conditions reconstructed from magnetic data.  The Integrated System Model (ISM)~\cite{Khalzov_2024} fits to time-dependent integrated simulations of the full compression.
The observed increase of electron temperatures during compression are well explained in these fits by a combination of compressional and Ohmic heating. 
Stability analyses and experimental data indicate the plasmas remain stable until deep into compression.  
The results form a basis for understanding the key physics governing compressional heating processes and inform continuing research into compressional heating using flux conserving metal liners.

\subsection{Magnetized target fusion}\label{sec:MTF}

The MTF concept dates back to the LINUS project built at the Naval Research Laboratory in the 1970s \cite{robson_linus_1973}.  LINUS was intended to compress plasma with a molten sodium-potassium liner driven by pressurized gas, but the testing that was done used water~\cite{osti_5723685}.  Due to a shift in focus toward tokamaks, research on plasma heating through magnetic compression subsequently became dominated by tokamak experiments, including ATC \cite{osti_4203067,BolPRL72,BolIAEA74, Ellis1985}, TOSCA \cite{Robinson_1985}, TFTR \cite{TFTR1985, Grove1985}, JET \cite{JET,JET2}, and TUMAN-3M \cite{askinazi2001}.  In particular, the ATC experiment was designed specifically for this purpose, producing significant compression ratios, and successfully demonstrating much of the basic concept. However, with the many advances in physics understanding and computational capability in the years since these experiments, it has become evident that the conventional tokamak configuration is not the optimal choice for magnetic compression in terms of efficiency, stability, and confinement.

At the turn of the millennium, dedicated MTF experiments began to be designed and built.  They took advantage of new progress in plasma physics and control electronics to advance the concept.
FRX-L~\cite{taccetti_frcx-l_pop2003}, a collaboration between the U.S. Air Force Research Laboratory and Los Alamos National Laboratory, generated and translated field-reversed configuration (FRC) plasmoids suitable for MTF injection and compression schemes.  The next machine, FRCHX~\cite{Wurden_IAEA2008}, used an electromagnetically driven flux-compression liner to rapidly implode an FRC plasma.  During the Plasma Compression Science (PCS) campaign~\cite{Howard_2025}, General Fusion compressed spheromaks and ST plasma configurations using aluminum liners.
Helion Energy collided FRC plasmas and compressed them with pulsed magnetic fields~\cite{Kirtley2023}.  MagLIF (Magnetized Liner Inertial Fusion) at Sandia National Laboratories used the Z pulsed-power machine to implode a cylindrical liner around axially-magnetized laser-preheated deuterium fuel~\cite{Yager-Elorriaga_NF2022}.
While diverse in design, approach, and scale, these experiments shared the goal of reaching fusion conditions through rapid compression of a transient pre-magnetized plasma.

\subsection{General Fusion MTF experiments}
Experiments at General Fusion (GF) have explored the viability of compressing a plasma with a spherical tokamak magnetic configuration~\cite{Howard_2025}. Because of their low aspect ratio, ST plasmas exhibit favorable confinement and stability properties \cite{peng_pop_2000, MOno_NF2001}. They operate at high toroidal beta $\beta_t$ (the ratio of plasma pressure to toroidal magnetic field pressure) and can sustain high plasma currents due to their low aspect ratio and natural shaping.  The energy confinement time, $\tau_E$, has a distinctive feature in ST plasmas: it depends strongly on the toroidal magnetic field and weakly on the plasma current \cite{Kurskiev_NF2022}.

Because of these features, General Fusion decided to pursue the spherical tokamak magnetic configuration as a promising approach to fueling a compact and low-cost fusion pilot plant~\cite{Laberge2019}.
In a sequence of subscale experiments, ST plasmas were compressed using chemically-driven implosions of aluminum liners \cite{Laberge2008, howard_development_2009, Froese2014, Howard_2025}.  These PCS experiments provided an understanding of plasma behavior and confinement scaling laws during rapid compression.

Concurrently, large-scale plasma generation was tested in the Plasma Injector 3 experiment (PI3)~\cite{Pi3_tauE_NucFus_2025}.  The goal of PI3 was to demonstrate that large (0.6--0.7\,m major radius) plasma targets that are suitable for compression to fusion temperatures could be generated with coaxial helicity injection (CHI)~\cite{Raman2014}.  CHI is a method of generating magnetized plasma at a location distant from the flux conserver where the plasma will settle and be compressed. Such separation is required to protect the plasma formation components from the liner compression system to enable reusability in a commercial reactor design.
The results from PI3 showed that high performance Ohmically heated ST plasmas can have good energy confinement time and high initial temperature, making them suitable as the target plasma in an MTF compression experiment.

Once large-scale plasma formation was demonstrated in PI3 and rapid plasma compression was demonstrated in PCS, the next stage in testing was to combine the two experiments in Lawson Machine 26 (LM26).

The rest of the paper describes LM26 and its first results. 
In Sec.~\ref{sec:LM26} the experimental setup is described in detail, including the geometry, plasma injector, lithium liner and its dynamics during compression, and the diagnostics suite.  In Sec.~\ref{sec:ExObserve} we highlight some of the key experimental observations stemming from direct measurements during compression experiments, including poloidal flux, density, X-ray emission and temperature increases, as well as phenomena such as X-ray crashes. 

In Sec.~\ref{sec:physics_analysis_modeling} we present the physics analysis and modeling that lead to the main conclusion of compressional heating, including the equilibrium reconstruction methods, magnetohydrodynamic (MHD) stability analyses, and quantitative analysis differentiating the compressional heating from Ohmic heating, taking into account transport losses.  Finally, in Sec.~\ref{sec:Conclude} we summarize the results and review our conclusions from the experimental outcomes.

\begin{figure*}
    \centering
    \includegraphics[width=1\linewidth]{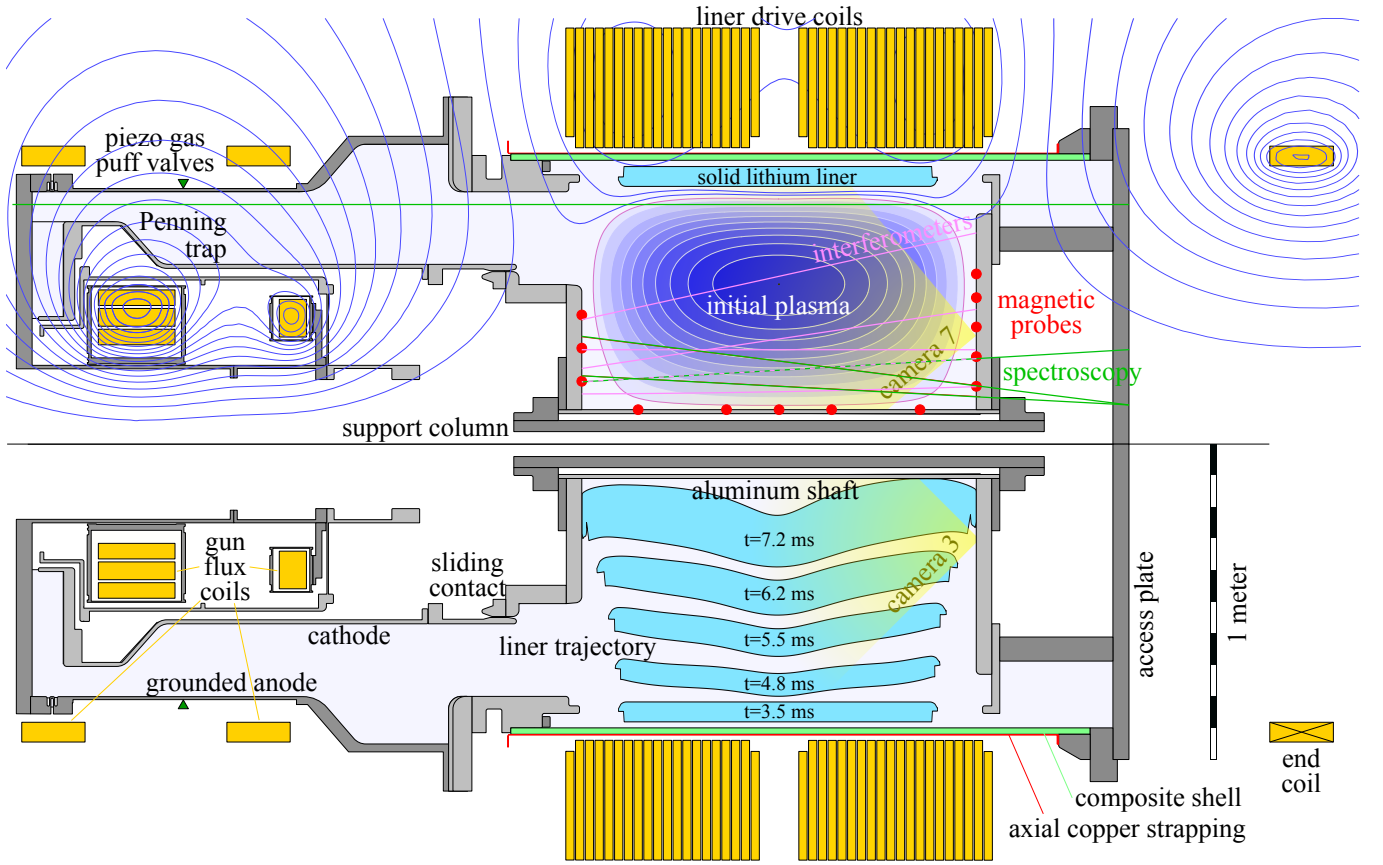}
    \caption{Schematic of LM26 experiment, with liner trajectory (lower half-view) and poloidal flux (upper half-view) shown for the LMC-9 shot as an example.}
    \label{fig:LM26_schematic}
\end{figure*}

\section{The LM26 experiment}\label{sec:LM26}

LM26 is a large-scale MTF machine built to demonstrate that compressional heating of an ST plasma can achieve fusion-relevant conditions. A Marshall gun forms, by CHI, a magnetized deuterium plasma inside a 1.6~m diameter hollow lithium cylinder within a composite vacuum vessel. A surrounding 36-turn theta-pinch coil generates a strong magnetic field to symmetrically compress the liner and the plasma it contains in less than 4~ms. The compression of the liner occurs without damaging any major structures, allowing the device to be rapidly restored for further experiments.

\subsection{Geometry}
A schematic of this first iteration of the LM26 experiment is shown in Fig.~\ref{fig:LM26_schematic}. The confinement region is roughly cylindrical, bounded on the inside by an aluminum shaft (10.8 cm radius) running along the axis of the machine, on the outside by a replaceable lithium liner, and bounded at each end by aluminum plates. The Marshall gun opening is a 23~cm concentric gap whose outer surface is flush with the liner.  
The lithium liner is surrounded by a composite vacuum vessel with stainless steel flanges mounting to the Marshall gun on one side and the removable diagnostic endplate on the other.  The compression coil is composed of an axial stack of 36 single-turn aluminum-plate coils with G10 fiberglass support components. It is mounted on a decoupled support structure that surrounds the vessel.  Each turn is designed to have low inductance and high mechanical strength.  To minimize the error fields produced by the coil, there are current-focusing features at the input gaps of each turn and the gaps are distributed toroidally around the machine. This design concept was first used on the Super-Magnetized Ring Test (SMRT) device~\cite{DunleaMagnetic_POP2020} at General Fusion and has proven effective at generating highly symmetric, fast rise-time, multi-Tesla poloidal compression fields.

The composite vacuum vessel is built around a primary cylinder (20~mm thick, 1.8~m ID) of multi-layer fiberglass and high-vacuum epoxy (EP42HT-2AO-1 Black) composite and is capable of high vacuum operation ($10^{-8}$ Torr range).  It is significantly more shock resistant and cost effective than glass or ceramic alternatives of comparable size. The vessel must be non-conductive to allow the poloidal compression field to penetrate it without attenuation. 
A set of axial copper straps are bonded to the outside of the composite vessel.  They are evenly distributed to provide a return path for current in the coaxial circuit that generates toroidal magnetic field. An array of ceramic blocks are attached to the inner surface of the vessel to center the lithium liner relative to the shaft.

\subsection{Plasma injector}

The ST plasmas in LM26 are produced via CHI using a Marshall gun, repurposed from the earlier PI3 device. A full description of the Marshall gun design, operation and power supply circuit parameters can be found in the previous work \cite{Pi3_tauE_NucFus_2025}.  
Here, we will focus on the modifications and upgrades made to implement the new configuration. 
During a compression shot in the LM26 experiment, the liner impacts the endplates, loading the structure with a large axial shock.  The gun was lengthened by 0.5~m to make room for a high-current sliding electrical contact, which mechanically decouples the Marshall gun from the compression region endplates.

\subsubsection{Power supplies}
The plasma formation system draws on several dedicated capacitor banks. The power to generate the plasma and drive the CHI process comes from a 2.7~mF, 25~kV, 0.8~MJ formation bank. The toroidal field is generated by current that runs along the central shaft and returns on copper straps outside the vacuum vessel.  The current is initially produced by a 30~mF, 10~kV, 1.5~MJ peaking bank and then resistive decay in the aluminum shaft is compensated by a 48~mF, 11~kV, 2.9~MJ sustain bank.  Shaft current can also be increased rapidly with a 42~mF, 12~kV, 3.0~MJ ramp bank.  It is subdivided into 24 modules which can be triggered in sequence to provide fine control of the ramp rate.

The purpose of the shaft ramp bank is to balance the toroidal field in the Marshall gun with that in the confinement region as the plasma is being compressed, until the gun is closed off when the liner contacts the endplates. In an ideal compression scenario, the shaft ramp should track the inverse of the coaxial inductance of the plasma confining volume to maintain the original current profile shape and prevent the plasma from being squeezed back into the gun~\cite{Howard_2025}.  However, we find that in the present cylindrical (rectangular cross section) geometry of LM26, the plasma is trapped within the confinement region better than in the PCS experiments, so less shaft ramp current is required to obtain reasonably good results.  As such, we typically run with the shaft ramp significantly below the inverse-inductance curve. 

\subsubsection{Control of plasma parameters}
The plasma is well described by a Grad--Shafranov equilibrium with a pre-compression total poloidal flux of 100--170~mWb.  The closed poloidal flux is due to the toroidal plasma current contained within the last closed flux surface (LCFS).  There is approximately 6--14~mWb of open ``buffer'' flux between the outside of the LCFS and the inner surface of the liner. 
Adjusting the rate of change of externally driven shaft current modifies the toroidal field at the edge of the plasma \cite{Pi3_tauE_NucFus_2025}.  This permits some control over the current density profile and is used to balance against the resistive decay of plasma current.
Setting the shaft-current slope above or below the slope of an unsustained shot causes an inductive decrease or increase of the edge plasma current, respectively. 

The density profile of the formed plasma can be varied from flat to peaked by adjusting plasma breakdown conditions in the gun.  A flat profile contains more total thermal energy than a peaked profile with the same core density and temperature. A peaked profile has reduced wall interaction. 
Experimental data for a typical high-performance plasma discharge are shown in Fig.~\ref{fig:Precompression}.

\subsection{Lithium liner manufacture and transfer}

The lithium liners used in LM26 are manufactured at General Fusion.  Manufacturing begins with lithium ingots that are melted in a steel mold, which is coated with boron nitride to prevent adhesion.
After solidifying and cooling to room temperature, the casting is removed from the mold and cleaned. It is mounted in a large vertical lathe, where it is formed to a rough size by rolling and machined to final dimensions with a single point cutter, leaving a clean, smooth plasma-facing surface. All these operations are performed under argon to prevent rapid corrosion of the lithium.  The completed liners have an outer radius of 880~mm to match the dimensions of ceramic centering blocks in the vacuum vessel, and have thicknesses ranging from 50--70~mm and lengths of 1--1.2 m. The liner mass has varied from 153 to 218 kg to explore different collapse velocities and impact loads.  

Transfer and positioning of the liner within the compression chamber of LM26 is conducted within large dry-argon filled polyethylene tubular bags, using a custom multi-axis motorized transfer and positioning system to place the liner within the chamber onto the centering blocks.  A large 2~m diameter gate valve uses a 1 in thick aluminum honeycomb panel as a sliding shutter that is raised and lowered by electric winches.  It serves as an air lock component needed for attaching and removing new bags during different stages of the lithium transfer and diagnostic shaft installation processes. The shutter is removed before plasma operations begin.  After the installation of the new liner under argon gas, the machine is pumped down to vacuum ahead of plasma formation.

\subsection{Operation of compression campaigns}

Prior to each compression, the plasma performance must be optimized.  A campaign of forming plasmas into the static liner is executed to condition the machine and improve the pre-compression plasma. Each plasma that is formed cleans residual contaminants from the liner and other plasma facing surfaces. Periodically, we evaporatively deposit thin layers of lithium on the plasma facing surfaces to further getter contaminants and reduce ion recycling \cite{Pi3_tauE_NucFus_2025}. 

\begin{figure}
    \centering
    \includegraphics[width=1\linewidth]{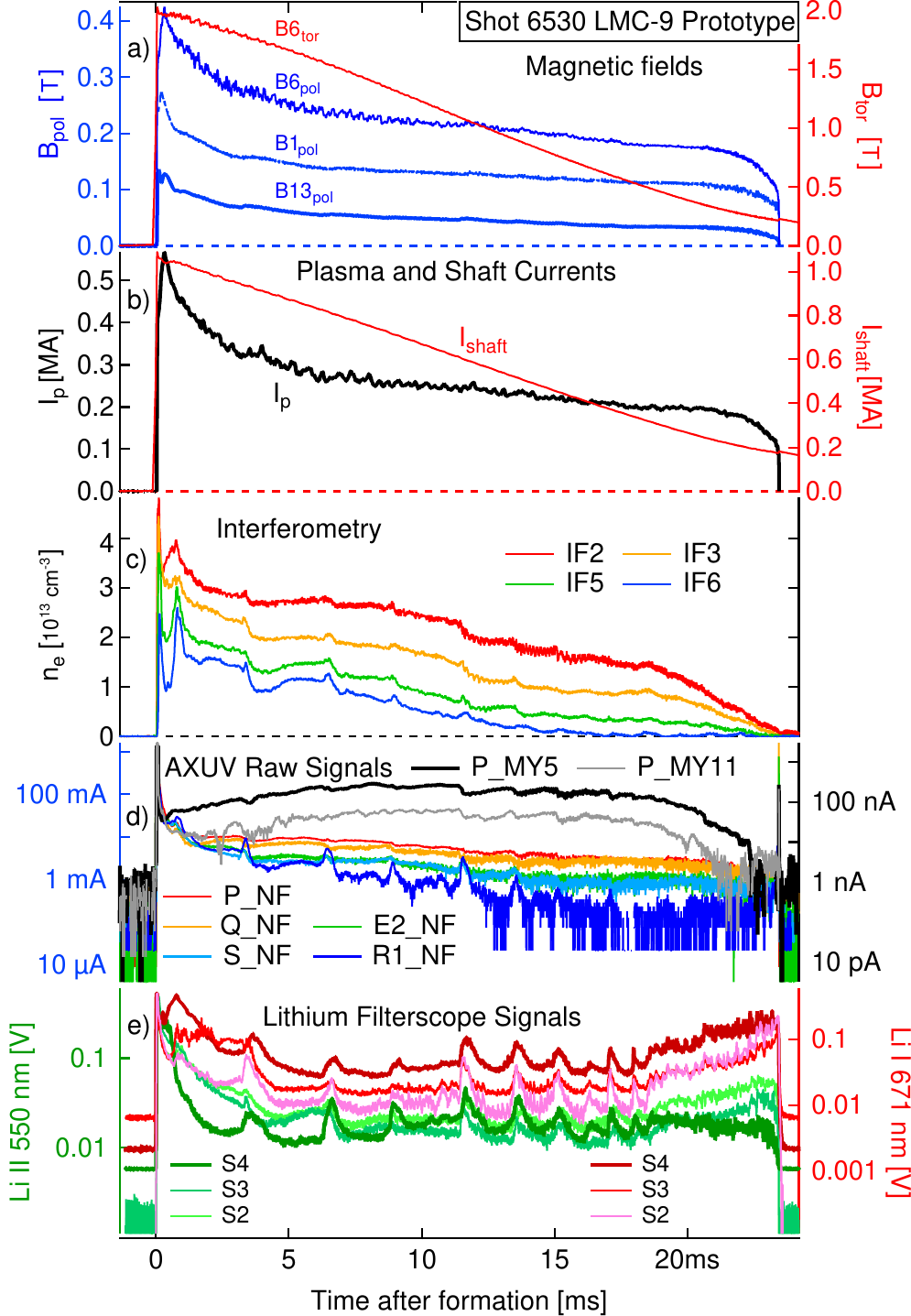}
    \caption{Example data from non-compression ST plasma formation shot 6530, which is the prototype for compression shot 6531 (LMC-9). a) Poloidal magnetic field at three positions (B1, B6, and B13 in Fig.~\ref{fig:DiagMap}) and toroidal field at B6 in red. b) Plasma current, $I_\text{p}$, and shaft current, $I_\text{shaft}$, determined by all edge poloidal and toroidal field measurements, respectively. c) Interferometer channels that measure line average electron density at four different radial locations.  d) AXUV signals which are described in detail in Fig.~\ref{fig:AXUVmap}.  e) Filterscope signals measuring Li\,I light at 671\,nm and Li\,II light at 550\,nm, both on three chords shown in Fig.~\ref{fig:DiagMap}.  See sections \ref{sec:diagoverview} and \ref{sec:ExObserve} for more details about plasma diagnostics used on LM26.}
    \label{fig:Precompression}
\end{figure}

Once the machine is sufficiently clean, the compression shot is designed by varying machine parameters to establish a baseline of plasma behavior and optimize the compressibility of the plasma target. In addition to optimizing parameters such as initial plasma temperatures and magnetic lifetime, a candidate plasma needs to remain quiescent during the 3~ms interval that compression will take place. An example of an optimized non-compression plasma can be found in Fig.~\ref{fig:Precompression}.

\subsection{Compression dynamics of solid lithium liners}

The lithium liner is electromagnetically accelerated radially inwards by a stack of coils in the manner of a theta pinch. The compression coil assembly contains 36 single-turn aluminum coils, which are driven in pairs by 18 isolated circuits. Each pair is located symmetrically across the mid-plane and wired in series to ensure the field generated is symmetric. Each isolated circuit is connected to a capacitor bank with a capacity of approximately 1 MJ (19.2~mF, up to 10.5 kV), for an overall capacity of 18~MJ across all circuits. When the banks are discharged into the compression coils, this produces a field around the lithium liner that briefly reaches 4.5~T (80 atm).

During liner implosion, a higher inward radial velocity is generated at the ends of the cylinder compared to the middle, which causes a converging crimp shape with a central pocket that contains the plasma during compression.  See the lower half of Fig.~\ref{fig:LM26_schematic} for examples of liner shape during the compression.
This shape is formed because the coil stack is longer than the liner and has a gap in the middle.  Additional shaping of the liner trajectory can be accomplished by connecting only select compression coils to the power supplies. The thickness of the liner increases during compression due to radial convergence. This thickening increases the stiffness of the liner, causing the liner to decelerate later in the compression. 

Once the liner contacts the shaft endplates at both ends it forms a closed coaxial circuit for the shaft current that had been, up to that point, driven externally. Now flowing in a closed poloidal loop, this current passively ramps up as the final stage of compression proceeds, with total toroidal magnetic flux within that loop being held nearly constant, but the inductance rapidly decreasing. The liner length is chosen based on the desired touchdown time and location, where shorter liners are used to delay the time of contact and the possibility of ejecta, and longer liners are used to close the coaxial current sooner.

Comparisons between shot parameters and key performance metrics for LMC-2 through LMC-11 are shown in Table~\ref{tab:shot_data}. LMC-1 was a commissioning shot and is not included in this comparison.

\begin{table*}[tbp]
\centering
\footnotesize
\begin{threeparttable}
\caption{Experimental shot parameters and performance for liner dynamics. Here $C_R$ is the liner radial compression ratio as defined in Sec.~\ref{sec:liner_implosion_performance}. }
\label{tab:shot_data}
\setlength{\tabcolsep}{6pt}
\begin{tabular}{l *{10}{c}}
\toprule
& \multicolumn{10}{c}{\textbf{LMC-}} \\
\cmidrule(lr){2-11}
\textbf{Metric / Parameter}
& \textbf{2}
& \textbf{3}
& \textbf{4}
& \textbf{5}
& \textbf{6}
& \textbf{7}
& \textbf{8}
& \textbf{9}
& \textbf{10}
& \textbf{11}\\
\midrule
Liner Thickness [mm]             & 69.3  & 69.5  & 63.2  & 72.0  & 62.2  & 54.3  & 70.6  & 63.1  & 65.2 & 54.6\\
Liner Length [mm]                & 1081  & 1033  & 1126  & 1218  & 1063  & 1060  & 1011  & 1012  & 1060  & 1013\\
Liner Mass [kg]                  & 204   & 184   & 193   & 218   & 182   & 162   & 195   & 176   & 186  &  153 \\
\midrule
Compression Cap Energy [MJ]      & 12.7  & 12.0  & 9.8   & 12.7  & 11.3  & 14.8  & 12.8  & 14.9  & 13.1 &  15.9\\
Peak Kinetic Energy\tnote{$\dagger$}~~[MJ] & 3.41  & 3.09  & 2.50  & 3.47  & 3.09  & 4.49  & 3.57  & 4.36  & 3.74 & 4.70 \\
Peak Liner Velocity\tnote{$\dagger$}~~[m/s] & 170   & 165   & 147   & 167   & 177   & 242   & 188   & 223   & 200 &  255\\
Compression Start Time~[ms] & 3.505   & 2.519   & 3.714   & 7.012   & 2.509   & 2.112   & 2.013   & 3.516   & 3.518 &  2.514\\
\midrule
$C_R$ at 1st Touchdown\tnote{$\dagger$}  & 1.60  & 1.55  & 2.05  & 1.10  & 1.65  & 2.05  & 2.20  & 2.75  & 2.00 & 3.20 \\
$C_R$ at 2nd Touchdown\tnote{$\dagger$}  & 1.60   & 1.70  & 2.20  & 1.45  & 2.05  & 2.90  & 2.30  & 2.90  & 2.00 &  3.30   \\
\midrule
Normalized $B_\mathrm{pol}$ Increase   & 2.5  & 4.0  & 2.7  & 1.0  & 6.0  & $7.0^\dagger$  & 7.0  & 9.1  & 6.6  & 10.8\\
$C_R$ at Max $B_\mathrm{pol}$\tnote{$\dagger$}  & 1.82   & 2.25  & 1.95 & 1.0  & 2.35  & $2.6^\dagger$  & 2.52  & 2.83  & 2.46  &  3.05 \\
\bottomrule
\end{tabular}
\begin{tablenotes}
    \small
    \item[$\dagger$] Estimates taken from best-matching simulations reconstructed from experimental data. Liner velocity taken at plasma mid-plane.
\end{tablenotes}
\end{threeparttable}
\end{table*}

\subsubsection{Liner trajectory and imaging diagnostics} \label{sec:linerdata}

Tracking the evolution of the liner shape during implosion is achieved using a set of diagnostics that measure liner velocity and position, which is required for plasma compression modeling and to assess the performance of the machine. 

Photon Doppler Velocimetry (PDV) is used to measure the velocity of the inner surface of the lithium liner wall. A 12-channel PDV measurement system is implemented using $1\mathrm{\,W}$ fiber-coupled $1550\mathrm{\,nm}$ lasers, $13\mathrm{\,GHz}$ photodetectors, and oscilloscopes sampling at $1\mathrm{\,GS/s}$. Optical collimators are located on the center shaft and directed radially outwards towards the inner surface of the liner. There is an axial array and a toroidal array, which are both shown in Fig.~\ref{fig:LMC9_snapshot_with_chords}. Raw PDV signals are transformed to velocity-time spectrograms~\cite{Sirmas2025_DPB}.
Using the measured initial liner radius, the velocity can be integrated to get the instantaneous position of the moving liner.  The final trajectories reconstructed from PDV input data are in Fig.~\ref{fig:EquatorialTrajectoryCompare}. 

Imaging of the plasma confinement region is performed using in-vacuum lens assemblies coupled to color and monochrome high-speed cameras. Experiments can include up to three wide (approximately 90$^\circ$) field of view (FOV) assemblies. A pair of calibrated wide FOV assemblies coupled to color cameras is used to provide stereoscopic reconstruction of plasma structures. See Fig.~\ref{fig:CameraData_LMC9} for views of the plasma during LMC-9. These imaging views are situated at the edge of the plasma cavity at radii that are cut off by the liner at compression ratios of approximately 2.5--3.5 depending on the view and liner shape, so this data is only available for earlier compression.

The Structured Light Reconstruction (SLR) system is used to measure the liner inner surface position and shape with finer toroidal resolution than the PDV and consists of a projected laser pattern imaged by a calibrated wide FOV lens assembly (Fig.~\ref{fig:LMC9_snapshot_with_chords}). A 15~W, 637~nm multimode fiber-coupled laser source is guided into the center shaft. Light emitted from the fiber is collimated and then reflected from a multi-faceted diamond-turned aluminum optic that creates four arcs, initially covering 60° to 80° in toroidal range, at four axial positions. Laser light scattered from the lithium liner is viewed through the imaging assembly, coupled to a monochrome high-speed camera via an imaging fiber bundle and fitted with a 636/10 nm bandpass filter to enhance contrast and block most other plasma light. A stationary landmarking laser beam in the field of view is tracked to facilitate image stabilization. As the lithium liner moves inward, the arc positions are tracked and processed to reconstruct the liner trajectory using methods as described in \cite{Sirmas2025_DPB} and \cite{Mangione2024}. SLR is suitable for assessing the uniformity of liner collapse and measurement of liner perturbations with toroidal mode number equal to or greater than $n=6$. See Fig.~\ref{fig:drr_shot9} for output data from liner symmetry analysis using SLR data. 

Depending on the liner trajectory and surface characteristics, impurities may be ejected by jetting when the liner impacts the endplates.  A narrow FOV in-vacuum lens assembly is used to observe the liner where it contacts the formation-side endplate. In some compression experiments, light emitted by the plasma is used to illuminate the FOV to observe the presence or absence of ejecta created from liner impact. In other experiments, a laser scattering system is implemented to detect ejecta by projecting a laser line from the shaft outwards parallel to the formation-side endplate surface and perpendicular to the narrow FOV observation plane. A filter rejects most of the visible plasma radiation enabling recording of the scattered laser light to provide qualitative evaluation of ejecta. This system is called the Laser Scattering from Ejecta (LSE) diagnostic.

\subsubsection{Liner trajectory modeling and liner reconstruction} \label{sec:linerRecon}

\begin{figure}
    \centering
    \includegraphics[width=1\linewidth]{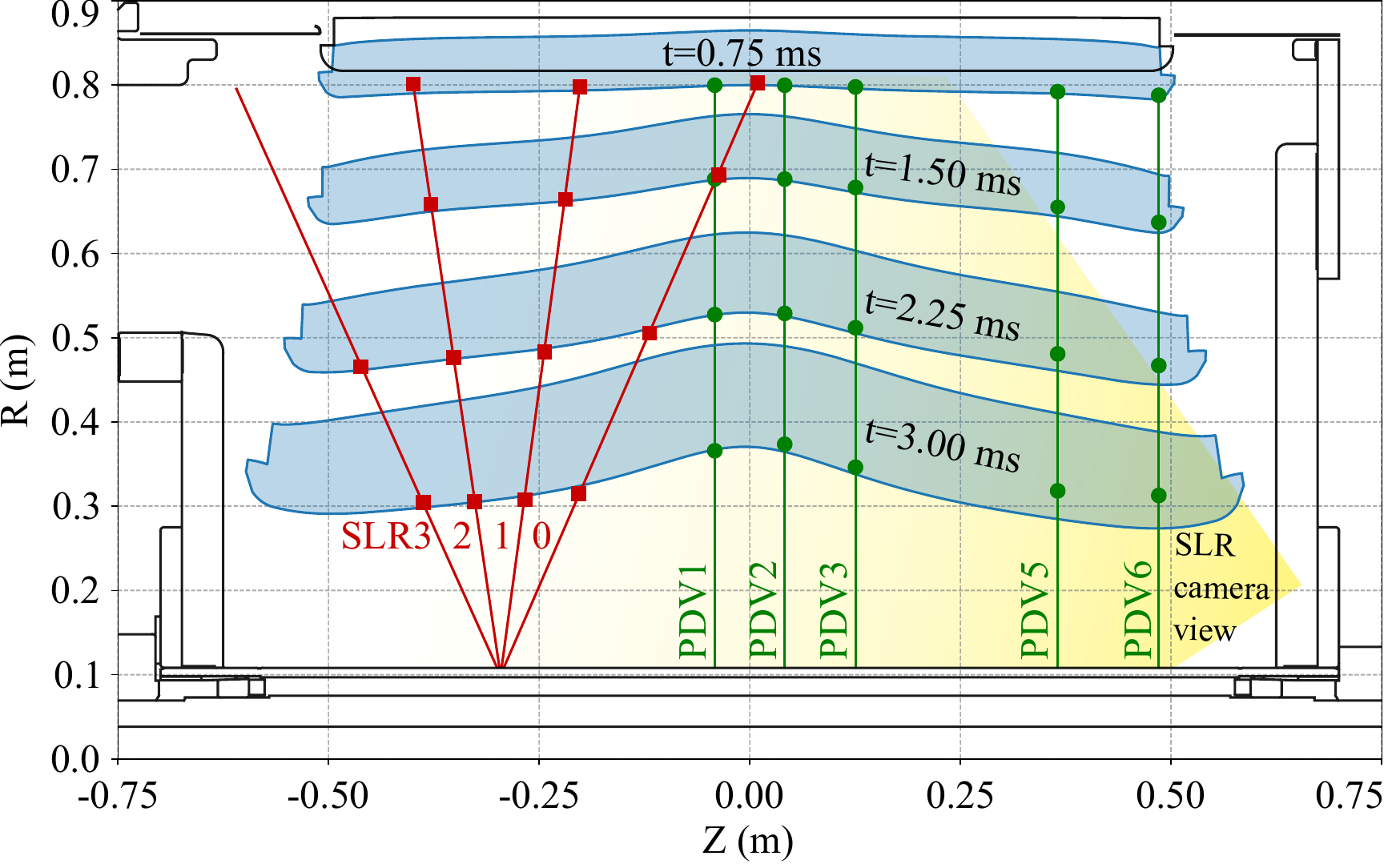}
    \caption{Fitted simulation trajectory for LMC-9 outlining inner and outer surfaces of the cross section of the liner as it approaches the machine axis (blue), PDV integrated position data points sampled from different toroidal angles (green), SLR reconstructed liner position, averaged over toroidal angle range of approximately 70--90 degrees (red). Time labels denote time since start of compression.}
    \label{fig:LMC9_snapshot_with_chords}
\end{figure}

Reconstruction of the liner trajectory for use in plasma compression modeling is performed using a combination of experimental measurements and simulations. Two-dimensional axisymmetric simulations are conducted using 
COMSOL Multiphysics, Ansys LS-DYNA, and a customized OpenFOAM solver. These models employ coupled electromagnetic, structural mechanics, and thermal solvers to capture the physics of the imploding liner. The simulation domain includes the liner, compression coils, and cavity geometries. Simulated coils are connected to external circuits that replicate the experimental setup, using measured voltages as inputs. Further details on the modeling can be found in \cite{Sirmas2025_DPB}, \cite{SuponitskyFluids2025}, and \cite{SuponitskyPamir2026}. Beyond liner profile reconstruction, the numerical tools also serve to optimize the shot conditions during the campaign, and evaluate the dynamic response of the overall assembly.

A sample trajectory obtained from simulation and compared with experimental measurements for \mbox{LMC-9} is shown in Fig.~\ref{fig:LMC9_snapshot_with_chords}, which includes liner position measurements from PDV and SLR. Measurements from PDV, shown by the green symbols, are obtained by using the known initial liner radius and integrating the measured velocity. For axial positions where PDV channels have toroidal redundancy, the average radial value is plotted.  The SLR data, indicated by the red markers, shows the radial liner position averaged over all toroidal positions of each laser arc at different compression times. Note that since the laser projection arcs are angled with respect to machine axis they provide liner radius at different axial positions as the compression proceeds. 

Trajectories obtained from COMSOL are used for plasma reconstruction. Nominal trajectories are obtained by applying the measured voltages into the model that contains the exact machine, liner dimensions, and baseline lithium material properties. The nominal trajectory is generally in agreement with the PDV and SLR measurements. Where necessary, the trajectory is refined through a Bayesian reconstruction method. The refinement is achieved by generating a database that sweeps through a physically reasonable range of parameters in the Johnson--Cook constitutive model for lithium. These parameters are meant to cover uncertainties in the metallurgy of the machined liners, specifically the coefficients related to strain hardening, strain rate hardening, and the initial yield stress.

Reconstruction of the liner trajectory includes a prediction for the distribution of the poloidal flux,  $\psi_0(R,Z,t)$, from the theta-pinch coils and the more slowly varying bias coils. Both the nonuniform currents induced in the liner and machine by these coils and advection as the liner moves are taken into account. However, the COMSOL simulations do not include plasma current.  This is justified by the fact that the plasma current and induced image currents only weakly affect the trajectory of the liner.  The effect of plasma current is accounted for in the next step of reconstruction, described in \ref{sec:EqRecon}.

\subsubsection{Liner implosion performance} \label{sec:liner_implosion_performance}

Key liner performance metrics are tracked for comparison between shots based on experimental measurements and supplemented by simulation results. Metrics include the maximum and time-resolved histories of the implosion velocity, as well as the timing, position, and velocity of the liner at plate touchdown. Some of these values can be found in Table~\ref{tab:shot_data}. Of particular interest is the trajectory at the mid-plane of the flux conserver, referred to as the \textit{equator}, with results generally presented in terms of the liner radial compression ratio at this location, $C_R(t) \equiv R_\mathrm{eq,0} / R_\mathrm{eq}(t)$, where $R_\mathrm{eq}(t)$ is the radius of the liner at the equator and $R_\mathrm{eq,0}=R_\mathrm{eq}(0)$. To better compare different shots, the results are also mapped in terms of $C_R$, rather than absolute time, normalizing out variations in implosion velocity. 

\begin{figure}
    \centering
    \includegraphics[width=\linewidth]{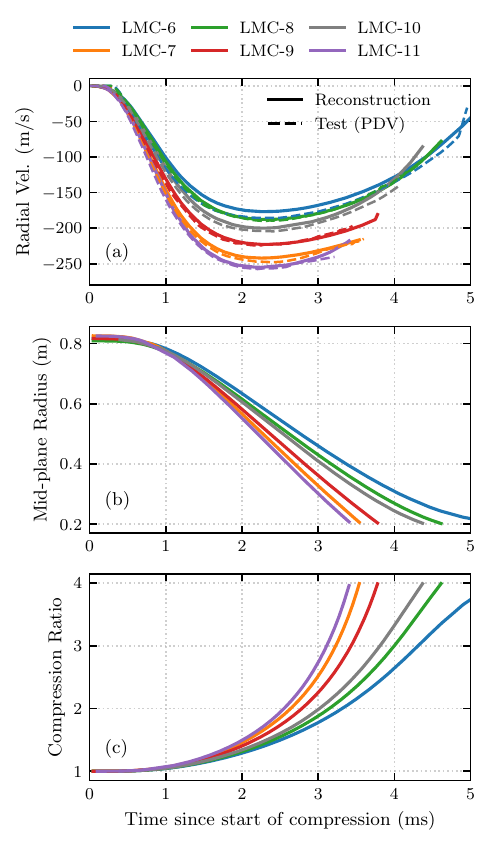}
    \caption{Comparison of reconstructed equatorial liner trajectories for shots LMC-6 through LMC-11, showing the evolution of (a) radial velocity, (b) radius,  and (c) compression ratio as functions of time since the start of compression. The reconstructed trajectories are obtained from simulations constrained by experimental PDV and SLR data. Dashed curves in the radial velocity plot are measurements from PDV, computed by taking the average PDV measurements within $\pm$50~mm of the equator.}
    \label{fig:EquatorialTrajectoryCompare}
\end{figure}

The trajectories at the equator for shots LMC-6 to LMC-11 are compared in Fig.~\ref{fig:EquatorialTrajectoryCompare}, with the evolution of radial velocity, radius, and radial compression ratio  up to $C_R=4$.    

LMC-11 was the fastest shot to date, utilizing a capacitor bank energy of 15.9\,MJ. For all cases, as shown in Table~\ref{tab:shot_data}, approximately 25--30\,\% of the energy supplied is transferred to the liner’s peak kinetic energy. For LMC-11, this leads to a peak radial velocity of approximately 255\,m/s, which was achieved by $C_R=1.5$. Higher velocities occur near the ends of the liner.

As the liner mass is increased, by increasing the length or thickness, and the energy supply is reduced, the trajectories become slower. \mbox{LMC-6} is an example of a slower compression, which reaches a lower peak radial velocity of 177\,m/s at the equator and achieves peak compression after 5 ms. With increased mass and lower peak velocities, these liners exhibit higher levels of deceleration prior to peak compression, eliminating impact with the shaft at this location, and not exceeding $C_R\approx4$.

Another key trajectory metric is the toroidal symmetry of the liner during compression. Evidence of asymmetry is present in all shots, as observed from the post-shot liner state. However, the final state provides limited information on the magnitude of asymmetry during implosion, because of the highly nonlinear nature of any buckling behavior. Additionally, contact between the liner and the shaft at peak compression modifies the inner surface topology, making it difficult to estimate the level of asymmetry present during implosion based on the final state. The limited number of toroidally redundant PDV channels also limits the ability to draw conclusions about liner symmetry.

Consequently, estimates on asymmetry are limited to using SLR data, which can be used as a representative estimate for the degree of symmetry due to its toroidal range spanning 70--90 degrees at varying axial position depending on compression ratio. A liner symmetry metric ($dR_\mathrm{max}={|R-\overline{R}|_\mathrm{max}}{/\,\overline{R}}$) is defined as the maximum variation of the reconstructed radius from the mean, divided by the mean radius. Because of the limited toroidal range, and uncertainty in the SLR reconstruction associated with calibration and alignment, it is not expected that distortions with toroidal mode number below $n=6$ can be reliably detected using this method. The current calibration alignment procedure is expected to calibrate out decentering of the liner associated with the $n=1$ mode. Due to resolution limitations and apparent width of the laser line signals it is also expected that high frequency modes will be detected to a reducing extent. 

SLR measurements are only available up to $C_R\approx2.5$ due to geometrical constraints limiting the depth to which the SLR signal is distinguishable and viewable, varying liner shapes obscuring the view at different compression ratios, and light from the plasma obscuring the SLR signal at different times. However, beyond $C_R\approx2.25$, close to the cutoff time of the diagnostic, fewer points are distinguishable, and measurements are not reliable for estimating $dR_\mathrm{max}$.

\begin{figure}
    \centering
    \includegraphics[width=1\linewidth]{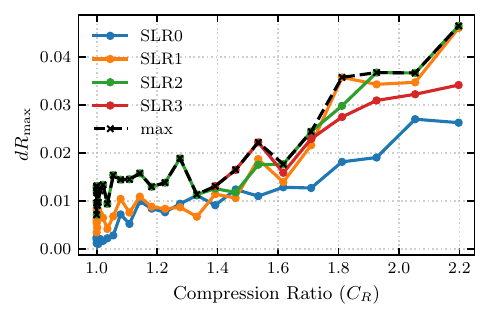}
    \caption{Measurements of liner symmetry ($dR_\mathrm{max}={|R-\overline{R}|_\mathrm{max}}{/\,\overline{R}}$) for LMC-9, obtained for different SLR chords (SLR0--3) and plotted against the radial compression ratio.}
    \label{fig:drr_shot9}
\end{figure}

For the representative case of LMC-9, $dR_\mathrm{max}$ is plotted in Fig.~\ref{fig:drr_shot9} for the different SLR chords, against compression ratio. Variation is due to the extraction uncertainty of the laser reflection lines from the image, and the varying toroidal range of discernible signal for extraction. The toroidal extractable range tends to reduce and the $dR_\mathrm{max}$ tends to increase as compression ratio increases, meaning that deviations  from the trend are more notable for later stages of compression ratio. It can be seen that $dR_\mathrm{max}$ remains below 5\,\% up to a compression ratio of~2.25, which is a general trend observed for other LMC shots as well. 

LMC-10 was designed to have the liner impact the endplates at an earlier compression ratio to test if there was ejecta from the liner created upon its impact with the plates. Using the laser scattering system, the liner was observed to travel across the narrow FOV during compression with no signs of ejecta or dust created upon impact traveling ahead of the liner.

\subsection{Overview of plasma diagnostics}\label{sec:diagoverview}
LM26 is equipped with a suite of diagnostics providing measurement of fundamental plasma parameters.  The flux conserver has Mirnov coils installed at eight unique radial positions on the endplates and five unique axial positions on the center shaft, all distributed across multiple toroidal angles to measure both poloidal and toroidal magnetic fields.  These are shown as red dots in Fig.~\ref{fig:LM26_schematic}. The sensors are installed inside metal wells with thin stainless steel covers, and a frequency-dependent sensitivity calibration is performed after installation to compensate the attenuation and filtering effects of the surrounding metal \cite{Pi3_tauE_NucFus_2025}.
Edge poloidal field data can be used to determine the total toroidal plasma current, and along with edge toroidal field data the profiles of current density and safety factor $q$ can be reconstructed. Toroidal modes up to $n=3$ can be determined with the array, providing information about transitions between MHD stable and unstable behavior. Magnetic field rise during compression is shown in Fig.~\ref{fig:LMC_Bpol_vs_CR}.\\

\begin{figure}
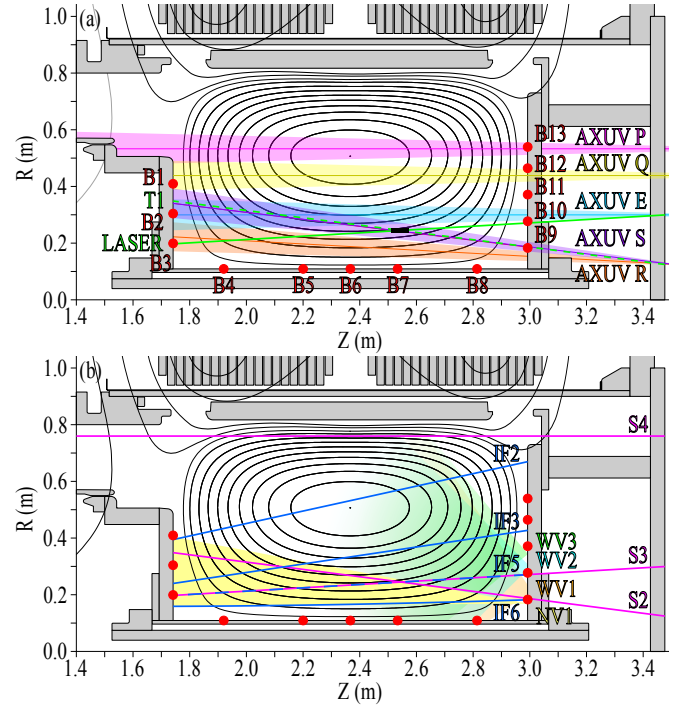

    \centering
    \includegraphics[width=1\linewidth]{images/LM26_LMC-11_diagnostics_AXUV_Mirnov_Thomson.pdf}
    \includegraphics[width=1\linewidth]{images/LM26_LMC-11_diagnostics_IF_IDS_Spect2.pdf}
    \caption{Location of diagnostics in LM26 overlaid on flux surfaces from an example uncompressed plasma. (a) Magnetic probes in red, Thomson scattering chords in green, and AXUV sight cones (P, Q, E, S, and R) in other colors. See Fig.~\ref{fig:AXUVmap} for more detail.  Thomson laser is the solid green line, T1 viewing chord is the dotted green line, and scattering volume is highlighted as black box. (b) Filterscope, survey spectrometer, and ion Doppler chords in magenta (S2-S4), interferometer chords in blue (IF2-IF6), and fast camera view cones in other colors (wide view WV1--3, narrow view NV1).  The mid-plane is at $Z=2.37$~m.}
    \label{fig:DiagMap}
\end{figure}

To measure plasma density, a two-color HeNe/$\mathrm{CO_2}$ heterodyne interferometer (IF) which measures plasma density along four chords is employed. As illustrated in Fig.~\ref{fig:DiagMap}, IF2/3/5/6 provide line average density measurement through the plasma. During compression the core of the plasma starts near IF2 and passes sequentially through each of the chords until it typically reaches the IF6 chord near the point of maximum compression. Due to access limitations, IF2 uses an internally mounted retroreflector located within the center electrode, as having a through-chord was not possible. The rest of the chords are through-beams with external optics.\\ 

\begin{figure}
    \centering
    \includegraphics[width=1\linewidth]{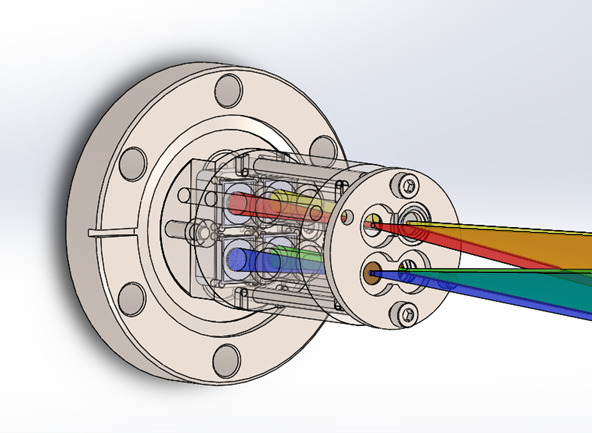}
    \caption{Detail of AXUV filtered photodiode array assembly. In the $2\times2$ array, side-adjacent diode pairs can have views of the plasma that overlap between $71.9\,\%$ and $83.4\,\%$ at the plasma's mid-plane, while diagonal pairs have between $62.5\,\%$ to $75.4\,\%$ overlapping area at the mid-plane. When gradients of plasma emissivity are small compared to the distance between non-overlapping components of the view cones, (such as when the core of the plasma is passing through the view), the ratio of filtered signals with different SXR energy cut-offs can provide 0D estimates of chord-averaged $T_e$. Ratios with significant gradients can still be used to constrain thermal models of the plasma via time-dependent reconstruction with the Integrated System Model (ISM).}
    \label{fig:AXUVDiagnosticView}
\end{figure}

\begin{figure}
    \centering
    \includegraphics[width=1\linewidth]{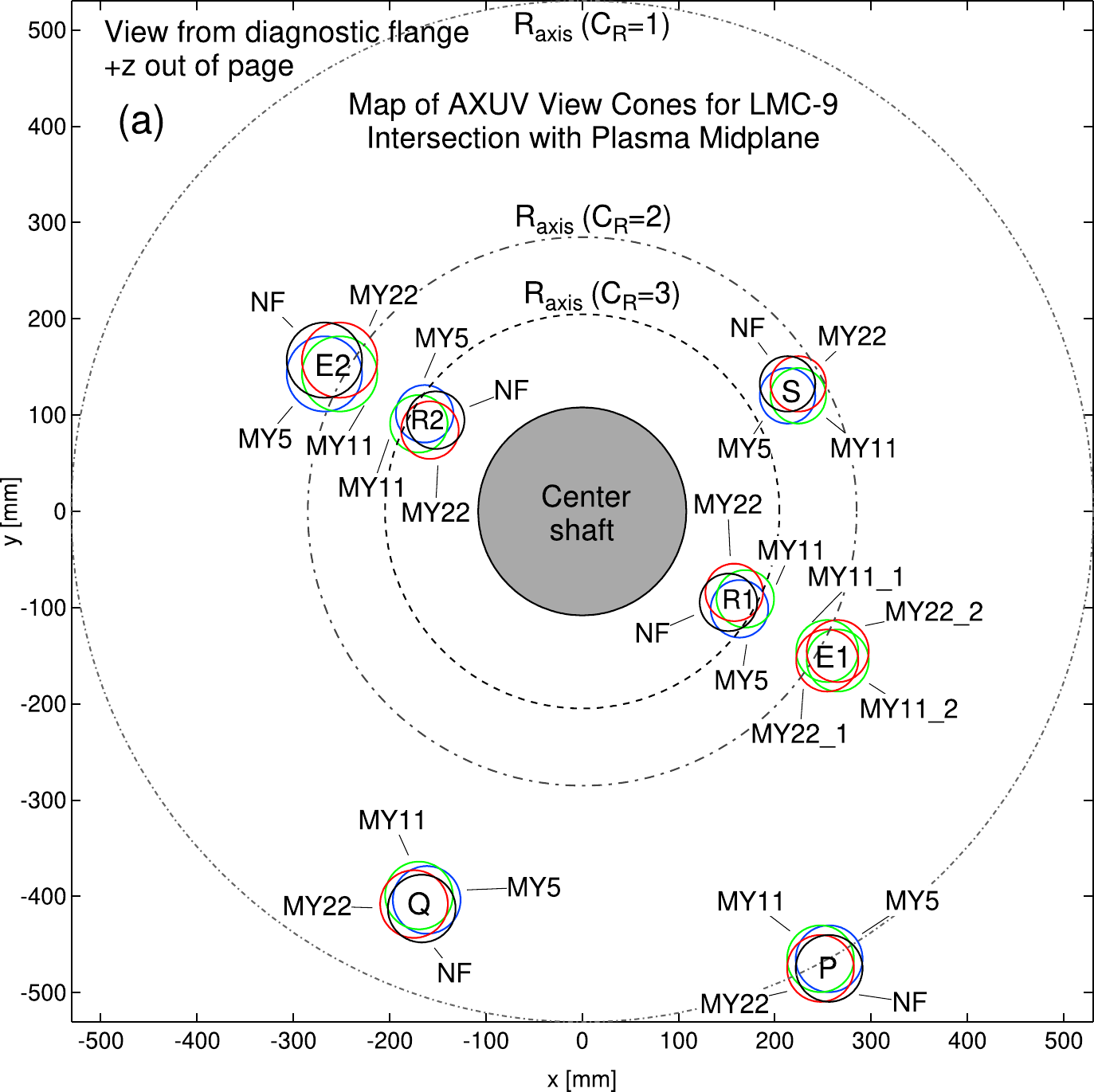}
    \includegraphics[width=1\linewidth]{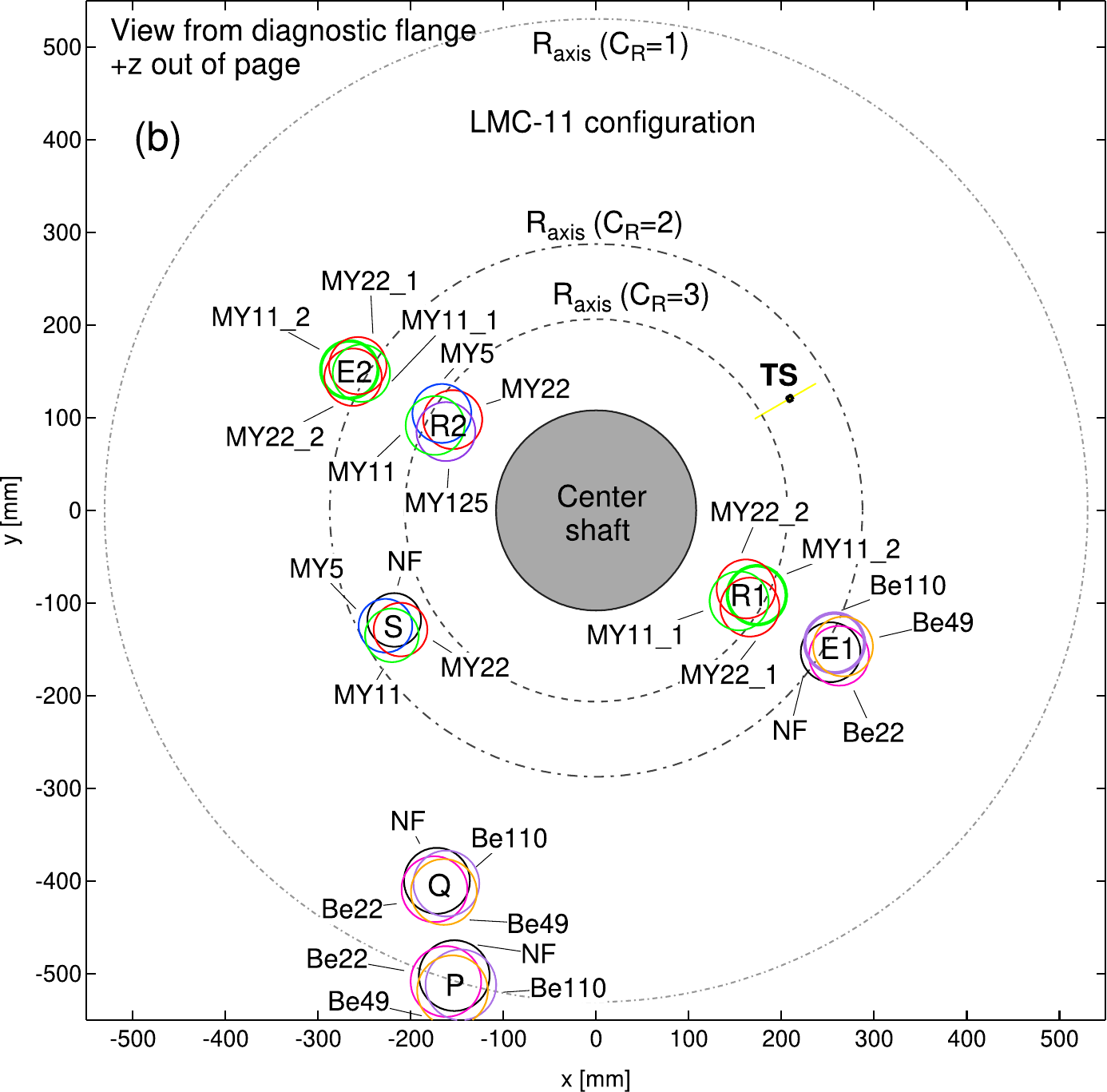}
    \caption{Maps of individual AXUV view cones as they intersect the mid-plane of the plasma are shown for the LMC-9 (a) and LMC-11 (b) configurations. Each campaign made a series of improvements to the AXUV detector electronics and filter combinations. The filter types are labeled as NF (black) for no filter, with Mylar filters MY5 (blue), MY11 (green), MY22 (red), and MY125 (lavender). Starting on LMC-10 beryllium filers were used (Be22, Be49, Be110). The number in the label indicates the filter thickness in {\textmu}m. When there are two of the same type of filter for gradient comparison (e.g. LMC-9 E1) the filter labels are enumerated 1, 2. For context, the radial locations of the magnetic axis of the plasma are shown for $C_R = 1, 2, 3$. On LMC-11 Thomson scattering was installed, laser path shown in yellow with scattering volume in black. }
    \label{fig:AXUVmap}
\end{figure}

Multi-color X-ray detectors are used to infer electron temperature ($T_{e}$) from the ratio of soft X-ray radiation incident on filtered absolute extreme ultraviolet (AXUV) diodes with different filter thickness and thus different cutoff energies in the soft X-ray (SXR) range, see Fig.~\ref{fig:AXUVDiagnosticView}. For two filters of the same material viewing a plasma with uniform $T_{e}$ and $n_{e}$, the ratio is a monotonic function of $T_{e}$ \cite{Pi3_tauE_NucFus_2025, Delgado2007,Howard_2025}. Seven of these detector arrays are installed in a spanning set of roughly axial view cones. Each detector consists of a $2\times2$ array of AXUV diodes each of which can be configured with any of a set of different thickness of aluminum-coated Mylar (5.47, 11.0, 22.1, or 125 {\textmu}m) or beryllium foil (21.6, 49.3, 110.0 {\textmu}m) or with no filter to collect all light from visible to hard X-ray as a broadband detector. These different filter types will be denoted MY5, MY11, MY22, MY125, Be22, Be49, Be110, and NF (no filter) throughout the rest of this paper. More details on the AXUV measurement position can be seen in Figs.~\ref{fig:DiagMap}, \ref{fig:AXUVmap}, and a more complete accounting of operational assumption, analysis methods and observations are described in Sec.~\ref{sec:AXUVresults}, and a full description of error analysis is given in Appendix~\ref{App:A}. \\

\begin{figure}
    \centering
    \includegraphics[width=1\linewidth]{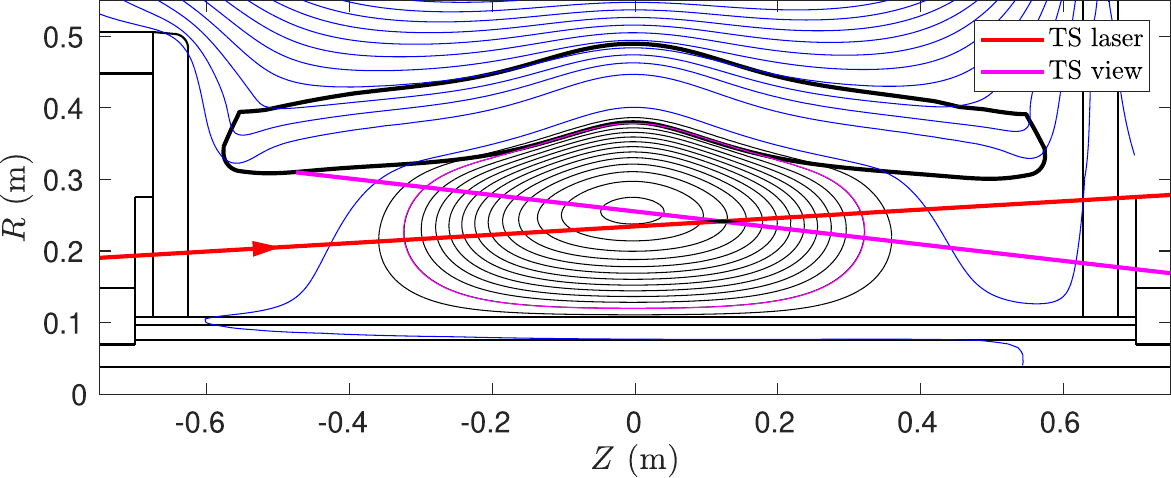}
    \caption{Thomson scattering laser and view line, with scattering volume at the intersection highlighted in black. Liner and poloidal flux contours shown for compressed plasma in LMC-11 at the time of the last Thomson pulse ($t=5.211\mathrm{\,ms}$).}
    \label{fig:ThomsonChords}
\end{figure}

A Thomson scattering (TS) $T_e$ measurement was added for LMC-11. The geometry is shown in Fig.~\ref{fig:ThomsonChords}. A 1064\,nm, 10\,ns beam is sent through the machine from gun to recovery side using a repurposed interferometer through-chord. The laser operates in burst mode which can deliver four 1.9\,J pulses separated by 250\,{\textmu}s each. The collection optics are mounted on a concrete pillar to provide mechanical isolation from machine motion. The 75\,mm diameter, $f=500$\,mm, plano-convex lens images a 3\,cm length of the 5\,mm diameter laser into a $3\times2$\,mm cross section, 30\,m long imaging fiber. The scattering volume is 1.9\,m from the lens and about 115\,mm off the cavity midplane towards the recovery flange. The scattering angle between the laser chord and view is about 10 degrees, which significantly narrows Thomson spectral broadening. For a 1\,keV plasma the broadening would extend to only 1040\,nm on the blue side of the laser line. The fiber carries the collected light to a UKAEA polychromator designed for low broadening measurements. Above about 350\,eV, the broadening is measurable in three polychromator channels. The polychromator spectral response was measured with a pulsed white light source, a monochromator, and an absolutely calibrated detector. 

The LM26 spectroscopy system contains five survey spectrometers and 20 filterscope channels. Survey spectrometers are employed to measure the emission lines in a range of 200--1000\,nm with a spectral resolution of 0.3--0.6\,nm/pixel and a temporal resolution of 0.5\,ms/frame. Filterscopes are used to measure the temporal evolution of emitted light within a set of wavelengths using narrow band-pass filters in front of avalanche photodiodes. The wavelengths measured include 523.2--523.8~nm (no associated lines), 548.5\,nm (Li\,II) 610.3\,nm (Li\,I), 656.1\,nm ($\mathrm{D\alpha}$), and 670.1\,nm (Li\,I). 
The location of the three filterscope chords (S2, S3, S4) can be seen on Fig.~\ref{fig:DiagMap}.

Two chords of ion Doppler spectroscopy (IDS) data are collected and used to measure the ion temperature of the plasma by analyzing C\,V impurity line radiation at $\lambda = 227$\,nm. Each system makes use of a high-resolution spectrometer (Horiba IHR500) with a 16-channel Hamamatsu photomultiplier tube linear array as a detector, each channel is digitized at 10\,MHz sampling rate. Doppler shift and broadening of the line can then be determined as a function of time in a processing code.

In addition to IDS, a deuterium ion temperature measurement is derived from neutron counting: a measured neutron yield, combined with the interferometric electron density measurements, is inverted through the Bosch--Hale reactivity \cite{Bosch_Hale_1992} of the $\mathrm{D(d,n){}^3He}$ fusion reaction to extract $T_{i}$ 
\cite{radichIonTemperatureInference2026a}.
Neutrons are counted with eight liquid scintillators (Eljen EJ-301 and EJ-309) coupled to photomultiplier tubes, with active volumes from 830 cm$^{3}$ to 3375 cm$^{3}$ placed at various distances from the plasma to ensure a large dynamic range of pulse rates. Each scintillator is housed in a steel enclosure with lead plates that suppress electronic noise and partially shield against the ambient gamma flux; the residual gamma component is rejected by pulse-shape discrimination (PSD) on the digitized waveforms. On compression shots, all detected pulses occurring more than 1~ms after formation have their PSD identification manually reviewed to ensure no noise spikes, gamma pulses with large noise fluctuations, pile-up pulses, or other pulse-shape altering features are counted as neutrons. 

\section{Experimental observations}\label{sec:ExObserve}

In the following sections, we discuss the key experimental observations and results obtained during compression, including increases in poloidal field, density, and temperature. 

\subsection{Poloidal field increase during compression}

The poloidal field $B_\mathrm{pol}$ that confines the plasma greatly increases during compression as the near-constant amount of poloidal flux is compressed into a smaller and smaller cross sectional area. Poloidal flux conservation within the plasma during compression can be a direct indicator of plasma core resistivity, and indirectly an indicator of temperature. For a given resistivity model, a relationship between the evolution of $T_e(t)$ and the poloidal flux $\psi(t)$ can be obtained  \cite{Howard_2025} if it is assumed that the current profile in the plasma remains invariant under compression. As our plasma is well explained by resistive MHD, significant (worst-case) cooling of the plasma always shows up immediately as a rapid resistive downturn in $\psi(t)$, which is observed as a drop in $B_\mathrm{pol}(t)$ at a fixed central point towards which the plasma is being compressed. The B6 Mirnov probes in the LM26 shaft, as shown in Fig.~\ref{fig:DiagMap}, are located at this central point. Whereas with the ideal, perfectly flux conserving case, the poloidal magnetic field will keep increasing, in the case of constant current profile this will be $B_\mathrm{pol}(t)/B_\mathrm{pol}(0) \sim C_R^2(t)$, as long as plasma continues to be hot enough to have a resistive decay time much longer than the remainder of the compression time.  

\begin{figure}
    \centering
    \includegraphics[width=1\linewidth]{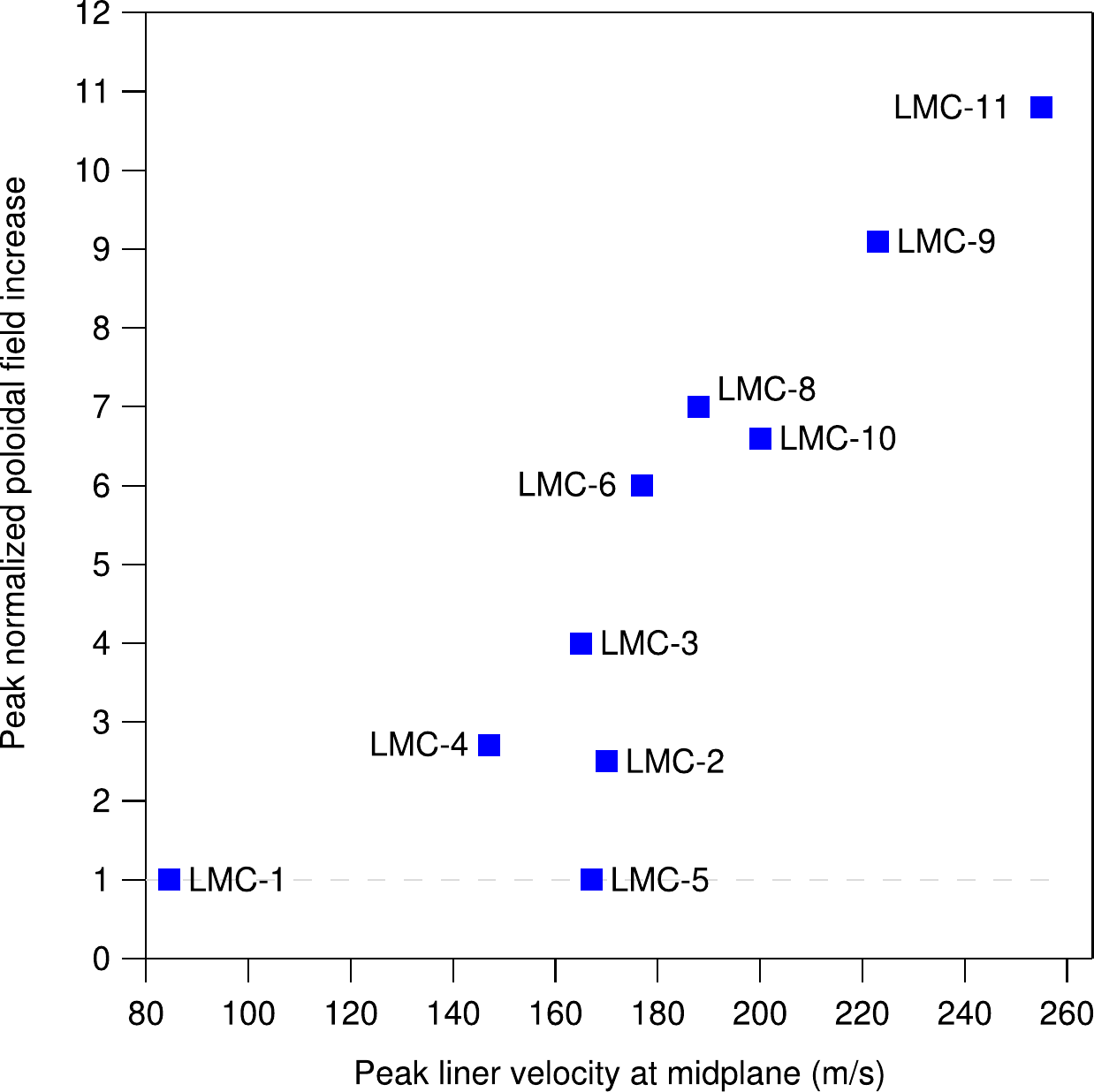}
    \caption{Trend of increase of peak $B_{pol}$ with increasing peak liner velocity at the mid-plane. The case of no $B_{pol}$ increase is the $y=1$ dashed line.}
    \label{fig:Bpolinc_vs_vliner}
\end{figure}

\begin{figure}
    \centering
    \includegraphics[width=1\linewidth]{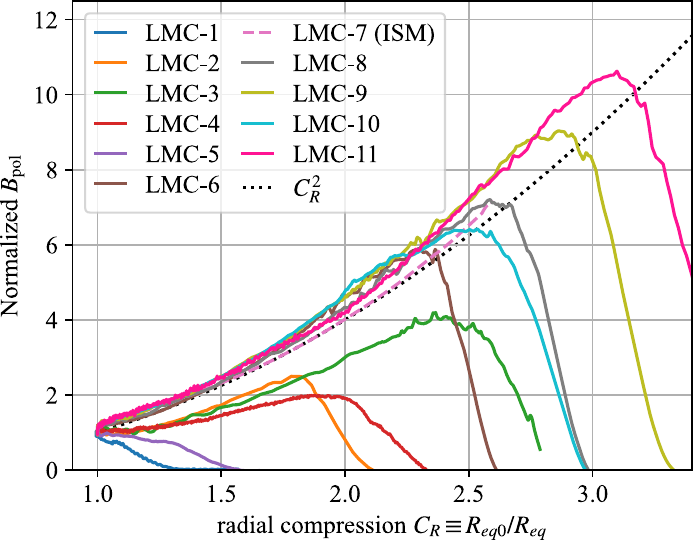}
    \caption{The poloidal field increase from the magnetic probes closest to the center of the shaft (B6 or B5, see Fig.~\ref{fig:DiagMap}) for all compression shots on LM26, as a function of compression ratio. For LMC-6 to LMC-10 the curves shown are the toroidal average of the multiple Mirnov signals at the B6 position.  The field rises, driven by compression while conserving poloidal flux until  near a maximum compression ratio.  The curve for LMC-7 is based on an ISM fit to the endplate magnetic probes because the shaft probe data was not available. The rise above the ideal flux conserving $C_R^2$ trend is due to a broadening of the internal plasma current profile during compression, seen in reconstruction in terms of the decrease of internal inductance $l_i$ shown in Fig.~\ref{fig:LM26_quants_shot9}c.}
    \label{fig:LMC_Bpol_vs_CR}
\end{figure}

What has been observed in the LM26 experiments is a general trend in improvement in poloidal flux conservation, as measured by the poloidal field rise relative to its starting value. This trend in improvement follows from achieving better initial plasma conditions and running at higher liner compression velocities, as highlighted in Fig.~\ref{fig:Bpolinc_vs_vliner}.  The poloidal field behavior is shown in detail in Fig.~\ref{fig:LMC_Bpol_vs_CR}. LMC-11 and LMC-9 had the highest poloidal field increases of any Li liner compressions to date and maintained their flux to a larger $C_R$ value before a final thermal crash. LMC-10 was similar to LMC-8, which had lower compression velocities. LMC-7 has some indications from AXUV and camera data that the plasma survived to significant $C_R$ values. However, shaft Mirnov data was not successfully recorded on LMC-7 due to a higher-than-normal transient electrical surge on the signal cables during the compression shot.  

The observation that most of the compression shots exceeded the ideal trend with $B_\mathrm{pol}(t)/B_\mathrm{pol}(0) > C_R^2(t)$, is explainable in terms of the current profile becoming more hollow during the shot, whereby the radial distribution of flux is more concentrated toward the edge, giving a higher magnetic field strength than if the original profile had been maintained.  This interpretation is supported by BPR and ISM analyses that indicate a decrease in internal inductance $l_i$ during the compression, see Fig.~\ref{fig:LM26_quants_shot9}c.

\subsection{Density increase during compression}

\begin{figure}
    \centering
    \includegraphics[width=1\linewidth]{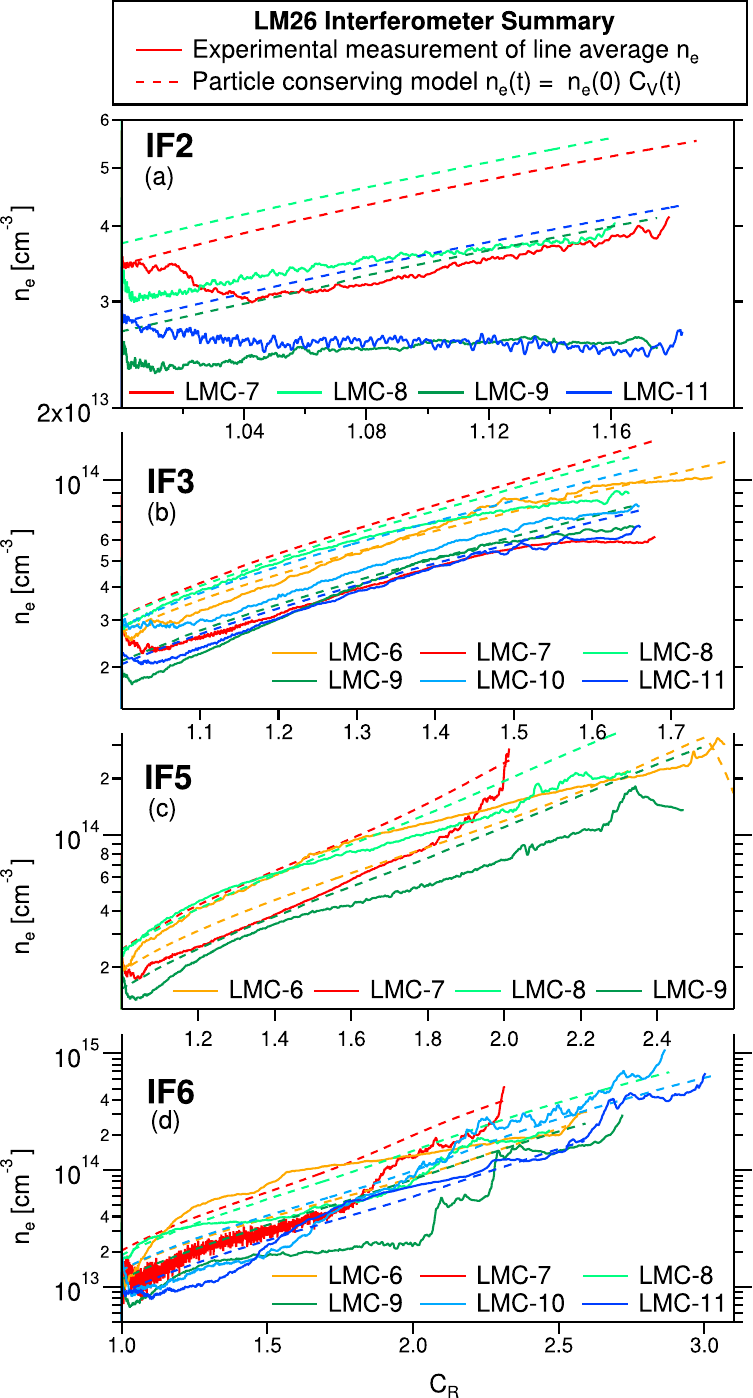}
    \caption{Vibration-compensated line average density of each IF chord vs compression ratio $C_R$ compared across relevant LM26 compression campaigns. The solid curves show experimental interferometer signals, while the dashed curves show the corresponding volumetric compression model $n_e(t) = n_e(t=t_{comp}) C_V(t) \sim n_e(C_R=1) C_R^3$, which serves as a simple guide to detect if the plasma seen by that IF chord has its density falling significantly below, or rising above the constant-inventory curve. Most, but not all shots begin with a drop in density below this simple model, followed by a rise near the end which could be due to impurity influx. $C_V$ is determined from an estimate of the plasma volume as a function of time during the compression, using a Taylor state magnetic model calculated as part of liner trajectory reconstruction. }
    \label{fig:IFsummary}
\end{figure}

The line average density is observed to increase during compression, see Fig.~\ref{fig:IFsummary}. 
For the IF6 chord, inboard of the magnetic axis at the start of compression, the line average density increases as the plasma is compressed and moved onto the chord.
The general behavior of density evolution during compression can be explained in terms of some variable amount of particle loss from the edge region in addition to nearly particle-conserving scaling with inverse volume near the core, followed late in compression by what is likely a net particle inventory increase from influx of impurities from the boundary near the end phase of compression. 
\begin{figure}
    \centering
    \includegraphics[width=\linewidth]{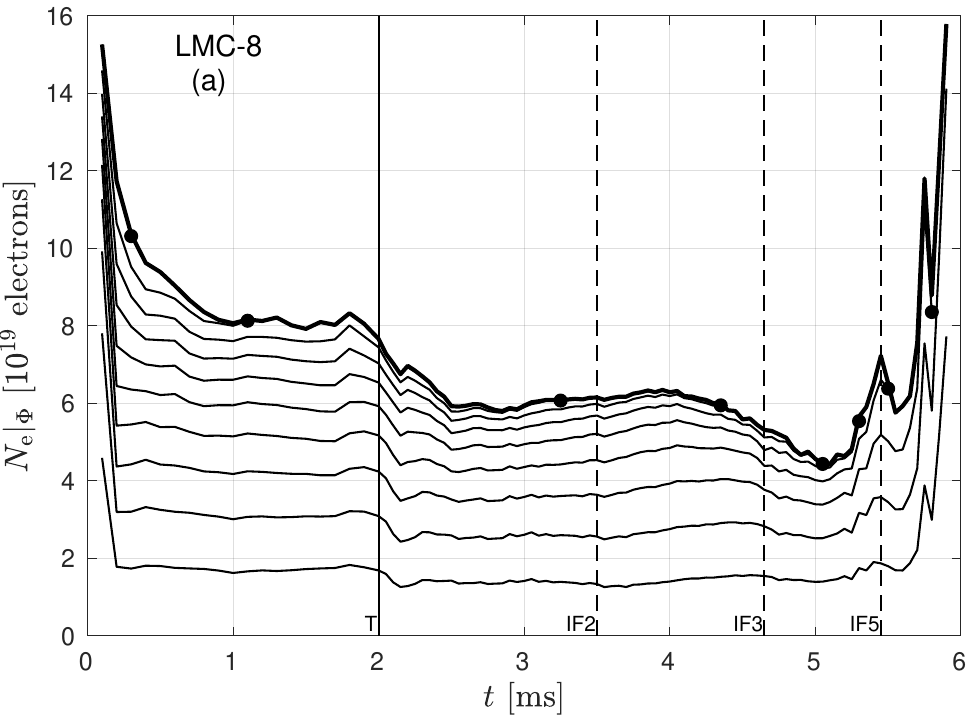}
    \includegraphics[width=\linewidth]{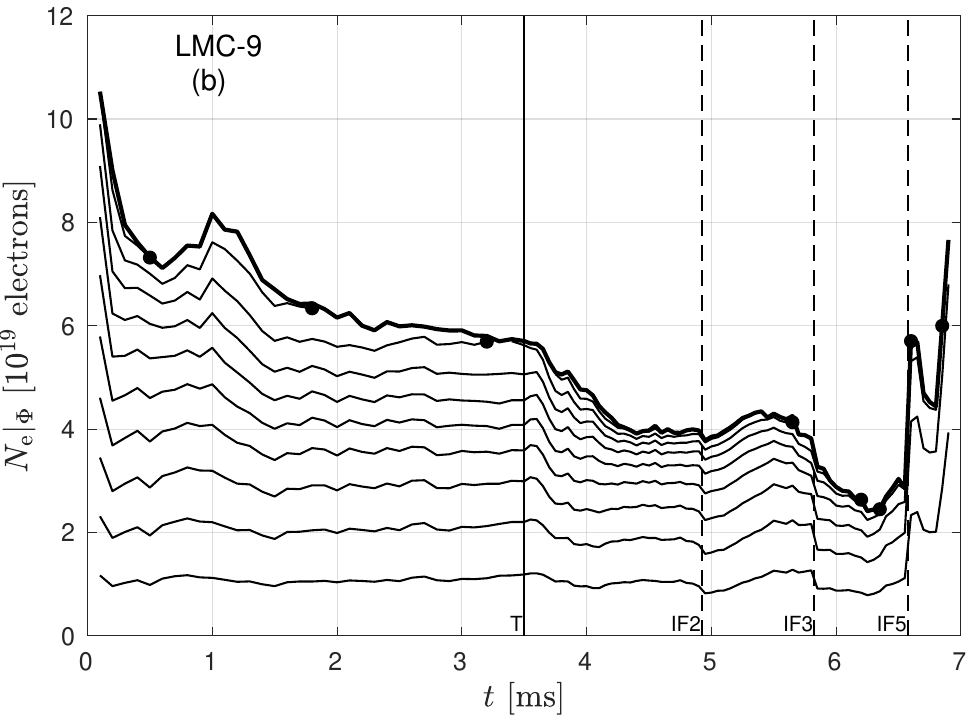}
    \includegraphics[width=\linewidth]{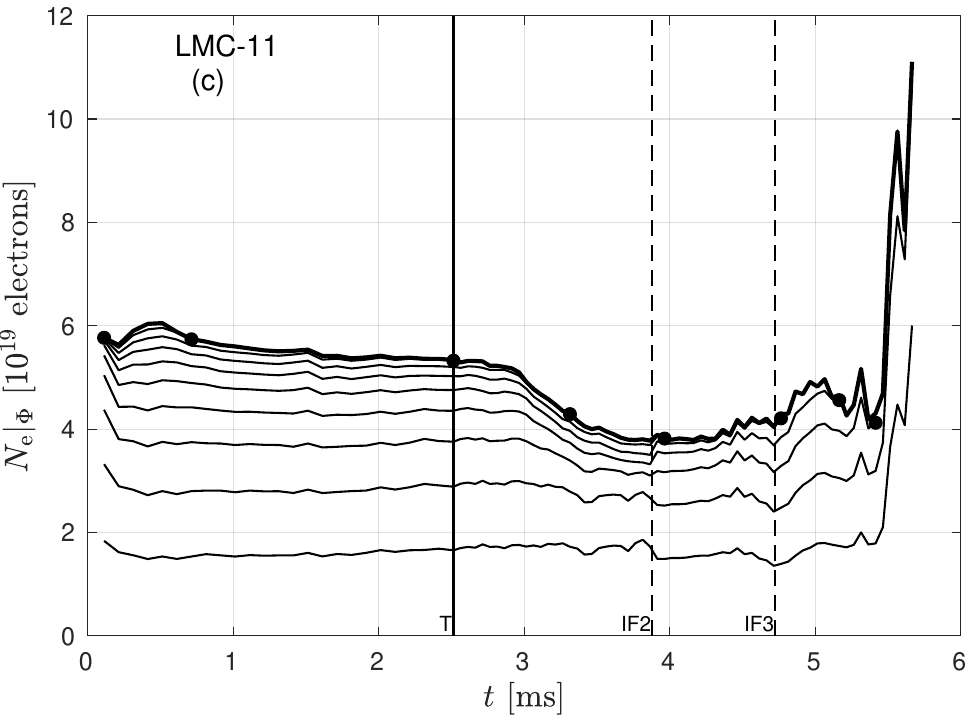}
    \caption{Flux surface resolved electron inventory versus time for LMC-8 (upper panel), LMC-9 (middle panel), and LMC-11 (lower panel). The vertical solid line labeled ``T'' indicates the compression trigger time, and the vertical dashed lines (IF2, IF3, and, except in LMC-11, IF5) indicate the times when interferometers are removed from the density fit.}
    \label{fig:lmc-8-inventories}
\end{figure}

The full set of data from interferometry provides information on the electron density profile and total electron inventory.  
The profile is obtained by fitting data from multiple interferometers making use of flux surface geometry reconstructed from Mirnov data (described in Sec.~\ref{sec:PsiBC}).
We illustrate this reconstruction in Fig.~\ref{fig:lmc-8-inventories}, which shows the flux surface resolved electron inventory versus time, as well as the total inventory contained within the LCFS, for LMC-8 (upper panel), LMC-9 (middle panel), and LMC-11 (lower panel).    Thin curves show total inventory within flux surfaces identified by constant toroidal flux $\Phi$, with spacing 10\,\% of the toroidal flux initially contained within the LCFS.  The envelope of these curves, the heavy curve, represents the total inventory within the LCFS.  Dots indicate times when curves terminate because their flux surface falls outside the LCFS, e.g., due to ``flux soak'', the resistive diffusion of magnetic flux into the liner and walls.
Compression trigger time, $2.0\mathrm{\,ms}$ for LMC-8, $3.5\mathrm{\,ms}$ for LMC-9, and $2.5\mathrm{\,ms}$ for LMC-11, is indicated by the vertical solid line labeled ``T''.  The times when interferometers are removed from the density fit are indicated by vertical dashed lines (IF2, IF3, and IF5).  LMC-11 does not have IF5 because that port was repurposed for the TS diagnostic.
As the liner moves inward it sequentially blocks the set of interferometer chords and so we exclude interferometer data after its cutoff time from the fitting process to prevent the result becoming unreliable on account of this.  Considering the case of LMC-8, we have four chords, spanning the plasma from core to edge. The figure shows that for LMC-8 the inventory settles down solidly after the CHI formation process, becoming quite constant after $t=1\mathrm{\,ms}$ ($t$ is the time after plasma formation).
A drop in inventory begins abruptly at $t=1.8\mathrm{\,ms}$, about $0.2\mathrm{\,ms}$ before compression is triggered.  Determination of the inventory distribution within the plasma is naturally of limited spatial resolution with only four interferometers, but the data does indicate that inventory loss occurs from the edge, and not globally.  After this drop the inventory is again quite stable, until a second inventory drop starts gradually at about $t=4\mathrm{\,ms}$.  
With only three interferometers the spatial resolution is reduced, but we can notice that losing the IF3
interferometer does not cause any discontinuity in the result.   Inventory starts to rise after $t=5\mathrm{\,ms}$, probably due to an influx of impurities at the edge of the plasma.  At this time we have reliable data from only two interferometers (IF5 and IF6)
and profile reconstruction becomes more uncertain.  After the IF5
cutoff we are left with only the IF6
interferometer and profile reconstruction is no longer possible.  In this final period we simply assume the profile shape is no longer changing and hold the profile shape determined just before IF5 is cut off.  
In the case of LMC-9 (middle panel in Fig.~\ref{fig:lmc-8-inventories}), the behavior during compression is similar, with an inventory loss shortly after compression is triggered and again later in compression.  In the case of LMC-11 (lower panel) there is a similar inventory loss event observed shortly after compression is triggered (there are too few interferometers to indicate density profile after the IF3 cutoff). 

The observation of this drop in electron inventory occurring at near compression trigger time seems to be mostly coincidental in nature since it is not limited to just compression shots and has a moderate degree of repeatability to it. Non-compression and LMC shots with similar system control parameters will tend to have an inventory drop event at around this time, and it even sometimes happens just before the compression is triggered, such as in LMC-8. So the logical conclusion is that this is a natural intrinsic behavior of the plasma and has little to do with the start of compression.  The density profile reconstruction indicates that the inventory is lost from the edge-most flux surfaces, and not from the plasma core. Further information about this plasma reorganization event can be determined from the AXUV data and analysis of the crash inversion radius in section \ref{sec:AXUV_Rinv}.

The increase in plasma density during compression (Fig.~\ref{fig:IFsummary}) is approximately a factor 10 before interpretation of interferometric measurements is complicated by the increase in inventory attributed to impurities entering the LCFS.  For example, the core density determined by interferometric reconstruction increases by a factor of \{10, 8, 10\} in LMC-8, LMC-9, and LMC-11, from density measured at 5.60\,ms, 6.55\,ms, and 5.36\,ms, respectively.

\subsection{AXUV-crash inversion radius}\label{sec:AXUV_Rinv}
When looking at the set of non-filtered AXUV channels, which span the radial extent of the inboard plasma, the effective total emissivity profile can be determined.  

\begin{figure}
    \centering
    \includegraphics[width=1\linewidth]{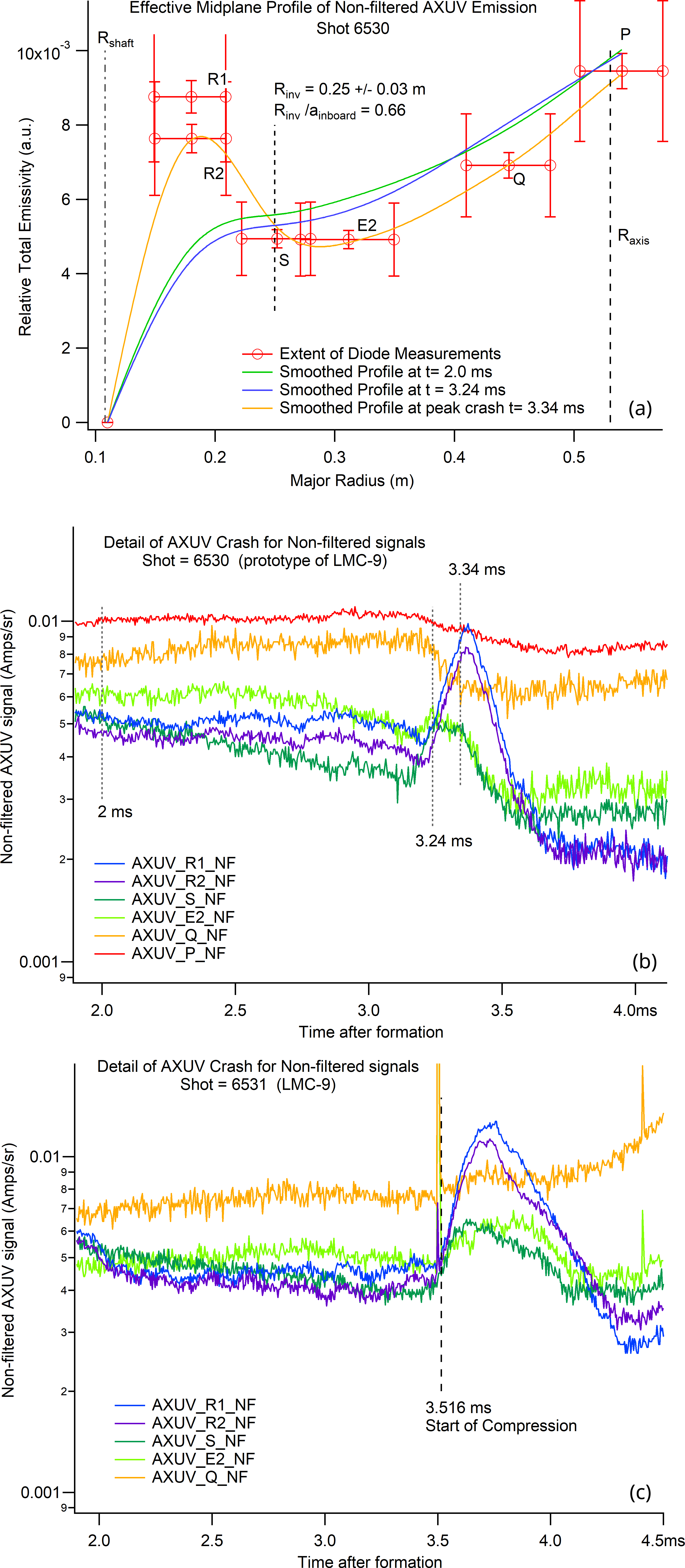}
    \caption{Observation of X-ray emissivity profile during a crash event.}
    \label{fig:AXUVcrashProf}
\end{figure}

In Fig.~\ref{fig:AXUVcrashProf}, there are radial profiles obtained at three different times in the plasma evolution, using a smoothing spline applied to the set of NF (no filter) AXUV signals on the LMC-9 prototype shot 6530, which used all the same control settings as LMC-9 (6531), except that 6530 did not have the compression driver banks or shaft ramp circuit triggered. 
The radial extent of each position of AXUV measurement is included in the graph as trios of three linked points, the center of which is located at the center of the mid-plane footprint of the view cone and is given a $10\,\%$ uncertainty, while the edge points of the view cone footprint are given $20\,\%$ uncertainties for the smoothing spline fit. 
The data from the AXUV NF measurements at $t = 3.34  \mathrm{\,ms}$ during the peak of the X-ray crash are shown in red, which then generates the solid orange smoothing spline profile. The green and blue profiles follow the same method but from using AXUV NF measurements taken at 2.0 and $3.24  \mathrm{\,ms}$ respectively. 
These sample times are shown as vertical lines in the middle graph of the raw signals for shot 6530. For the first $1  \mathrm{\,ms}$ after plasma formation the total emissivity profile is somewhat hollow, but then by $2  \mathrm{\,ms}$ has become centrally peaked and stays peaked for the remainder of the discharge except when the periodic AXUV crashes are happening (see Fig.~\ref{fig:Precompression} for indications of periodic behavior on multiple diagnostics). 
During the crash event regions near the core decrease in X-ray brightness, while regions at larger minor radius than the inversion radius will temporarily increase in brightness. For this first crash beginning at $t = 3.25 \mathrm{\,ms}$, the inversion radius is $R_\mathrm{inv} = 0.25 \pm 0.03\mathrm{\,m}$, which corresponds to a normalized minor radius of $r/a=0.66$. 
This can be compared to available BPR data for shot 6530 (see Sec.~\ref{sec:BPR} for methods) where the reconstructed safety factor $q(R)$  value at $t=3\mathrm{\,ms}$ is found to be $q(0.25\mathrm{\,m}) = 3.3 \substack{+1.0 \\ -0.8}$, while after the crash at $t = 4\mathrm{\,ms}$ it has dropped a moderate amount to $q(0.25\mathrm{\,m}) = 2.7 \substack{+0.7 \\ -0.6}$, so it is plausible that this inversion radius might be near the $q = 3$ surface. 
This MHD event is most likely related to the loss of electron inventory from the outer edge of the plasma just after compression begins on LMC-9 (see Fig.~\ref{fig:lmc-8-inventories}) and other shots. 
It also coincides with an $n=0$ increase of all $B_\mathrm{pol}$ signals corresponding to a small net increase in total plasma current $I_\text{p}$, see Fig.~\ref{fig:Precompression}.

The nature of this plasma reorganization is of considerable interest as it may constrain the initial conditions of the plasma just prior to the start of the compression. Notably, the reproducibility of the plasma formation method was sufficiently precise that when shot 6531 was initiated it returned to a similar state, with the same first crash possibly delayed by only 300\,{\textmu}s, but was then triggered to begin with the action of the compression field directly modifying the plasma boundary through the lobes of flux fringing around the ends of the liner (see Fig.~\ref{fig:psibc-geometry}). In this case there was still an increase in brightness in the outer region of the plasma, followed by a drop, however the measurement closest to the core (AXUV Q) only increases in brightness in this early compression phase before significant motion of the Li liner has happened, possibly due to the direct magnetic compression effects. For shot 6531 there was no data recorded from the AXUV P array due to rapid mechanical damage to the diode electronics from the liner impact shock. 

\subsection{Increase of X-ray emission and \texorpdfstring{$T_e$}{Te}}\label{sec:AXUVresults}

\begin{figure}
    \centering
    \includegraphics[width=1\linewidth]{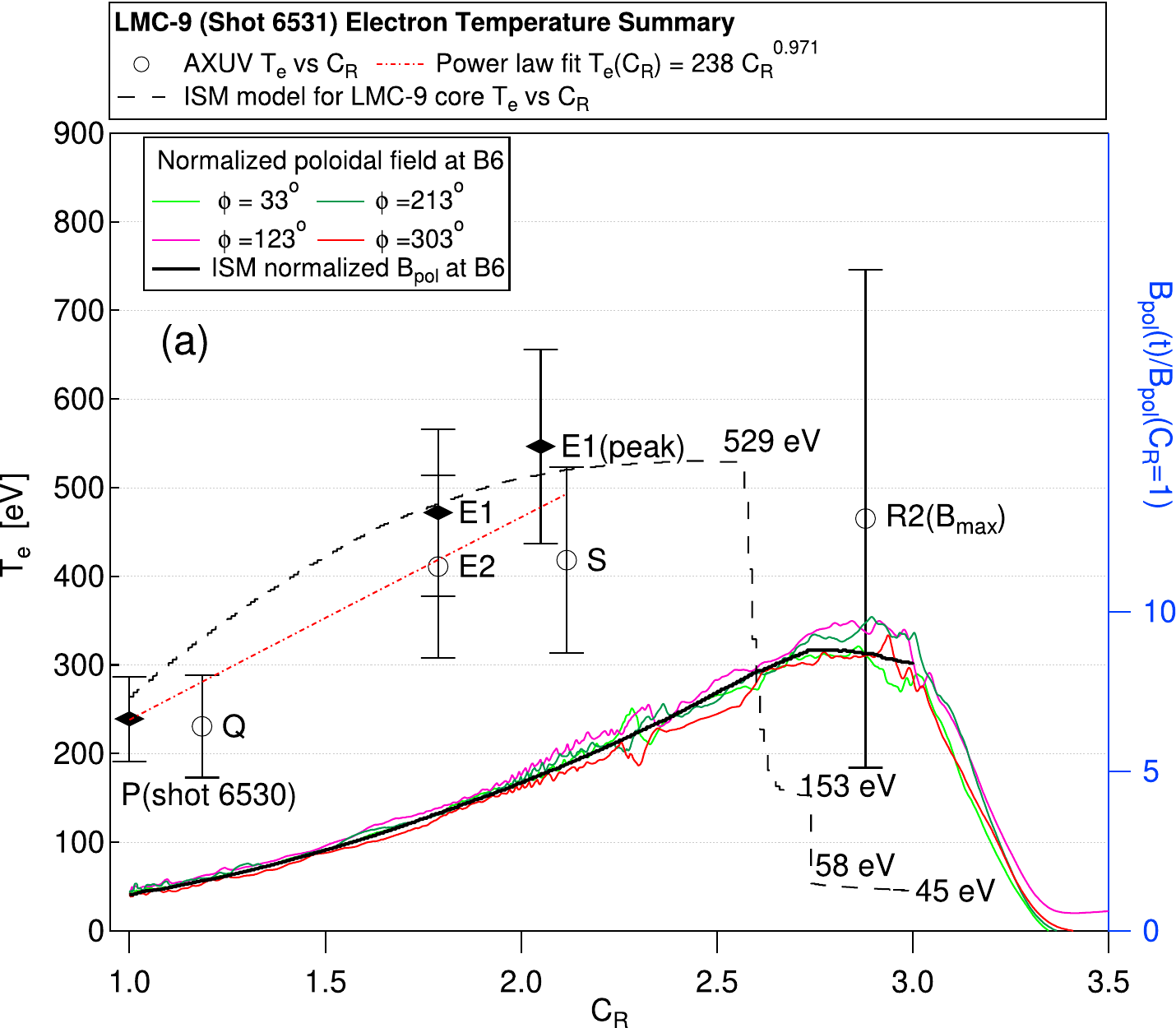}
    \includegraphics[width=1\linewidth]{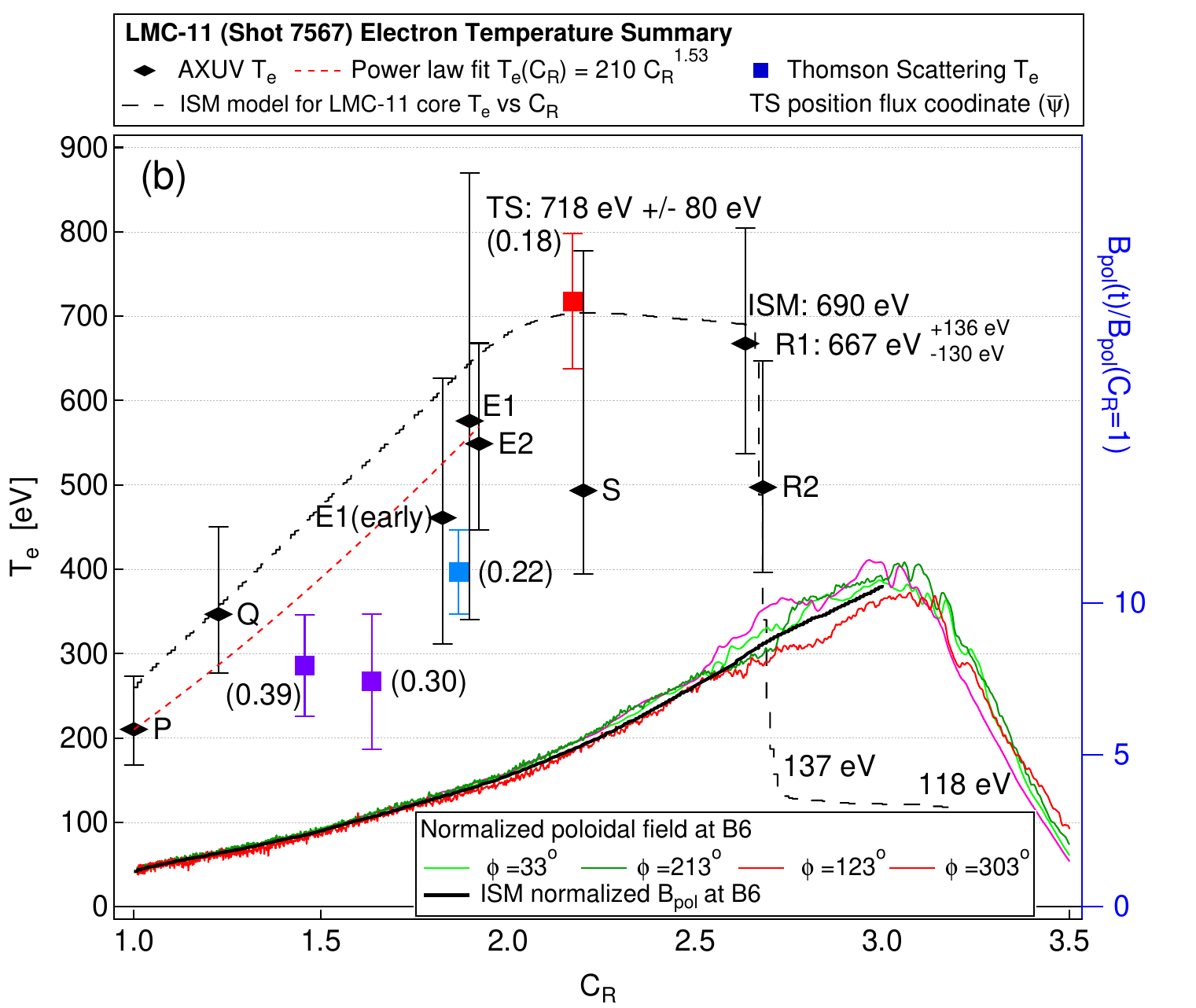}
   \caption{Electron temperature rise for LMC-9 (a) and LMC-11 (b) as a function of $C_R$, with AXUV ratio measurements of $T_e$ compared to ISM best-fit model. Thomson scattering measurements of $T_e$ are included for LMC-11 with ($\bar{\psi}$) values of the TS measurement position shown in point labels.  For most of the AXUV data points the values are taken when the magnetic axis passes the average center of that view cone. Several exceptions for this exist, the E1(peak), E1(early) and R1 and R2 data taken at alternative times where valid data exists when not viewing through the core. See main text for further discussion. AXUV $T_e$ data with (\scalebox{0.8}{$ \blacktriangleleft\!\blacktriangleright$}) markers have been calibrated for diode variability, data with $(\circ)$ markers use the manufacturer's responsivity value.  Also shown are $B_\mathrm{pol}$ measurements at B6, and the corresponding synthetic signal in solid black from the best-fit ISM simulation, the ISM core $T_e$ values are shown as a dashed black curve. The sharp drop in temperature near $C_R = 2.57\text{--}2.66$ is required in both models to obtain a magnetic field that rolls off at the observed peak. There is also a power law fit shown in dashed red.}
    \label{fig:axuv-Te-vs-CR}
\end{figure}

Plasma electron temperature estimates on LM26 are inferred from measuring the plasma's soft X-ray spectrum with a coarsely resolved basis set of different transmission filters, detecting the photons that have enough energy to pass through the filter with absolutely calibrated X-ray photodiodes (AXUV), using a generalization of previously existing methods \cite{Pi3_tauE_NucFus_2025}.   The temperature estimate is calculated from a lookup table of filtered AXUV diode ratios as a function of a 0D value for $T_e$. The ratio-$T_e$ tables are constructed using a simple radiation model where emissivity as a function of photon energy $E$ is assumed to be proportional to $\exp{(-E/T_e)}$, and this is multiplied by the transmission curve $T_k(E)$ for the $k$-th filter and then integrated to give a dimensionless synthetic diode signal. Physical scalar coefficients in the emissivity involving plasma density will cancel out when taking the ratio of diodes looking at a spatially homogeneous plasma, and so a table of dimensionless diode ratios as a function of $T_e$ is sufficient.  

The filtered diodes are mounted in-vacuum within a closely packed $2\times2$ rectangular array (12.2\,mm $\times$ 11\,mm), with each diode having a collection region 5\,mm in diameter. This set of four filtered diodes views the plasma through an aperture in the aluminum inner cavity wall plate, with a distance to the aperture ranging between 482\,mm and 765\,mm, depending on which port the diode package is mounted.  
This arrangement produces a tight cluster of mostly-overlapping view cones for each of the four diodes. Depending on the geometric optics of each AXUV port, side-adjacent diode pairs can have views with between $71.9\,\%$ and $83.4\,\%$ of areal overlap at the plasma's mid-plane, while diagonal pairs have between $62.5\,\%$ to $75.4\,\%$ overlapping area at the mid-plane. As such, taking ratios of diode signals (thick filter/thin filter) approximately cancels all dependence on density and overall brightness, due to mostly observing the same cone of plasma with the pairs of diodes being compared. 
However, the accuracy of this approximation will depend on the length scale of the gradient in plasma emissivity that is being observed by this $2\times2$ cluster of views, as well as the accuracy of the filter thickness used and overall responsivity of the photodiodes being well characterized.  
The most accurate measurement will occur when the magnetic axis of an axisymmetric plasma with a centrally peaked temperature profile is passing through the central view of a filtered diode array, then the measurement is taking place at the ``top'' of the hill where there is zero average gradient, and so ratios between diode signals will represent a measurement of a common core temperature. 

As much as possible, we focus our analysis on the $T_e$ values measured through the core of the plasma to reduce sensitivity to gradients. However, there are indications of temperature rise even at points away from the magnetic axis; these are also presented here as further evidence of global heating of the plasma, observed as the sampling location is shifting relative to the plasma core. 
A summary of electron temperature data from AXUV and TS is shown in Fig.~\ref{fig:axuv-Te-vs-CR}, where a clear trend of core electron temperature rise is observable during compression up to $C_R = 2.5$ in both the LMC-9 and LMC-11 data. LMC-6 and LMC-7 also showed preliminary evidence of heating, but these shots only had a single AXUV ratio pair at a single location to provide the peak measurement. Significant effort was made on later shots to collect higher quality data at more locations with redundancy to show a consistent repeatable set of results. Analysis to interpret this increase in terms of compressional heating versus an increase in Ohmic heating is described in Sec.~\ref{sec:CompressHeatConfine}.

Off-core measurements can be made relatively insensitive to gradient effects by using a diagonally symmetric filter set of two thickness values. For example, on the LMC-9 campaign the E1 AXUV array was set up with aluminized-Mylar filters arranged as \mbox{$\{\{ 11\text{\,{\textmu}m}, 22\text{\,{\textmu}m}\},\{22\text{\,{\textmu}m}, 11\text{\,{\textmu}m}\}\}$}, which allows for direct measurement of the radial variation of the MY11 signals, and possible toroidal variation of the MY22 signals.
To get an improved-accuracy ratio for temperature estimation with the E1 module it is possible to take the average of both MY11 signals, which will have a concentric view to the average of both MY22 signals, producing a ratio that is less sensitive to plasma gradients.  See Fig.~\ref{fig:AXUVmap} for the orientation of diode view cones and filter selections used for LMC-9.

\begin{figure}
    \centering
 \includegraphics[width=1\linewidth]{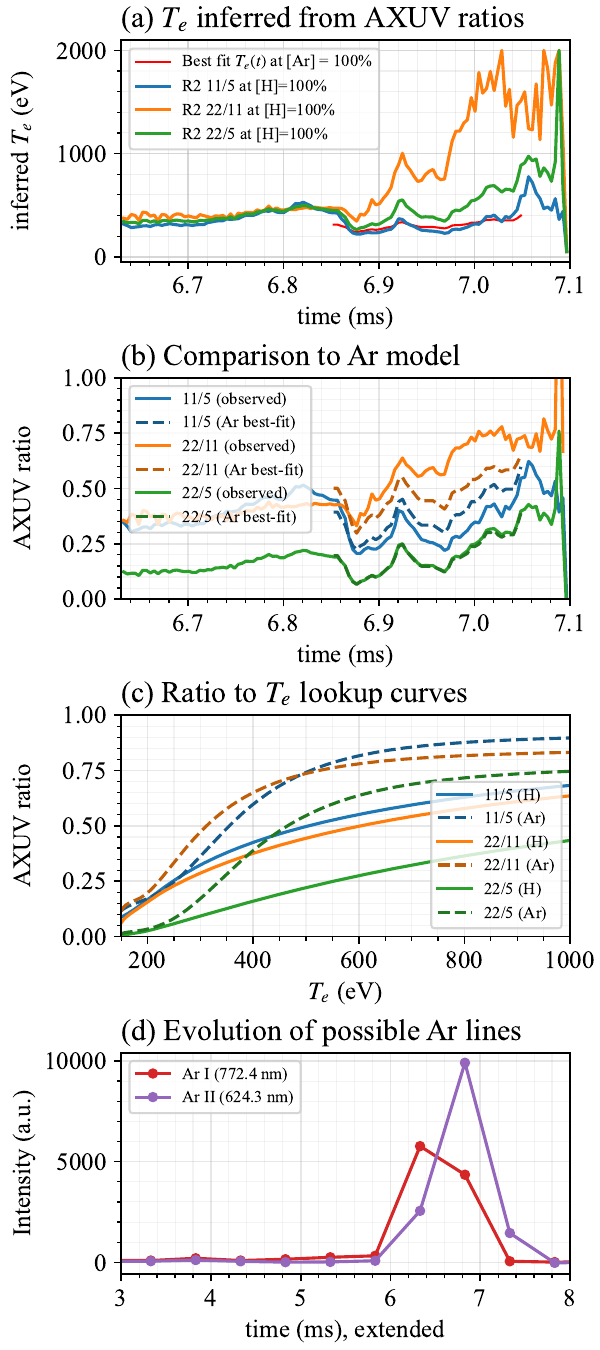}
    \caption{Possible contribution to AXUV R2 ratio discrepancy observed on LMC-9 due to the ingress of Ar, late in compression. Spectral line intensity is shown from spectrometer 2 positioned at the $R = 125$\,mm port. The best-fit Ar model for $T_e$ at late time (red) shown in (a) ranges from 240 to 400 eV, which is on the low end of the outputs of the H-only radiation model, and gives an improved match to the three divergent ratios in (b) with the dashed curves from the Ar model (c), and (d) shows Ar data.}
    \label{fig:ArgonModel}
\end{figure}

In the final phase of compression, after the magnetic field has begun to rapidly drop and non-axisymmetric modes are larger in amplitude, the ratios of the three distinct pairs (MY11/MY5, MY22/MY5, MY22/MY11) will very strongly disagree in estimating $T_e$ from the disparate ratios---in particular for the R1 and R2 arrays at smallest radius (see Fig.~\ref{fig:ArgonModel} for analysis of effect of Ar on the R2 data). This occurs before the magnetic axis has passed through the center of their views, and the length scale of displacement between non-overlapping components (30\,mm) becomes a significant fraction of the plasma emissivity gradient ($\sim$170\,mm for total emissivity across the minor radius and $\sim$70\,mm for the core region hot enough to generate photons that pass the 13\,{\textmu}m filter). 
During this time, toroidal variations in emissivity due to observed $n=1$ modes could also contribute to ratio errors. 
The three estimates of $T_e$ from the R2 array taken at the peak of the poloidal magnetic field disagree strongly with each other, and the most plausible interpretation is that the true core temperature at that moment is on the lower end of those values, more likely agreeing with the ISM thermal model that matches the resistive downturn of the $B_\mathrm{pol}$ signals. 

In addition to gradient effects, another factor which must be carefully considered is the fact that certain impurities can distort the relationship between filtered ratios and electron temperature. Although higher energy photons getting through the filters are mostly continuum radiation from  recombination and bremsstrahlung emission, it is possible that line emission from certain impurities could be sufficiently bright that the filtered ratios will start to deviate from what is expected with the simple radiation model lookup curve. 

To address this, the total emission from various potential impurities that pass through each filter can be modeled from a set of FLYCHK \cite{chung_flychk_2005} simulations. This is done over a range of simulated $T_e$ values and impurity concentrations to determine a trend for the error that arises from interpreting measured ratios with the simple radiation model. As described in Appendix \ref{App:A}, this can provide error bounds on $T_e$ estimates for specified upper limits on the concentrations of different plausible impurities that might be present in the LM26 system. 

The most pertinent case for LMC-9 is that of argon impurities possibly entering the plasma in the final phase of compression, which could explain much of the disagreement in the R2 AXUV ratios (see Fig.~\ref{fig:ArgonModel}).  The best-match model, from FLYCHK data using $100\,\%$ Ar plasma composition, gives a lower $T_e$ estimate similar to the ISM result. Visible spectral lines associated with Ar impurities are observed to increase at late time during LMC-9 in the survey spectroscopy data.  This model does not perfectly match the observed R2 ratios, so other effects due to plasma gradients, uncorrected differences in diode responsivity, or filter imperfections are likely to be contributing factors.  

The overall result across multiple LM26 campaigns is that electron temperature will rise during compression of this class of Ohmically heated ST plasmas if the peak liner velocity is sufficiently fast (greater than $175~\text{m/s}$) and the plasma remains MHD stable. These overall temperature increases are shown in Table~\ref{tab:AXUV-Te_summary} for LMC-6, 7, 9, 10, and 11 for which there was sufficient AXUV data to determine electron temperature. Based on the resistive evolution of the poloidal magnetic flux, we can also conclude that LMC-8 must have had a similar temperature rise as was observed on LMC-10, given the strong similarity in magnetic evolution and liner velocity on the two shots.

\begin{table}
    \centering
    \caption{Electron temperature inferred from filtered AXUV ratios during compression, measured at the start of compression (at $t_\mathrm{initial}$) and at peak $T_e$ (at $t_\mathrm{final}$) before the plasma appears to crash.  Times in milliseconds and $T_e$ in eV.  These estimates use a lookup table based on the simple radiation model. The error bars account for statistical variation in equivalent measurements and uncertainty in filter thickness. For simplicity they are symmetric averages of the asymmetric upper and lower uncertainties. Effects of possible variations in plasma composition and gradient effects, as well as the presence of electrical pickup effects have been used as criteria to exclude data from this analysis.  
    LMC-8 data is absent because the AXUV electronics were damaged by compression shock.
    Only shots with peak liner velocity greater than $175\mathrm{\,m/s}$ are shown.  }
    \label{tab:AXUV-Te_summary}
    \begin{tabular}{lccccc}
        Shot & $t_\mathrm{initial}$ & $T_e(t_\mathrm{initial})$ & $t_\mathrm{final}$ & $T_e(t_\mathrm{final})$\\ 
        \hline
        \rule{0pt}{2ex}%
        LMC-6 & 2.35  &  $220\pm 151$ & 5.70  & $634 \pm 127$   \\
        LMC-7 & 2.09   & $266\pm 144$  & 4.72  & $688 \pm 138$  \\
        LMC-9 & 3.50   & $239\pm 57$ &  6.38  & $546\pm 100$\\
        LMC-10  & 3.45  & $226\pm 51$ & 6.51  & $429 \pm \enspace85$  \\
        LMC-11  & 2.49  & $210\pm 53$ & 5.47  & $667 \pm 134$  \\
    \end{tabular}
\end{table}

\subsection{Thomson scattering measurements}
\label{sec:TSresults}

\begin{figure}
    \centering   \includegraphics[width=0.85\linewidth]{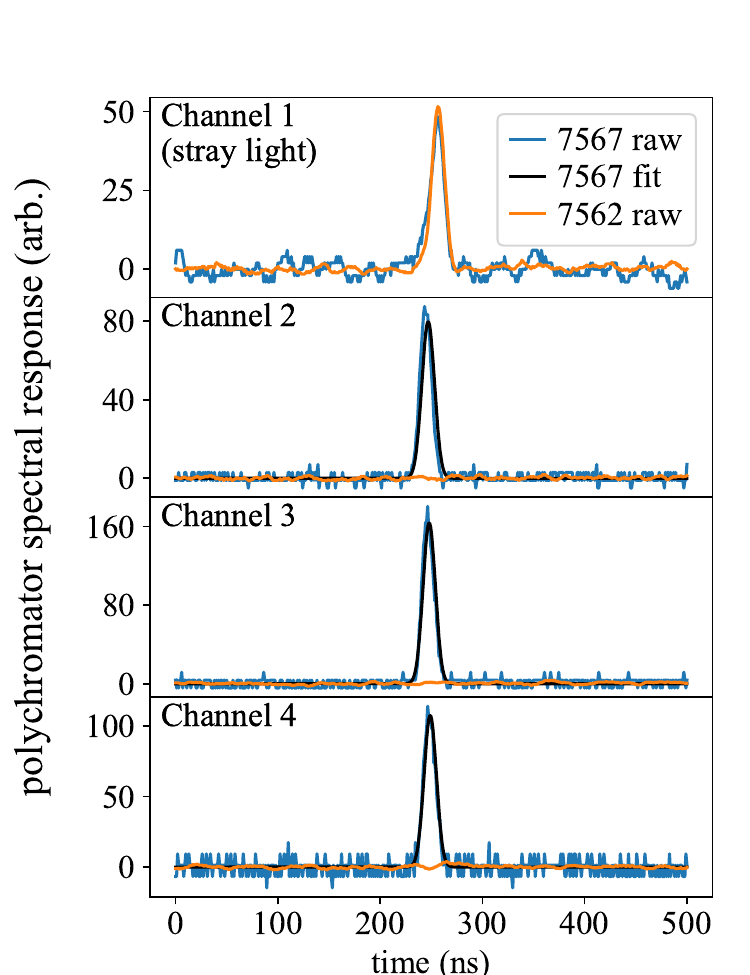}
    \caption{Raw and Gaussian-fit polychromator data for laser pulse 4 for LMC-11 (7567) and a no-plasma vacuum test shot (7562).}
    \label{fig:TS_raw_pulse_4_lmc11}
\end{figure}

For LMC-11, $T_e$ was measured by Thomson scattering at a single location and at four points in time during compression. The laser trigger was timed such that the fourth pulse occurred about 100\,{\textmu}s before liner cutoff of the laser on the recovery side. Raw polychromator data for the fourth pulse during LMC-11 is shown in Fig.~\ref{fig:TS_raw_pulse_4_lmc11} alongside a vacuum (no-plasma) test shot. For LMC-11, the oscilloscope's channel gain was set to a conservative level to avoid saturation during compression and so digitization noise was apparent on channels 2--4.

 Stray light at $1064\mathrm{\,nm}$ was measured by the dedicated polychromator spectral channel 1. Stray light was a negligible source of error and was at a similar level during the compression shot as it was when shooting into vacuum (Fig.~\ref{fig:TS_raw_pulse_4_lmc11}).

\begin{figure}
    \centering
    \includegraphics[width=\linewidth]{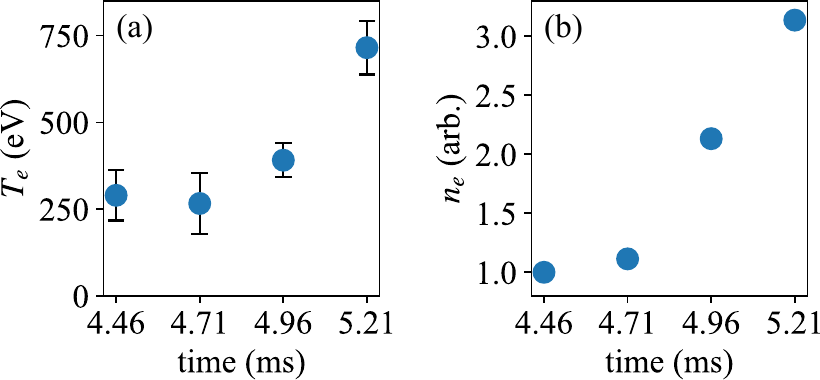}
    \caption{Thomson scattering measurements of (a) electron temperature and (b) relative density as a function of time after formation, taken during compression of LMC-11. 
    The same temperature data are shown versus compression ratio in Fig.~\ref{fig:axuv-Te-vs-CR}.
    The temperature at the fourth pulse is $718\pm80$ eV.}
    \label{fig:thomson_Te_ne}
\end{figure}

\begin{figure}
    \centering
    \includegraphics[width=\linewidth]{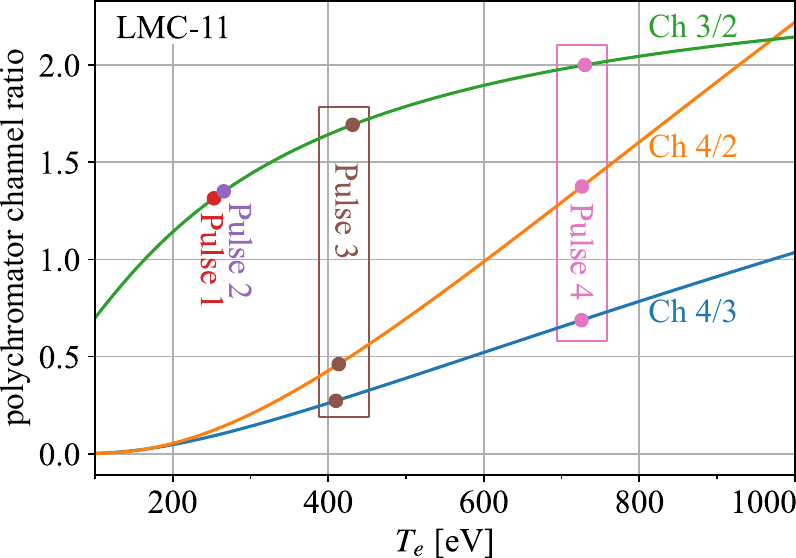}
    \caption{Expected polychromator channel signal ratios for a range of temperatures. Vertical dashed lines are the measurements for LMC-11 and their implied temperature. Pulses 1 and 2 only show the channel 3/2 ratio because channel 4 was in the noise.}
    \label{fig:thomson_poly_ratios}
\end{figure}

The LMC-11 Thomson $T_e$ and relative $n_e$ measurements are shown in Fig.~\ref{fig:thomson_Te_ne}. 
The relative residuals indicated a good fit, which can also be seen from the agreement of the polychromator signal ratios in Fig.~\ref{fig:thomson_poly_ratios}. 

The calculated uncertainty in $T_e$ takes into account background plasma shot noise, Thomson shot noise, intrinsic electrical noise, and the scattering angle. The scattering angle was estimated to be $9.9\pm 0.5\deg$ and, according to a sensitivity analysis, contributes a $10\,\%$ uncertainty to $T_e$. More details on the error analysis can be found in Appendix \ref{app:thomson}.

\subsection{Neutron and gamma rates during compression}
\label{sec:neutron-and-gamma-rates}

In this section we discuss the observations of neutrons and gammas from the LMC shots with deuterium plasma.
It is worth noting that test discharges of LM26 performed with hydrogen instead of deuterium do not generate any scintillator pulses, indicating that the observed gammas and neutrons are a consequence of D-D fusion.

The neutrons counted on the liquid scintillators are converted into a plasma neutron yield (neutrons emitted per unit time) using a per-detector detection probability. The probability is the product of an inverse-square geometric factor and an intrinsic efficiency. The geometric factor scales as $1/d_s^2$, where $d_s$ is the distance from the plasma center to the detector; across the array $d_s$ ranges from 4.48~m to 10.48~m. Some detectors were moved closer to the plasma between LMCs to improve sensitivity to lower-yield discharges. The detector intrinsic efficiency $\epsilon_s^{\mathrm{MCNP}}$, computed from MCNP simulations performed by the UKAEA~\cite{radichIonTemperatureInference2026a}, is $0.565$ for the smaller EJ-309 detectors, $0.612$ for the EJ-301 detectors, and $0.633$ for the single larger EJ-309 detector. The total efficiency of detecting neutrons ($\epsilon_\mathrm{det}$) is thus estimated as: 

\begin{equation}
\epsilon_\mathrm{det} = \sum_{s} \epsilon^{\mathrm{MCNP}}_{s}\frac{A_s}{4\pi d_s^2},
\end{equation}
where $s$ represents each of the eight detectors, and $A_s$ is the cross sectional area of the scintillator. $\epsilon_\mathrm{det}$ ranges from $2.2\times10^{-4}$ for LMC-7, LMC-8, and LMC-9, to $2.0\times 10^{-4}$ for LMC-11. These shots are selected because they show the greatest evidence of increase in yield post-compression.

The effect of lab geometry on neutron scattering and efficiency of measurement are not considered in this model other than two bounding assumptions for the attenuation of the neutron signal by the lithium liner: one conservative case (``no Li'') in which the liner causes no attenuation, and one (``max Li'') in which the signal is attenuated by $e^{-d_\mathrm{Li}/\lambda}$, where $\lambda \approx 10$~cm is the mean free path of 2.45~MeV D-D neutrons in lithium and $d_\mathrm{Li}$ is the lithium path length along the direct line of sight to the scintillator.  In LMC-9, for example, $\epsilon_\mathrm{det}$ drops from $2.2\times 10^{-4}$ before the liner begins to enter the lines of sight, to $0.6\times 10^{-4}$ at $7\mathrm{\,ms}$ in the case of maximum attenuation.

Because individual neutron detections are discrete and their rate varies by orders of magnitude across a discharge (frequent during the early plasma but sparse just before compression and after termination of plasma), the rate is reconstructed without fixed-width time bins, which would otherwise wash out the early structure or leave the late-time signal dominated by empty bins. Instead we use an adaptive estimate in which the local averaging timescale is set by the data themselves: each detected neutron in the aggregate across all scintillators contributes a Gaussian kernel whose width $h_i$ is the time interval to its $k{=}7$th nearest-neighboring detection, so the estimate is sharp where detections are dense and smoothly broadened where they are sparse \cite{breiman1977variable}. The detected count rate ($\hat{\lambda}(t)$) is: 
\begin{equation}
    \hat{\lambda}(t)=\sum_i h_i^{-1}\phi\big((t-t_i)/h_i\big),
    \label{eq:detected-count-rate}
\end{equation}
where the $t_i$ are the detection times and $\phi$ is the unit Gaussian distribution. The statistical uncertainty at each time is set by the effective number of detections contributing locally $(n_\mathrm{eff})$, which is calculated:
\begin{equation}
n_\mathrm{eff} = \frac{\left( \sum_i h_i^{-1}\phi\big((t-t_i)/h_i\big) \right)^2}{\sum_i \big(h_i^{-1}\phi\big((t-t_i)/h_i\big)\big)^2}.
\end{equation}
 From this, a $90\,\%$ confidence band is obtained from Poisson counting statistics. 
Dividing $\hat{\lambda}(t)$ by $\epsilon_\mathrm{det}$ 
converts the detected rate into the plasma neutron yield; the two bounding liner-attenuation assumptions then translate into a corresponding band on the yield that widens as the liner enters the lines of sight.

\begin{figure}
    \centering
    \includegraphics[width=\linewidth]{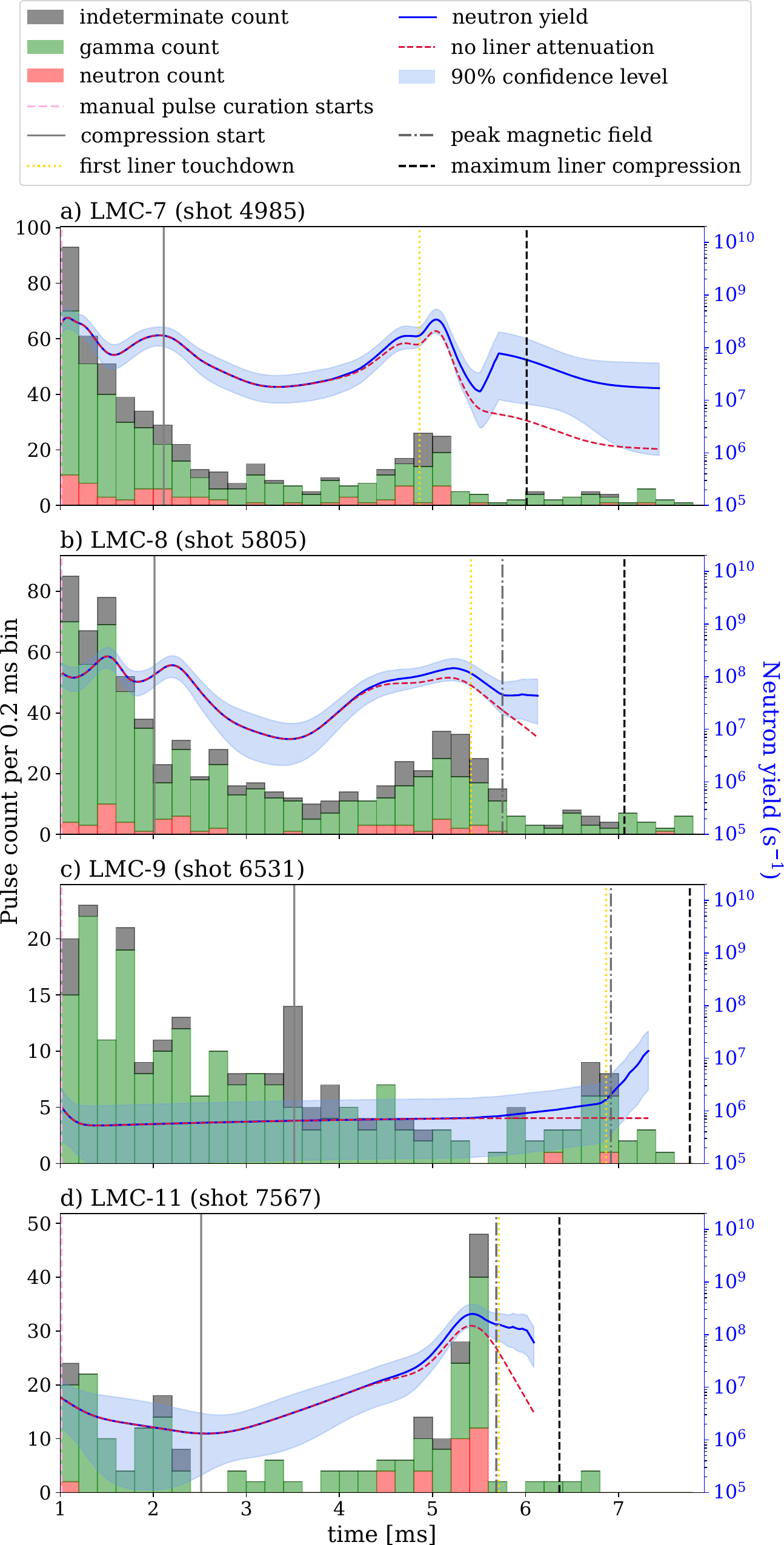}
    \caption{Neutron yield and scintillator signals (stacked histograms of counts from all scintillators) as a function of time for LM26 compression shots LMC-7 (a), LMC-8 (b), LMC-9 (c), and LMC-11 (d).  Data after the first $1\mathrm{\,ms}$ is shown, when ST magnetic equilibrium has become fully established. Pulses which have either a secondary pulse or large noise fluctuation on the tail preventing the normal PSD algorithm from working, but are obviously real particle-interaction based pulses, are labeled indeterminate.
    }
    \label{fig:LMC89}
\end{figure}

The neutron diagnostic indicates that the yield is high early in the discharge, falls more or less quickly depending on the shot, and then in the best performing shots, exhibits a modest increase during compression.  Fig.~\ref{fig:LMC89} shows the pulse time data for LMC-7, LMC-8, LMC-9, and LMC-11, starting $1\mathrm{\,ms}$ after plasma formation, when accurate pulse identification becomes feasible.   In all four shots an increase in neutron and gamma yield is evident during compression.

 It is possible to obtain a continuous determination of $T_i(t)$ from the yield curve obtained from scintillator counts via (\ref{eq:detected-count-rate}), and the assumption that the ions have had sufficient time to thermalize to a Maxwellian distribution, with additional knowledge of the deuterium density profile. The result is plotted as the blue band in Fig.~\ref{fig:lmc-8-doppler-Ti} for LMC-8, LMC-9, and LMC-11 in comparison to ion Doppler data.  It is again shown in comparison to a simple model in Fig.~\ref{fig:lmc-9-thermal-model} for LMC-9 and LMC-11.  The method, similar to that used in \cite{Howard_2025, Pi3_tauE_NucFus_2025, radichIonTemperatureInference2026a}, is as follows.

To obtain a $T_i(t)$ curve,
the continuous yield data including liner attenuation (from Fig.~\ref{fig:LMC89}) is converted to a temperature by adjusting the core $T_i$ in a D-D fusion forward model to match the yield at each time in the reconstruction. The neutron yield is calculated by integrating the Bosch--Hale reactivity \cite{Bosch_Hale_1992} over the volume of the plasma, with profiles of density $n_e(\bar\psi)$ taken from the interferometer fit, and with an assumed linear $T_i$ profile of the form $T_i\propto 1-\bar\psi$.  
A temperature confidence band is found by adjusting the core $T_i$ to reproduce the yield values at the upper and lower limits of the yield confidence band.  
In the yield calculation, the plasma density profile is not corrected for the possibility of impurity influx, which may underestimate $T_i$ at later times.
For the fuel dilution factor $f_D=n_d/n_e$ we use a constant value $f_D=0.7$.  This value is arbitrary but, in our opinion, reasonable to represent the average behavior during early compression.  In order for this value to be consistent with the  $Z_\mathrm{eff}$ and $Z_\mathrm{avg}$ values (also arbitrary but reasonable) assumed for the ISM simulations, there should be a fraction of protons in the plasma, as could be expected to be generated from residual water, hydrocarbons, or hydroxides in the machine.

The resulting $T_i(t)$ confidence band is plotted in Fig.~\ref{fig:lmc-8-doppler-Ti}.  It is worth noting that the confidence band for temperature is much tighter than that for yield because of the fast dependence of reactivity on temperature.  
The figure shows that the ion temperature is high immediately after plasma formation.  
This is probably due to dissipation of magnetic energy by the initial reorganization of helicity after injection, and likely also has a contribution from the stagnation of the fast MHD flow into the chamber.
The fall of ion temperature after formation may be due to e-i collisions (see Section~\ref{sec:simpleIons}), which will be especially effective before the electrons become Ohmically heated after the formation of closed flux surfaces.  It is also possible that the ions are initially not thermalized (non-Maxwellian distribution) and that at least part of the apparent temperature drop is due to reduction of an enhanced reactivity by thermalization of the ion velocity distribution, as discussed in \cite{radichIonTemperatureInference2026a}.  
Here we continue with the working assumption of a Maxwellian distribution.  For reference, the ion collision time in LMC-11 at the start of compression is $\tau_i\approx 0.3\mathrm{\,ms}$.

An alternative, simpler method of estimating $T_i$ during compression is possible directly from the scintillator neutron histogram.
Table~\ref{tab:isothermal-Ti} shows the result of this method to estimate the ion temperature that would produce the observed neutron signal from the compressed plasma, using the histogram data shown in Fig.~\ref{fig:LMC89}.  To simplify the analysis, we assume the conservative case of isothermal compression with the same linear $T_i(\bar{\psi})$ profile as before.
In this case the increase of neutron yield with time is due to the increase of density.  The density in the core of the plasma is assumed to increase inversely with the volume of a small flux surface of constant toroidal flux, with density referenced to the interferometer density at the time compression is triggered.  
In this way we avoid mistaking the impurity influx at the edge for an increase in deuterium density in the core.     
The varying attenuation of emitted neutrons by the collapsing lithium liner coming between the plasma and the scintillators is taken into account.
The probability distribution for temperature is calculated as $P(T_i)\propto\prod_{t,s} P(k_{t,s}|\lambda_{t,s})$ where the product is over bin times $t$ and scintillators $s$, $k_{t,s}$ is the observed count, $\lambda_{t,s}$ is the expected count rate given the temperature $T_i$ and the plasma configuration at time $t$, and $P(k|\lambda)=\lambda^k e^{-\lambda}/k!$ is the Poisson distribution.  In this calculation the bin width is $0.1\mathrm{\,ms}$.  For error bars we give the range of $T_i$ where $P(T_i)$ exceeds $e^{-1}$ of its peak value.  
Expectation values for $T_i$ from this analysis are shown in Table~\ref{tab:isothermal-Ti} and also represented graphically in Fig.~\ref{fig:lmc-9-thermal-model} together with with the continuous $T_i(t)$ curve.

\begin{table}
    \centering
    \caption{Ion temperature inferred from neutron data during compression, assuming isothermal compression.  The analysis window is $t_\mathrm{min}<t<t_\mathrm{max}$ (times in ms), the number of neutron events in the window is $N$, and the inferred ion temperature is $T_i$ (in eV), with error bars given by Poisson statistics.  Analysis is performed with worst case Li attenuation (max Li) and ignoring Li attenuation (no Li).}
    \label{tab:isothermal-Ti}
    \begin{tabular}{lccccc}
        Shot & $t_\mathrm{min}$ & $t_\mathrm{max}$ & $N$ & $T_i$(max Li) & $T_i$(no Li)\\ 
        \hline
        \rule{0pt}{2ex}%
        LMC-7 & 4.0 & 5.3  & 21 & $391\pm 16$ & $378\pm 15$ \\
        LMC-8 & 4.0 & 5.8  & 21 & $382\pm 15$ & $368\pm 14$ \\
        LMC-9 & 4.0 & 7.0  & 2  & $278\pm 32$ & $269\pm 31$ \\
        LMC-11 & 4.0 & 5.6 & 15 & $411\pm 20$ & $397\pm 19$
    \end{tabular}
\end{table}

The scintillator signals also provide evidence of both prompt and delayed gamma production, relative to neutron production. The prompt gammas are evident because neutrons and gammas both rise when the emission rate increases, both at formation and during compression.  Prompt gammas are to be expected as a byproduct of D-D fusion: gammas from reactions when energetic fusion protons, ${}^3$H, and ${}^3$He hit the wall, and capture gammas emitted when fusion neutrons thermalize and are absorbed in the machine and surrounding material (neutrons will thermalize in 20--50\,{\textmu}s and then be captured in 100--500\,{\textmu}s). The evidence for delayed gamma production is strongest in LMC-9, because no clear neutrons, but many gammas, are detected in the interval between $1\mathrm{\,ms}$ and $6\mathrm{\,ms}$, before neutrons associated with compression appear.  During this interval the gamma yield decays with a half-life of $t_{1/2}=1.5\pm 0.2\mathrm{\,ms}$. 
These delayed gammas may result from a confined fraction of charged fusion products leaving the plasma and impacting the wall.  There may also be a gamma contribution from decay of neutron-activated isotopes or isomers, a mechanism that probably accounts for the long tail of gamma emission observed even after the discharge is over.  But the brightness of the faster gamma decay before compression is difficult to explain this way. 

The pre-compression scintillator signal observed in LMC-7 and LMC-8 is different from that in LMC-9, with more gamma counts, a faster decay ($t_{1/2}\simeq 0.6\mbox{-}0.9\mathrm{\,ms}$), and neutrons observed as well as gammas ($N_\gamma/N_n\simeq 6\text{--}11$).  

\begin{table}
    \centering
    \caption{Relative enhancement of curated scintillator counts (neutrons, gammas, and indeterminate pulses) during compression.  The pre-compression decay is fit to an exponential $\propto e^{-t/\tau}$ over $1\mathrm{\,ms}<t<t_\mathrm{comp}+1.5\mathrm{\,ms}$ (after formation and before the compression starts, see Table~\ref{tab:shot_data}), where $\tau_d$ is the fitted decay time constant (half-life $t_{1/2}=\tau_d\ln 2$), and is extrapolated across the compression window $t_\mathrm{comp}+1.5\mathrm{\,ms}<t<t_\mathrm{comp}+5\mathrm{\,ms}$.  $N$ is the measured count of all types of radiation in the window, $N_\mathrm{exp}$ the count predicted by the extrapolated decay, and $R=N/N_\mathrm{exp}$ the enhancement factor.  Uncertainties on $\tau$ and $N_\mathrm{exp}$ are propagated from the Poisson-weighted fit covariance; the uncertainty on $R$ additionally includes the Poisson error on $N$.}
    \label{tab:comp_enhancement}
    \begin{tabular}{lccccc}
        Shot  & $\tau_d$ (ms) & $N$ & $N_\mathrm{exp}$ & $R$\\
        \hline
        \rule{0pt}{2ex}%
        LMC-7   & $0.93\pm 0.05$ & 142 & $24.6\pm 4.4$  & $5.8\pm 1.1$ \\
        LMC-8  & $1.16\pm 0.11$ & 249 & $59.8\pm 12.8$ & $4.2\pm 0.9$ \\
        LMC-9   & $2.42\pm 0.36$ &  43 & $37.5\pm 9.2$  & $1.1\pm 0.3$ \\
        LMC-11  & $1.47\pm 0.31$ & 132 & $20.2\pm 9.7$  & $6.5\pm 3.2$
    \end{tabular}
\end{table}
Overall, the general property of the radiation field having more gammas than direct neutrons, yet being roughly proportional, and having zero gammas emitted when forming hydrogen-only plasmas, implies that both PSD-identified neutrons and the prompt and delayed gammas serve as an indication of the D-D fusion rate on LM26 deuterium plasmas.  The rise of the more abundant and easily detected gammas during compression on high performing shots (Fig.~\ref{fig:LMC89}) provides some degree of supportive evidence that the measured increase in PSD-identified neutron counts is indeed from an enhanced fusion rate in the plasma during compression. Table \ref{tab:comp_enhancement} shows the relative increase during the compression window compared to the window immediately preceding. LMC-11 is consistent with showing the largest relative increase of all LMCs.

\subsection{Ion Doppler C V temperature} \label{sec:IonDoppler}

\begin{figure}
    \centering
    \includegraphics[width=\linewidth]{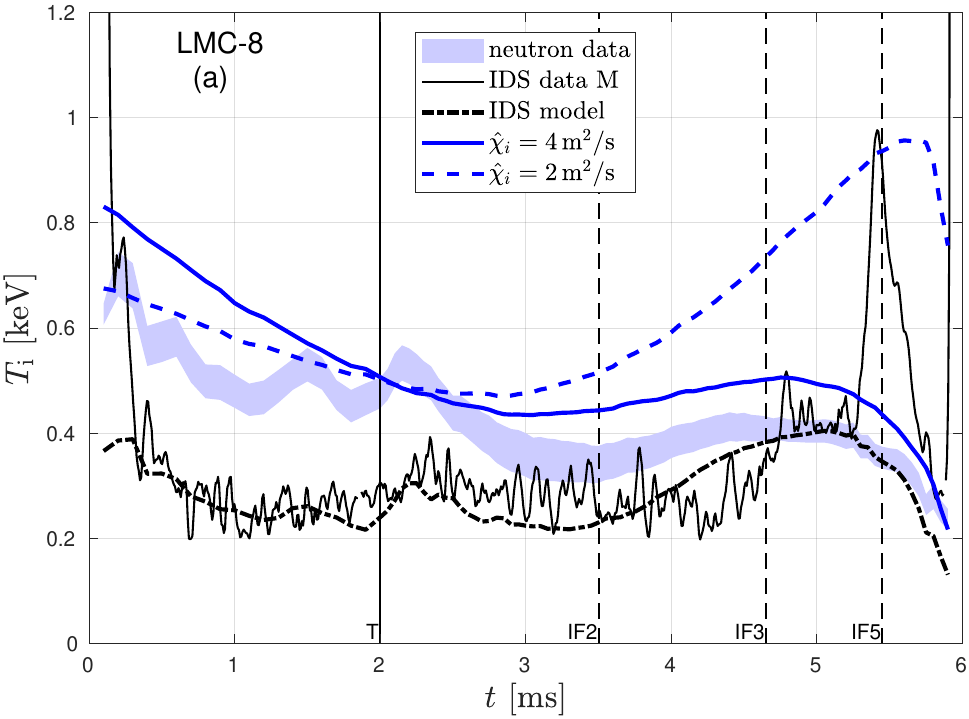}
    \includegraphics[width=\linewidth]{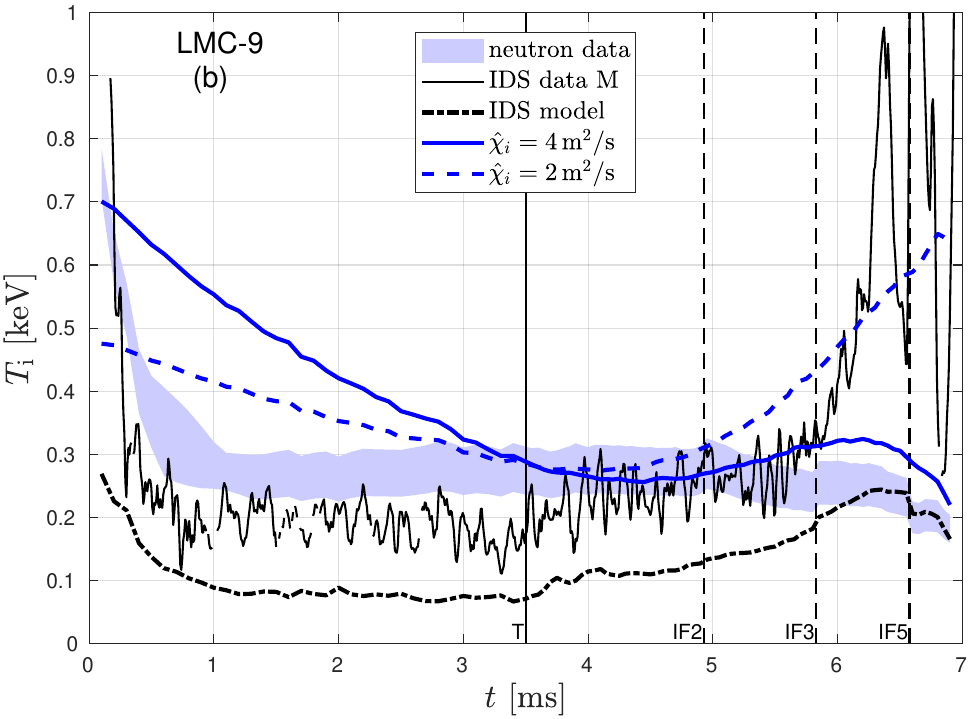}
    \includegraphics[width=\linewidth]{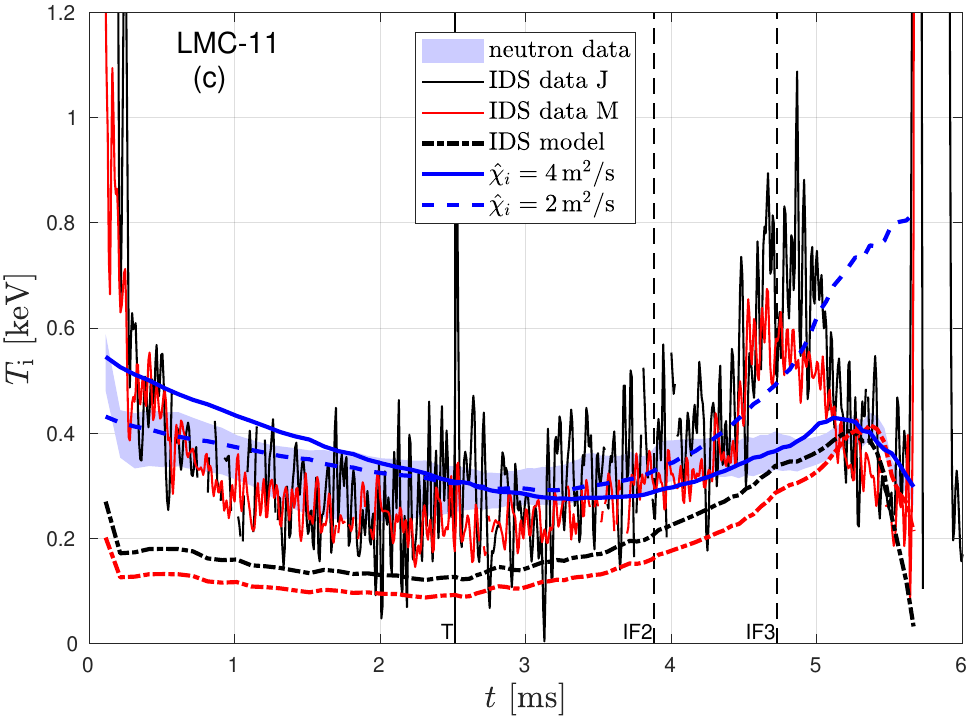}
    \caption{Ion temperature obtained by neutron yield (blue confidence band) and by IDS (black curve, and red curve in LMC-11) for LMC-8 (upper panel), LMC-9 (middle panel), and LMC-11 (lower panel).   Blue solid and dashed curves are from Model A (Sec.~\ref{sec:simpleIons}). 
    Dot-dashed curve(s) indicates IDS $T_i$ trend based on neutron $T_i$.  The vertical solid line labeled ``T'' indicates the compression trigger time, and the vertical dashed lines (IF2, IF3, and, except in LMC-11, IF5) indicate the times when interferometers are removed from the density fit.
    }
    \label{fig:lmc-8-doppler-Ti}
\end{figure}

The IDS diagnostic \cite{Howard_2025, Pi3_tauE_NucFus_2025} gives a C\,V ion temperature evolution shown in Fig.~\ref{fig:lmc-8-doppler-Ti} (solid black and red traces) for compression shots LMC-8, LMC-9, and LMC-11. 
The IDS data for the carbon ion temperature is found from a Gaussian fit to the $\lambda = 227\, \mathrm{nm}$ spectral line shape, with instrumental broadening subtracted in quadrature before applying the Doppler broadening formula for $T_i$.  Raw signal of the C\,V line is well above the noise floor throughout compression. 

For comparison, we also show deuterium ion temperatures from neutron data (blue bands) and curves from two cases ($\hat\chi_i = 4\mathrm{\,m^2/s}$,  $\hat\chi_i = 2\mathrm{\,m^2/s}$) of the simple zero-dimensional transport-only compression model used to estimate the level of ion heat transport required to explain the evolution of ion temperature indicated by neutron yield, as will be described in Sec.~\ref{sec:simpleIons}.
Synthetic IDS model data are shown by the black and red dot-dashed curves assume a fixed temperature profile shape ($T_i\propto 1-\bar\psi$) with core $T_i$ determined by neutron measurement (the value of the dot-dashed curve indicates the $T_i(\bar\psi)$ at the minimum value of $\bar\psi$ along the IDS view). This assumes the hottest part of the chord is also the brightest by a large enough factor that it dominates the spectral line. This simple model does not take into account the expected abundance of the C\,V ionization state as a function of local $T_e$, and disagreement with measured IDS data would most likely be explainable that the actual $T_i$ profile is spatially broader than the linear model assumed for this analysis.  

In LMC-8 the IDS $T_i$ is nearly constant during most of compression, only rising late in time, while LMC-9 and LMC-11 begin to rise gradually almost immediately after the compression start time (labeled ``T'' for trigger time in Fig.~\ref{fig:lmc-8-doppler-Ti}). The initial rise is explainable in terms of some amount of modest compressional heating compounded with the IDS chords measuring deeper into the plasma as the compression moves the plasma radially inward towards the chords.  

In LMC-9 (middle panel of Fig.~\ref{fig:lmc-8-doppler-Ti}) the C\,V ion temperature shows a near continuous rise to a maximum value near the end of compression.

In LMC-11 (lower panel of Fig.~\ref{fig:lmc-8-doppler-Ti}), there are useful signals from two IDS channels, J and M, with views along the S2 and S3 diagnostic chords (see Fig.~\ref{fig:DiagMap}) which intercept the equatorial plane at $0.266\mathrm{\,m}$ and $0.236\mathrm{\,m}$, respectively.  The IDS $T_i$ reaches a maximum in both channels around $t\approx 4.9\mathrm{\,ms}$ before falling at late time. 
At the time of this maximum, according to BPR, the magnetic axis is at $R\simeq 0.314\mathrm{\,m}$, still outside the radius of the channel views.  The magnetic axis is estimated to cross the views at $5.11\mathrm{\,ms}$ and $5.36\mathrm{\,ms}$.  At the time of the IDS $T_i$ peak there is no corresponding peak on the neutron $T_i$ and the neutron $T_i$ is lower than the IDS $T_i$.  Thus the IDS $T_i$ peak may indicate a preferential heating of C\,V over D.

Overall, the trends observed with neutron-inferred $T_i$ are more consistent with the case $\hat\chi_i = 4\mathrm{\,m^2/s}$ than the more optimistic case of $2\mathrm{\,m^2/s}$, when compared to the 0D transport model, Model A.

The final phase of compression contains a narrow spike in ion Doppler $T_i$ on LMC-8 and LMC-9, starting at 5.2\,ms and 6.2\,ms, respectively, reaching nearly 1\,keV, however this spike up is absent from LMC-11 data.  The spike is too fast to be compressional heating.   We attribute this spike in ion Doppler $T_i$ to heating of the carbon ions by magnetic reconnection, presaged by MHD activity observed in the Mirnov probe signals as shown, for example, in Fig.~\ref{fig:ISM_stability_8n9}.  The neutron data shows no  simultaneous increase in deuterium temperature. Such preferential heating of carbon ions relative to deuterium ions when reconnection occurs has been observed, for example, in MST \cite{gangadharaIonHeatingReconnection2008,mageeIonEnergizationTearing}, with deuterium temperature remaining more stable the higher the density.

\subsection{Observations with fast-cameras}\label{sec:fastcameras}

\begin{figure*}
    \centering
    \includegraphics[width=1\linewidth]{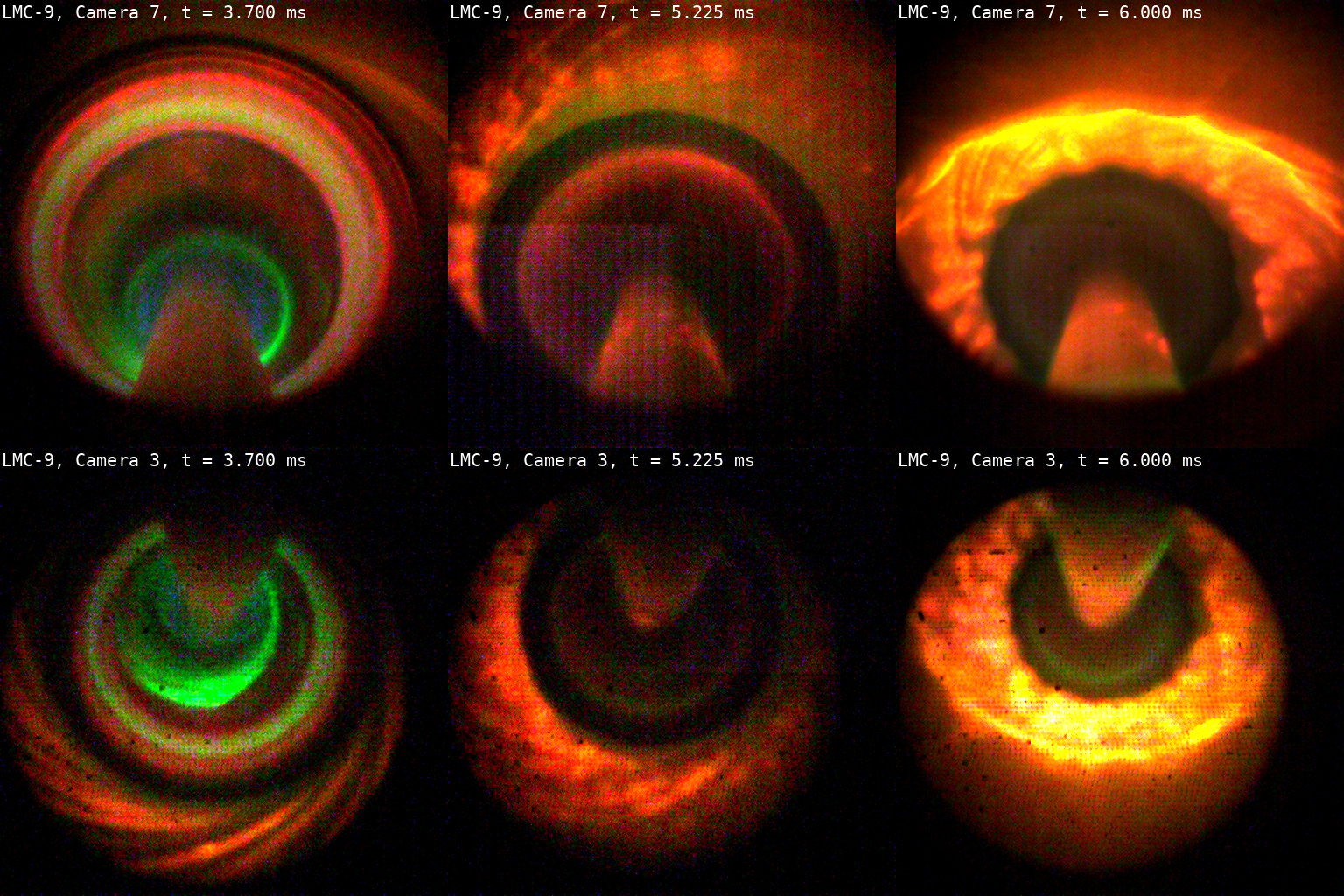}
    \caption{Fast color-camera observations of visible light emission from near the edge of the plasma during the LMC-9 compression. Magnetic filaments can be seen, as well as the onset of toroidal-mode buckling of the liner. The green filament in the left frame is from Li\,II ions emitted from the center electrode. Red light is typically from the $\mathrm{D\alpha}$ line or from neutral Li at $671\mathrm{\,nm}$. The poloidal extent of camera 3 (WV2) and camera 7 (WV3) view angles are indicated in Fig.~\ref{fig:LM26_schematic} as yellow cones.}
    \label{fig:CameraData_LMC9}
\end{figure*}

For most LMC campaigns we had one to three high-speed RGB color video cameras viewing the plasma behavior through wide view internal fish-eye optics that could record 1 frame per 25\,{\textmu}s in the compression cavity (up to two camera views), and when present an additional camera that viewed forward through the Marshall gun section that could record 1 frame per 3.5\,{\textmu}s for studying the gas breakdown and CHI bubble-out process \cite{Hooper2012, Howard_2025} within the gun section.  Frames from two of these cameras at three times during LMC-9 are shown in Fig.~\ref{fig:CameraData_LMC9}.  
These high speed videos can provide global information about the plasma discharge that line averaged or point-wise individual diagnostic signals can not directly provide. 
The visible light is mostly emitted from the plasma near the edge where low ionization states can exist. This can indicate the location, degree of intensity, and atomic number of impurities involved in plasma-wall interaction.
Noticeable on most frames are red filaments on the liner inner surface, which are best explained as high mode number electron density fluctuations at the LCFS, which have radial displacements into the ambient field of neutral D emitted nearly uniformly from the wall, lighting up with red $\mathrm{D\alpha}$ emission where there are both significant free electrons and neutral deuterium. Outside the LCFS the electron density is almost zero in comparison and so appears dark, while deep inside the plasma there are too few neutrals to observe with the camera. 

\section{Physics analysis and modeling}\label{sec:physics_analysis_modeling}
\subsection{Plasma behavior during compression}

\begin{figure}
    \centering
    \includegraphics[width=\linewidth]{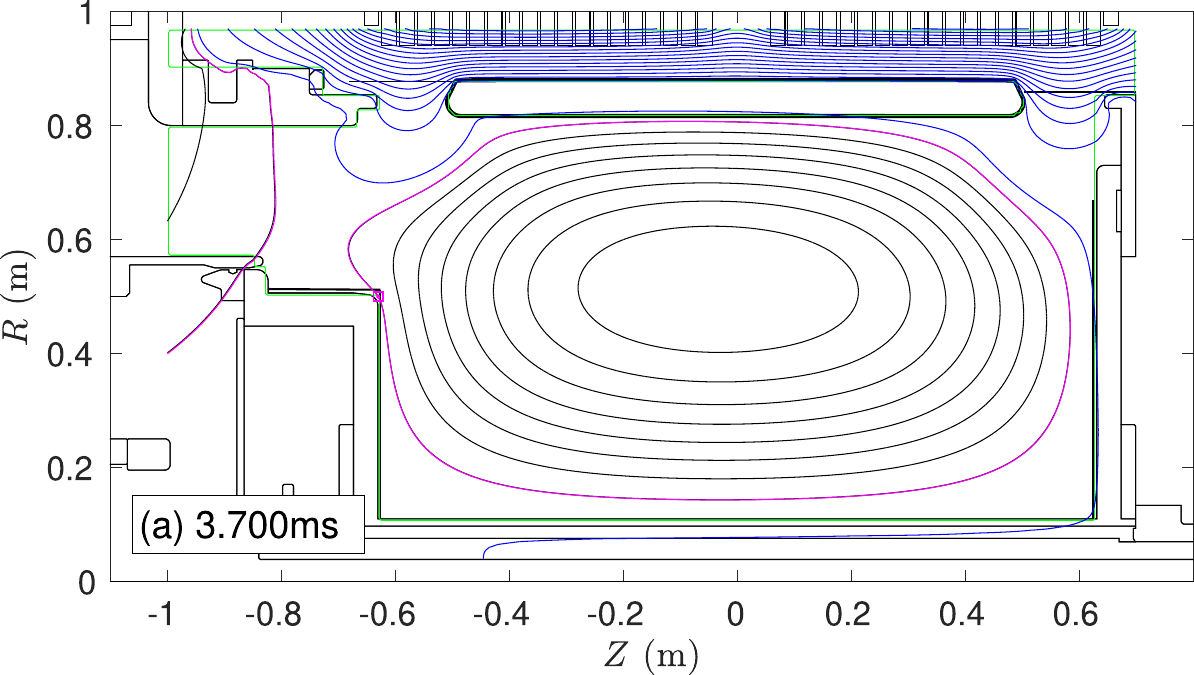}    \includegraphics[width=\linewidth]{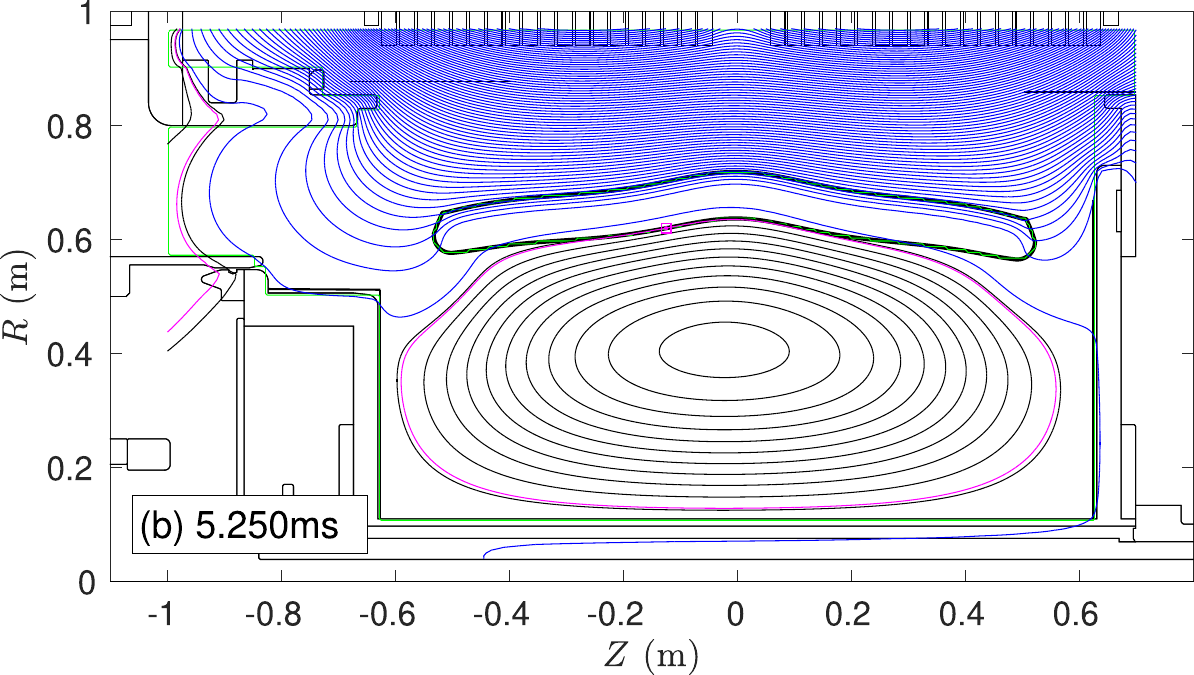}
    \includegraphics[width=\linewidth]{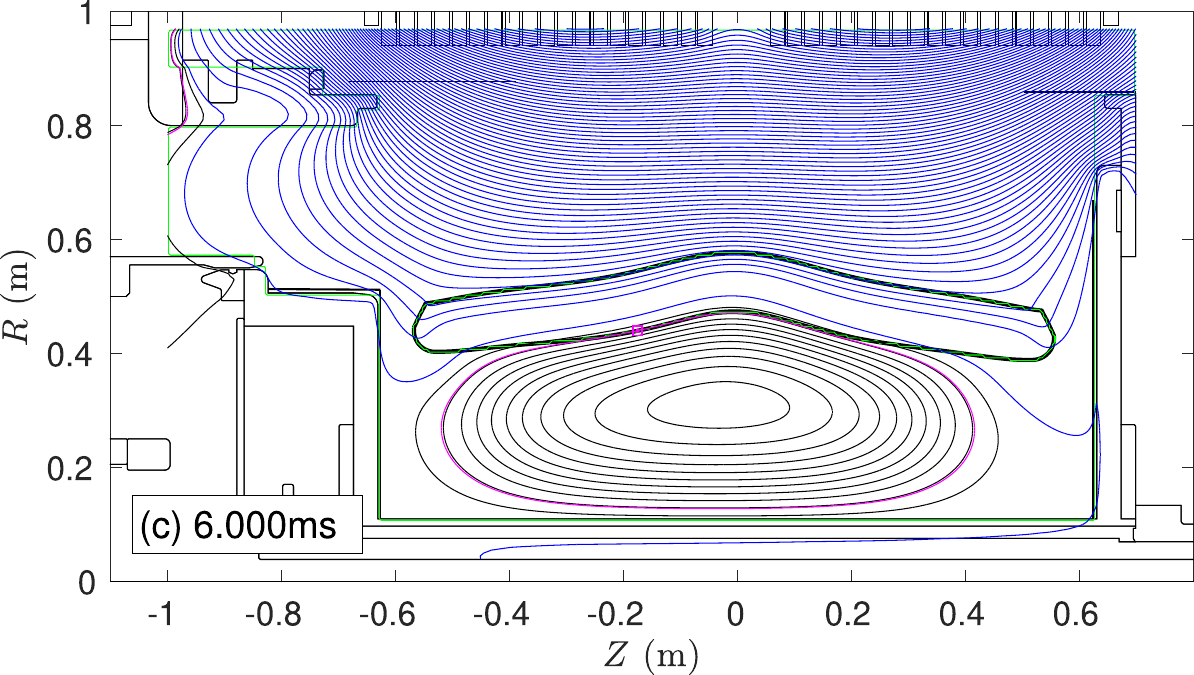}
    \caption{Example geometry during a PsiBC reconstruction of flux soak during compression of LMC-9, corresponding approximately to the times of the images shown in Fig.~\ref{fig:CameraData_LMC9}. 
    }
    \label{fig:psibc-geometry}
\end{figure}

\begin{figure*}
    \centering
    \includegraphics[width=\linewidth]{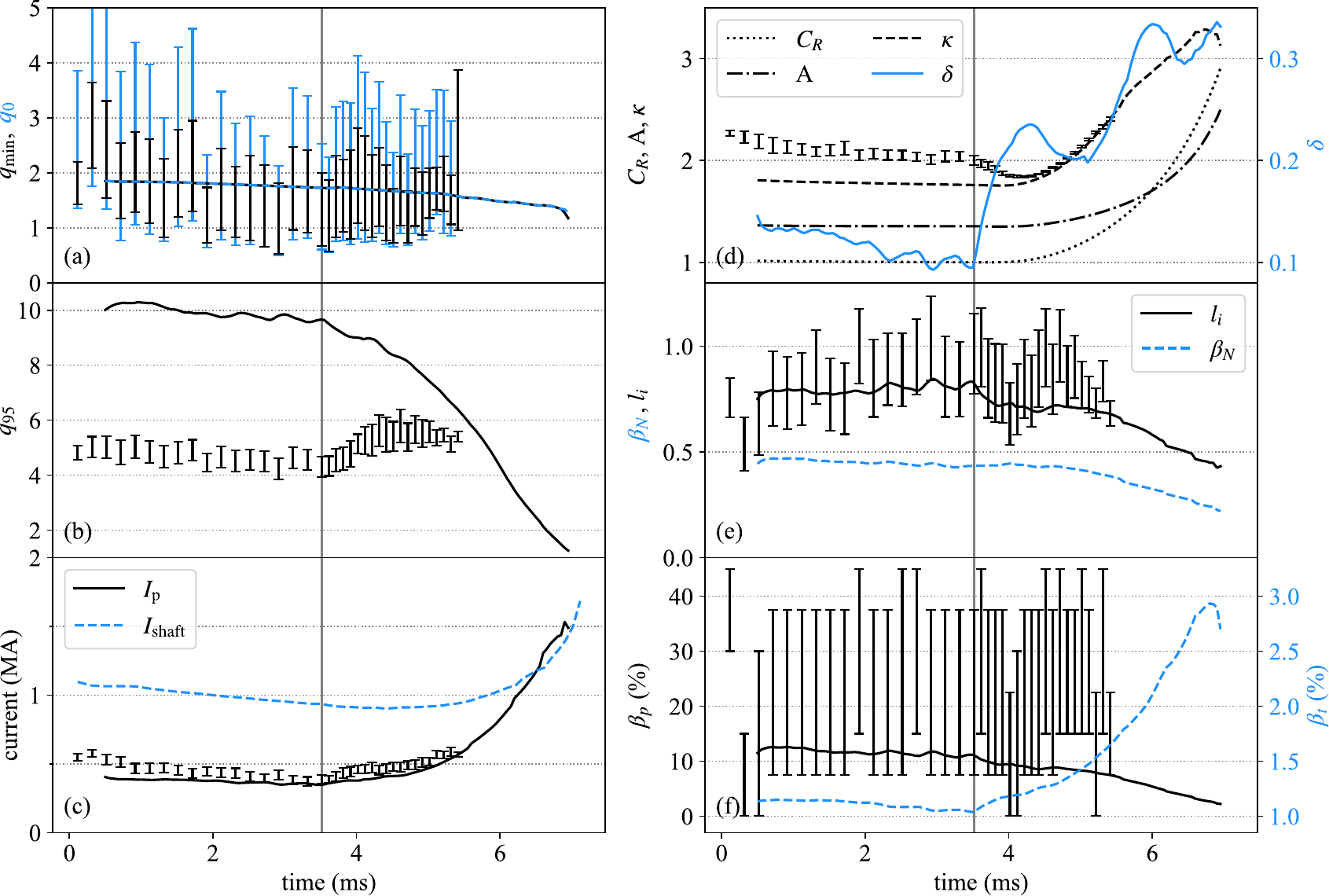}
    \caption{Geometric and plasma equilibrium parameters as a function of time after formation in LMC-9.  Compression starts at 3.5~ms, indicated by vertical lines. Curves are sourced from a time-dependent ISM fit.  Error bars are sourced from BPR reconstruction and show 1-$\sigma$ range.  (a) Safety factor minimum, $q_\mathrm{min}$ in black, and on axis $q_0$ in blue. (b) Safety factor at 95\,\% flux surface, $q_{95}$.  The ISM model does not include the Marshall gun, so the two models differ until the liner closes off the gun.  (c) Plasma current, $I_\text{p}$, and shaft current, $I_{\mathrm{shaft}}$. (d) Elongation, $\kappa$, aspect ratio, $A\equiv R/a$, and compression ratio, $C_R\equiv R_0/R(t)$, where $R_0$ is the initial major radial location of the outboard mid-plane and $R(t)$ is its value as a function of time. The elongation for the BPR model is initially higher because the plasma extends to the X-point in the gun.  The triangularity, $\delta$, is in blue with its axis on the right. (e) Internal inductance, $l_i$, and the Troyon metric $\beta_N\equiv \beta_t/(I_\text{p}/aB)$, where $I_\text{p}$ is the plasma current shown on the right axis in blue, $B$ is the toroidal magnetic field, $a$ is the minor radius. (f) The poloidal $\beta_p\equiv \mu_0 P/B_\mathrm{pol}^2$, with $P$ the pressure on axis, and the toroidal $\beta_t\equiv \mu_0 P/B^2$.}
    \label{fig:LM26_quants_shot9}
\end{figure*}

The plasma geometry changes dramatically during compression, as evidenced by the liner shape and poloidal magnetic flux function $\psi(R,Z)$, reconstructed at three times during LMC-9 shown in Fig.~\ref{fig:psibc-geometry}.  Reconstructions combine data on the liner location, magnetic fields, density and temperature to find fits to the equilibrium state as discussed below.  Starting with a liner inner radius of 80 cm, a plasma major radius of $\sim$47 cm, and a minor radius of $\sim$33 cm, the plasma is squeezed to a final minor radius of roughly 8.6 cm and a major radius of $\sim$20 cm at full compression, representing a $\sim$50-fold reduction in plasma volume.  Geometric and plasma equilibrium parameters from an ISM fit to LMC-9 are shown as curves in Fig.~\ref{fig:LM26_quants_shot9} as a function of time after formation, where the compression trigger at 3.5 ms is indicated by the vertical gray lines.

The safety factor profile $q$ is characterized by $q_\mathrm{min}$ in Fig.~\ref{fig:LM26_quants_shot9}a and $q_{95}$ in Fig.~\ref{fig:LM26_quants_shot9}b.  The safety factor profile is monotonic throughout, so $q_0=q_\mathrm{min}$ and both are well preserved with $1.3<q_\mathrm{min}<1.8$ for most of compression.  In contrast, the near edge values decrease as flux surfaces diffuse into the edge during compression.  Initially, the plasma is diverted and $q_{95}\sim 5$.  When the compression coils are energized, the compression field squeezes around the liner and separates the plasma from the gun flux, which leads to a brief increase in $q_{95}$.  The plasma then becomes limited and as the poloidal flux diffuses into the liner, the edge $q$ falls rapidly.  Once $q_{95}$ drops to less than $3$, $n=1$ magnetic instabilities appear (see Fig.~\ref{fig:axuv-Te-vs-CR} for the occurrence of magnetic instability on LMC-9 and LMC-11).

The plasma current $I_\text{p}$ is shown in Fig.~\ref{fig:LM26_quants_shot9}c.  It starts at 500 kA, half the shaft current, and slowly decreases by 20\,\% until compression starts.  Then it grows to 1.5 MA and becomes larger than the shaft current.  The aspect ratio $A\equiv R/a$, elongation $\kappa$, compression ratio $R_0/R$, and triangularity $\delta$ are shown in Fig.~\ref{fig:LM26_quants_shot9}d.  Triangularity increases from the initially rectangular cross section with $\delta\sim 0.1$ to a more D-shaped plasma with $\delta\sim 0.3$.  The early-compression increase in triangularity at 3.7~ms is driven by the flux from the compression coils pushing around the ends of the liner, while the second increase at 5.5~ms occurs as the liner compresses the plasma against the shaft (Figs.~\ref{fig:LM26_schematic}, \ref{fig:psibc-geometry}).  Elongation and aspect ratio steadily increase from initial values of $1.8$ and $1.4$ respectively to $3.4$ and $2.5$ by the end of compression.  The fixed radius of the shaft is the primary driver to increasing $A$, while the increase in $\kappa$ reflects the cylindrical compression produced by the liner trajectory.  Though toroidal $\beta$ increases, as shown in Fig.~\ref{fig:LM26_quants_shot9}f, the normalized $\beta$ slightly decreases through compression as shown in Fig.~\ref{fig:LM26_quants_shot9}e, and remains well below the Troyon limit of $4l_i$, heuristically indicative of $\beta$ limits in tokamaks.  This suggests any observed MHD instabilities are likely to be current driven.

\subsection{Equilibrium reconstruction}\label{sec:EqRecon}

We reconstruct the evolving plasma equilibrium state in the experiment by two separate methods: reconstructions of sequential time slices, and fitting to time-dependent integrated simulations of the full compression. Both methods rely on the fact that the velocity of the liner is much slower than the Alv\'en velocity, so the plasma is always in quasi-equilibrium and, therefore, governed by the Grad--Shafranov (GS) equation~\cite{Grad1958HYDROMAGNETICEA}.

\subsubsection{Diffusion of poloidal flux} \label{sec:PsiBC}

The poloidal flux sourced by plasma current diffuses into the liner and other metal during the discharge.  This results in time-dependent magnetic boundary conditions that are reconstructed using the PsiBC code.
PsiBC takes as input the COMSOL liner reconstruction result: time-dependent geometry and poloidal field $\psi_0(R,Z,t)$ due to compression coil and bias coil currents and the currents they induce in the metal.  Then, using the Mirnov data from experiment, PsiBC reconstructs the field generated by plasma current, including the effect of skin currents induced in the metal by plasma current, by fitting a GS equilibrium at each time point. Between time points, the diffusion of the magnetic field into the metal is simulated while holding the plasma current density constant.  And, as the liner moves, the magnetic flux in the liner is advected.  In this manner, PsiBC produces the total time-dependent field $\psi(R,Z,t)=\psi_0(R,Z,t)+\delta\psi(R,Z,t)$, where $\delta\psi(R,Z,t)$ is the additional poloidal flux due to plasma current and eddy currents induced in the metal by the plasma.  This total field is saved for use as a boundary condition in subsequent plasma reconstruction using the Bayesian Plasma Reconstruction (BPR) code~\cite{Howard_2025}, which is described in the next section.
A more detailed description of the PsiBC code is given in Appendix \ref{app:psibc}.

We find that useful preliminary interpretation of the plasma data can be done using the magnetic flux surface geometry produced by PsiBC, despite the simplicity of its plasma model.  Interferometer data, for example, can be fit to characterize the density profile and electron inventory, as was shown in Fig.~\ref{fig:lmc-8-inventories}.

\subsubsection{Bayesian plasma reconstruction}
\label{sec:BPR}
More complete and flexible reconstruction of the plasma state is performed with the BPR code.  It employs Bayesian inference to match the magnetic probes to tabulated plasma equilibria~\cite{Howard_2025, Pi3_tauE_NucFus_2025}.  For selected points in time throughout the shot, we precalculate a multi-dimensional lookup table of GS equilibria based on the changing wall geometry and poloidal flux boundary conditions calculated by PsiBC at the surface of the flux conserver.  
Each equilibrium in the table is weighted by its fit quality, defined as the normalized likelihood $w = \exp(-\chi^2/2)$, where $\chi^2$ is the sum of squared normalized residuals between the forward-modeled magnetic fields and Mirnov probe measurements. That is, $\chi^2=\sum_{probes}\left((B_m-B_s)/\sigma\right)^2$.
These weights are normalized to sum to unity across the table and used to compute posterior statistics for each output parameter, such as poloidal flux or $q$ profile.

The GS equilibria are calculated with the Flagships code~\cite{BayesianRecon_APS2022}, assuming no toroidal current outside the LCFS.  The fitting space is regularly sampled over six parameters, which typically include the plasma current, current density profile, and $\beta_\mathrm{pol}$. The plasma current profile is constrained to be of the form~\footnote{This current profile equation is attributed to William M. Nevins in the CORSICA code base.}
\begin{equation}
f(\bar{\psi}) = \exp\left[ a \bar{\psi} + b \bar{\psi}^2 - (y+a+b) \bar{\psi}^{n} \right],
\end{equation}
where the free parameters are $a$, $b$, $y$, and $n$.  The behaviour of the current density near the core is determined by $a$ and $b$, the current density at the LCFS is determined by $y$, and the internal inductance of the plasma is determined by $n$.

While BPR can produce detailed multi-dimensional probability density functions, Fig.~\ref{fig:LM26_quants_shot9} shows the error for LMC-9 presented more simply. 
The error bars for each output parameter span the 16th–84th percentile range of the posterior distribution, where each equilibrium is weighted by its normalized fit quality such that all weights sum to unity.
The time-dependent ISM fit curves in the plot are described in the next section, but it is clear that in most cases they are enclosed within the error region calculated by BPR.  The two exceptions are the safety factor at the 95\,\% flux surface, $q_{95}$ in Fig.~\ref{fig:LM26_quants_shot9}b, and the elongation, $\kappa$ in Fig.~\ref{fig:LM26_quants_shot9}d, which differ because only the BPR model includes the gun region.  The BPR model includes the gun flux, so the plasma becomes diverted and extends into the gun to an X-point, which results in greater elongation and a smaller $q_{95}$.  This difference between the BPR and ISM models becomes less significant as the liner closes off the gun region during compression. 

\subsubsection{Time-dependent reconstruction} \label{sec:time-dep-recon}

Bayesian methods offer a rigorous probabilistic approach to reconstructing instants in time independently.  This can be highly advantageous for understanding how the accuracy of the fitting evolves in time, and what is driving uncertainty, such as mode activity.  In steady state scenarios, such approaches can be sufficient.  However, in the dynamic environment of an MTF compression, the inclusion of time-dependent physics to the reconstruction process can facilitate understanding from an entirely different perspective.  The combination of these two approaches is often necessary to fully understand the experimental outcome. 

The Integrated System Model (ISM) is a comprehensive code framework developed by GF that provides time-dependent reconstructions and predictions for the behavior of magnetized target fusion systems \cite{khalzov_ism_2021, khalzov_ism-plasma_2023, Khalzov_2024}. The plasma component of the framework, ISM-plasma \cite{khalzov_ism-plasma_2023}, is a 1.5D plasma transport code based on the assumption that the evolution of a magnetically confined plasma can be represented as a sequence of 2D axisymmetric Grad--Shafranov equilibria linked through 1D transport processes (resistive and thermal diffusion, etc.). 
ISM includes plasma evolution with resistive diffusion of magnetic fields into the surrounding structure, self-consistent motion of the metal liner, and mechanisms for driving the liner. 
For reconstructions of the plasma, the liner motion is specified from experimental data via the liner trajectory reconstruction process (Sec.~\ref{sec:linerRecon}). Synthetic diagnostic outputs from ISM allow direct comparison with plasma experimental data, such as magnetic probes, interferometers, and AXUV filter ratios, which are used to constrain the thermal confinement in the plasma via a $\chi^2$ approach.

In the studies described here, ISM-plasma uses a two-temperature model, with electrons at temperature $T_e(\psi)$ and all ion species assumed to be in thermal equilibrium at temperature $T_i(\psi)$. Both temperatures are assumed to be 10\,eV at the LCFS, as modifying this value did not have a significant effect on the results.
Thermal exchange between electrons and ions is included. Thermal transport is modeled by simplified perpendicular thermal diffusivities for electrons and ions, $\hat\chi_e$ and $\hat\chi_i$ respectively. For the purposes of reconstruction, these are effective diffusivities which lump together all thermal transport processes including radiation.
Magnetic flux and particles are gradually lost from the edge of the plasma domain as the magnetic field resistively diffuses across the plasma or soaks into the liner. Particles are otherwise conserved in this model. Further details of the model are described in Appendix~\ref{app:ISM}.

\subsection{MHD stability}\label{sec:stability}

In all axisymmetric toroidally-confined plasmas, as higher performance plasma states are reached, instabilities can appear suddenly and lead to disruptive events where most of the confined energy is rapidly lost~\cite{gerhardt2013disruptions,gerhardt2013detection,deVries2011,pautasso2014disruption,eidietis2015itpa}.  Disruptive events have been observed in previous compression experiments at GF \cite{Howard_2025}, evidently caused by instabilities driven by magnetohydrodynamic (MHD) limits.  As the plasma is compressed, the geometry of the flux conserving walls and plasma properties change, potentially carrying the plasma across a stability boundary.  

The key driving factors governing the stability of the plasma are the cross-sectional shape, safety factor profile, current density profile, temperature profile, and density profile. These are all affected by operating parameters and internal plasma processes. Operating parameters, such as the liner trajectory, shaft current over time, plasma formation power, and timings of the gas puff, plasma formation, and start of compression, influence the plasma shape, initial conditions, and boundary conditions. Plasma processes, such as resistive transport, flux loss into the boundary, and impurity transport, determine the evolution of the profiles. Plasma stability can be maintained well into compression with carefully selected operating parameters.

The stability of equilibria reconstructed by ISM from experimental data are analyzed with the Resistive DCON code (RDCON) \cite{glasser-2016-dcon,glasser-2016-rdcon,Brennan2020,Brennan2021}, and the NIMROD extended MHD code \cite{Sovinec04,brennan2005categorization}.  RDCON calculates both the ideal MHD perturbed potential energy, $\delta W$, and the resistive MHD stability based on the method of matched asymptotic expansions in full toroidal geometry.   
In this analysis, the linear form of NIMROD is used to evolve MHD perturbations over an equilibrium state, including resistivity and viscosity throughout the plasma.  Both codes implement a small vacuum region surrounding the plasma. 

If the $\delta W$ calculated in RDCON of any mode is negative, this represents a lower energy state with the perturbations, and the equilibrium is ideal unstable.  If the equilibrium is ideal stable, it can still be unstable to slow-growing resistive modes.  The free energy available to a non-ideal (i.e. resistive) mode given some resistivity in a surface layer is described by the local stability index $\Delta^\prime$, calculated from the resonant response at each layer.  

In general, at each surface in the plasma where the safety factor matches the ratio of integer poloidal and toroidal mode numbers, $q=m/n$, a resonance occurs in the ($m$,$n$) component of the MHD mode.  Such surfaces will form resonant layers in the plasma, with highly localized currents sensitive to the resistivity of the plasma.  The resistivity allows for magnetic reconnection and interchange in the layer, allowing the global mode to grow at a rate determined by the physics in the layers. 
Depending on the growth rate, structure and location of the instability, it could be observed in experiment, with consequences ranging from disruptive to benign. 

\begin{figure}
    \centering
    \includegraphics[width=\linewidth]{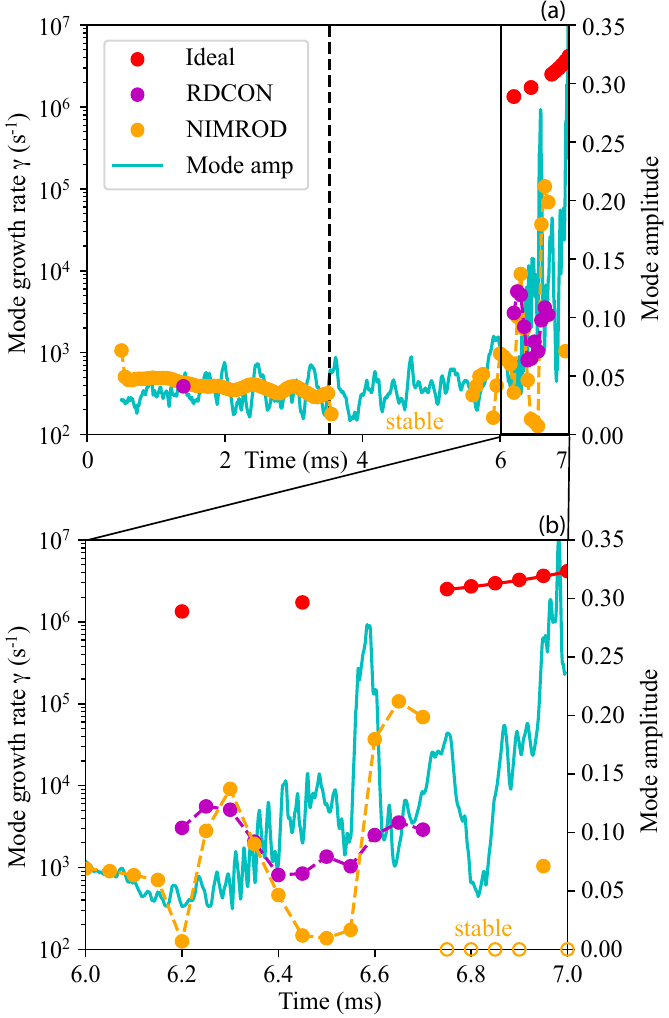}
    \caption{The resistive MHD growth rates of the $n=1$ mode in LMC-9 from RDCON (magenta), from NIMROD (orange), 10\,\% of the inverse Alfv\'en time when ideal instability is found (red), and the experimentally determined mode amplitude from magnetic probe measurements 90$^\circ$ apart (cyan), for the full discharge (a) and deep in compression (b).  Dashed lines connect sequential unstable points, with gaps where they are found stable. Very early the mode is marginal with one low growth rate found by RDCON and a steady low growth rate found by NIMROD, but evidently too low and short lived to manifest.  Through most of compression the plasma remains stable. Deep in compression the plasma is found unstable just as an instability grows and the shot terminates.}
    \label{fig:ISM_stability_8n9}
\end{figure}

 Overall, analyses have been conducted on every compression shot, and indicate that the plasma profiles at the start of compression, as well as how quickly the shaft current is ramped during the early compression before trapping the toroidal flux, are critical to subsequent stability.  The results from a stability analysis of the ISM reconstructions of LMC-9 are summarized in Fig.~\ref{fig:ISM_stability_8n9}.

The safety factor $q_{a}$ at the LCFS drops significantly during LMC shots from initial values $q_a\sim 20$ to $q_a<4$ near the end of compression as the plasma begins to terminate.  Flux loss to the boundary reduces the edge $q$ after compression starts, while the core plasma $q$ remains relatively steady, slowly decreasing, indicating good core confinement.  Initiating the plasma with higher edge $q$ and ramping the shaft current during compression both counter the effect of flux loss to the boundary.  As compression begins, the liner must first traverse the radial gap to the endplates to make contact and complete the poloidal circuit that traps the toroidal flux.  During this traversal, the shaft current is ramped to mimic the trapping of toroidal flux.  The ideal ramping scenario would track the inverse inductance $I_\mathrm{shaft}(t)\propto L_0/L(t)$ where $L_0$ is the plasma inductance at the start of compression ($L=\int_\mathrm{LCFS} R^{-1}\,dR\,dZ$).  With insufficient ramping, the $q$ profile in the edge region can fall rapidly, causing global MHD instability.  With sufficient shaft current ramping, the $q$ profile and stability can be maintained, though it is not necessarily advantageous to track the ideal ramping rate exactly.  In practice, lagging the ideal rate induces edge current in the plasma, which heats the edge plasma via Ohmic heating, possibly improving performance.  The shots presented here all lag the ideal rate to some degree.  

Before compression, the plasmas are typically found to be stable to disruptive ideal and resistive MHD instabilities in experiment and analyses.  In some cases, early very slow-growing resistive modes ($\gamma\lesssim 2\times 10^{3}\mathrm{\,s^{-1}}$) are found in stability analyses that are either too slow to be observed on the timescale of the experiment or are, in fact, stable.  In Fig.~\ref{fig:ISM_stability_8n9}, one early resistive instability is calculated to be weakly unstable by RDCON, with growth rate $\sim 4\times 10^2\mathrm{\,s^{-1}}$, while NIMROD finds a similar steady low growth rate.  This growth rate is too low to manifest on the experimental timescale, where $10^{4}\mathrm{\,s^{-1}}$ or above is a concern, and would likely nonlinearly saturate at small size.  Because $q_\mathrm{min}\sim 1.85$ in the reconstruction, as shown in Fig.~\ref{fig:LM26_quants_shot9}, the $q=2$ surface is relatively close to the magnetic axis in low shear, with variable contribution to the stability as the equilibrium changes.  The $q=3$ surface is more steady, and is a stabilizing influence.  The combination causes an intermittent instability for RDCON in Fig.~\ref{fig:ISM_stability_8n9}.  NIMROD's model includes viscosity and is more accurate for these plasmas.  However, there are not significant signals above the noise level on the magnetic probes that would suggest growing instabilities, as evidenced by the relatively flat $n=1$ mode amplitude until $\sim$6\,ms.

Shortly after compression starts, the flux from the compression coils pushes through the gaps at the axial ends of the liner, which pushes the flux surfaces radially inward in both regions.  This effect can be seen on the top plot of Fig.~\ref{fig:psibc-geometry} and it typically has the effect of increasing the triangularity just as the liner has begun to move, when $C_R\sim 1$, as shown in Fig.~\ref{fig:LM26_quants_shot9}.  The triangularity during compression falls in the range $0.1<\delta<0.4$, which tends to improve stability, as can be seen by the stabilizing influence on NIMROD in Fig.~\ref{fig:ISM_stability_8n9}.  Although, triangularity can be destabilizing if it increases further~\cite{DovRhodes}.  An interesting aspect of these shots is that they are conducted in positive triangularity, whereas previous studies~\cite{Brennan2020,Brennan2021}, and the target configuration for upcoming experiments in LM26 are in negative triangularity.  However, even in these negative triangularity cases, deep in compression the triangularity can reverse to positive as the plasma is compressed against the center shaft.  

Later in compression, as the edge shear in $q$ is reduced, resistive instabilities are found unstable in some cases, with growth rates $\gtrsim 10^4\mathrm{\,s^{-1}}$.  Above this growth rate the mode can $e$-fold more than ten times in a millisecond, and therefore may become an observed instability as its growth is fast enough to surpass the linear regime and transition to a nonlinear state within the timescale of the experiment.  Even if the mode is marginally stable, or too slow-growing to be observed, such marginal modes are highly susceptible to error fields, which can drive an instability in a manner nearly indistinguishable from spontaneous instabilities in practice~\cite{finn2015error}.  In Fig.~\ref{fig:ISM_stability_8n9}, late in compression, the plasma is found ideal unstable by RDCON as indicated by the red dots.  These markers are placed at 10\,\% of the inverse Alfv\'en time $0.1/\tau_A$ to indicate an upper bound for what is expected of the ideal growth rates of the mode.  In two of these cases at 6.2\,ms and 6.45\,ms, the $\delta W$ briefly turns slightly negative as the edge $q$ of the plasma approaches $3$ and then $2$ respectively.  The resistive stability growth rates from RDCON and NIMROD are initially in good agreement, increasing as the $q=3$ edge is crossed, though the RDCON growth rates are significantly below the NIMROD result as the edge $q$ approaches $2$.  In these final moments, the RDCON assumption of localized current layers at the $q=2$ surface are less accurate than the NIMROD model, which makes no such assumption, as the temperature drops and the resistive layers become too wide.  

Approaching 7\,ms, the $q=2$ surface leaves the plasma, and RDCON finds ideal instability as the plasma terminates. The red dots in Fig.~\ref{fig:ISM_stability_8n9} are only markers of the upper bounds expected of ideal modes, while without resonant surfaces any unstable mode is driven by non-resonant mechanisms internally, and may be related to the external kink \cite{turnbull2016external}.  If an ideal growth rate were calculated for these times, it may in fact be significantly lower than the markers.  NIMROD finds most of the times from 6.75\,ms to 7\,ms as stable, except for a slow growing case at 6.9\,ms.  NIMROD's model includes resistivity, viscosity and adiabatic pressure, which may be responsible for the discrepancy.  In any case, at the mode amplitudes observed in experiment, these instabilities are nonlinear and therefore the linear picture is only partially relevant, though their evolution may include linear drive \cite{Brennan2003,brennan2005categorization}.

Though the total plasma current is well constrained, it remains a challenge to reconstruct the detailed structure of the current profile, in particular the initial hollowness at the start of compression.  Hollowness in the initial current profile is primarily driven by the CHI formation process, where current is driven on outer flux surfaces and relaxes minor-radially inward.  How hollow the current profile is can directly affect stability because it determines $q_\mathrm{min}$ as well as the current gradient across surfaces, two important factors to stability.  

Note the BPR reconstructions of LMC-9 indicate a higher initial $q_\mathrm{min}\gtrsim 2$ than the best fit from ISM, where $q_\mathrm{min}\sim 1.5$ initially.
In fact, reconstructions from either BPR or ISM with a $q_\mathrm{min}$ value in close agreement with the other would show little difference in $\chi^2$ from their best fits, as the minima in $\chi^2$ are somewhat shallow.  This fact also manifests as a significant variability in $q_\mathrm{min}$ as a function of time reconstructed with BPR, while the ISM values are constrained to vary smoothly by the time-dependent physics of the integrated simulation.  However, in experiment and stability analyses, $q_\mathrm{min}$ can strongly affect the resistive stability, especially when the central value is near a $m/n$ resonance.   A point of observation in experiment is that there is no evidence of characteristic transient instability as $q_\mathrm{min}=2$ is crossed in either LMC-8 or LMC-9.  This suggests $1.7<q_\mathrm{min}<2$ early in these shots, as per $\chi^2$ of the reconstructions.  Thus, the case shown in Fig.~\ref{fig:LM26_quants_shot9} and \ref{fig:ISM_stability_8n9} is chosen taking both the ISM and BPR fits into account, with $q_\mathrm{min}\sim 1.85$.

As discussed above, stability analyses indicate instability late in compression as $q$ edge falls to $q_a<4$, and ideal instability is found should $q_a$ cross $2$, as is typical in tokamak discharges \cite{turnbull2016external}.  
In the experiment, MHD activity is often observed to begin growing at an earlier time than the stability analyses indicate. 
One aspect of the onset of instability is that they are first detected at nearly the same compression ratio $C_R\sim 1.8$ in shots where they are observed, except for LMC-11 where the mode appears at $C_R\sim 2.2$. 
This fact suggests a mechanism based on plasma geometry, although other interpretations can not be ruled out.  Buckling of the liner during compression, as discussed in Sec.~\ref{sec:fastcameras} could lead to magnetic perturbations as the current flowing through the liner becomes displaced.  As discussed above, error field penetration can drive instabilities that can be difficult to distinguish from spontaneous instabilities \cite{finn2015error}.  This is especially true near marginality for the mode, where a finite frequency to the mode can appear, which translates into a propagating mode stemming from a static error field.  Indeed related error field interactions involving finite frequencies can persist into the nonlinear regime with large islands \cite{fitzpatrick2014linear}.

The stability analyses offer the ability to explore modifications of the experimental conditions which can improve stability, by building understanding of the driving mechanisms involved.  Shortly after formation, as $q_\mathrm{min}$ decreases, assuming a fixed profile shape, the locations of the $q=2$ and $3$ surfaces, along with all others, are moved outwards toward the edge.  As the surfaces move outward, the shear in $q$ increases and the distance to the conducting wall decreases, and in a simplified view of Furth, Rutherford and Selberg \cite{FRS}, the plasma becomes less easily driven unstable.  Also, the shaft ramp current, which primarily supports the near edge current profile, is more effective for near edge surfaces.  However, the temperature generally decreases during this pre-compression period, having been driven up by the initial formation process, and now reliant solely on Ohmic heating to counter transport losses.  As such, stability concerns must be balanced against the advantages of starting compression with sufficient plasma performance.  Overall, the goal is to achieve a compressed plasma state that achieves high confinement and remains stable through compression. 

\subsection{Simple models for ion temperature}
\label{sec:simpleIons}

As shown in Secs.~\ref{sec:AXUVresults}--\ref{sec:IonDoppler}, the ion temperature does not heat as much as electron temperature during compression.
Our initial interpretation of the ion temperature evolution is based on the preliminary plasma reconstruction provided by PsiBC.
We start by using a simple zero-dimensional (0D) compressional heating model (described in Ref.~\cite{Howard_2025}), which we call ``Model A'':
\begin{equation}
    \frac{T_i(t)}{T_i(0)}=\left[\frac{V(0)}{V(t)}\right]^{2/3}\!\exp\left(-\hat\chi_i\!\int_0^t\!  k(t')^2\,dt'\right)
    \label{eq:model-1}
\end{equation}
where $\hat\chi_i$ is an adjustable constant and $k(t)=2.4/a(t)$ with $a$ the minor radius, where 2.4 the first zero of the Bessel function $J_0$ \cite{Howard_2025}.  For the volume $V(t)$ we use the reconstructed volume of a small flux surface with constant toroidal flux.  The basic assumption underlying this model is that all ion thermal losses can be lumped into an energy confinement time $\tau_{E,i}$ that scales during compression as $a^2$.  The constant $\hat\chi_i$ can be interpreted as an effective thermal diffusivity, and serves as a way to characterize $\tau_{E,i}$.
In Fig.~\ref{fig:lmc-8-doppler-Ti} we compare this model to reconstructed $T_i$ and find that a value $\hat\chi_i\simeq 4\mathrm{\,m^2/s}$ is sufficient to prevent significant compressional ion heating, while $\hat\chi_i\simeq 2\mathrm{\,m^2/s}$ would result in a significant temperature rise.  
This simple model does not distinguish collisional electron-ion energy exchange (Equation~\ref{eq:e-i}) from cooling of the ions due to actual ion heat transport.  The collisional energy exchange depends on both the collision frequency and the temperature difference between electrons and ions. Therefore, collisional cooling of ions should diminish and even become a heating term for the ions as the electron temperature rises to exceed the ion temperature during compression as shown by the AXUV measurements. The absence of a significant ion temperature increase therefore indicates that ion heat transport is a dominant loss mechanism during the compression phase of the experiment.

\begin{figure}
    \centering
    \includegraphics[width=\linewidth]{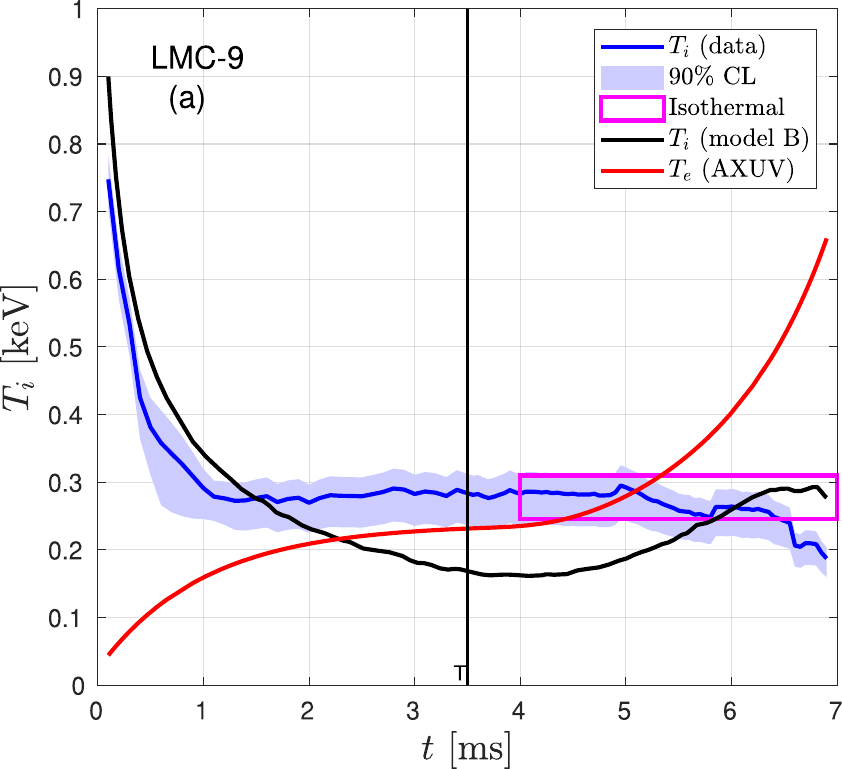}
        \includegraphics[width=\linewidth]{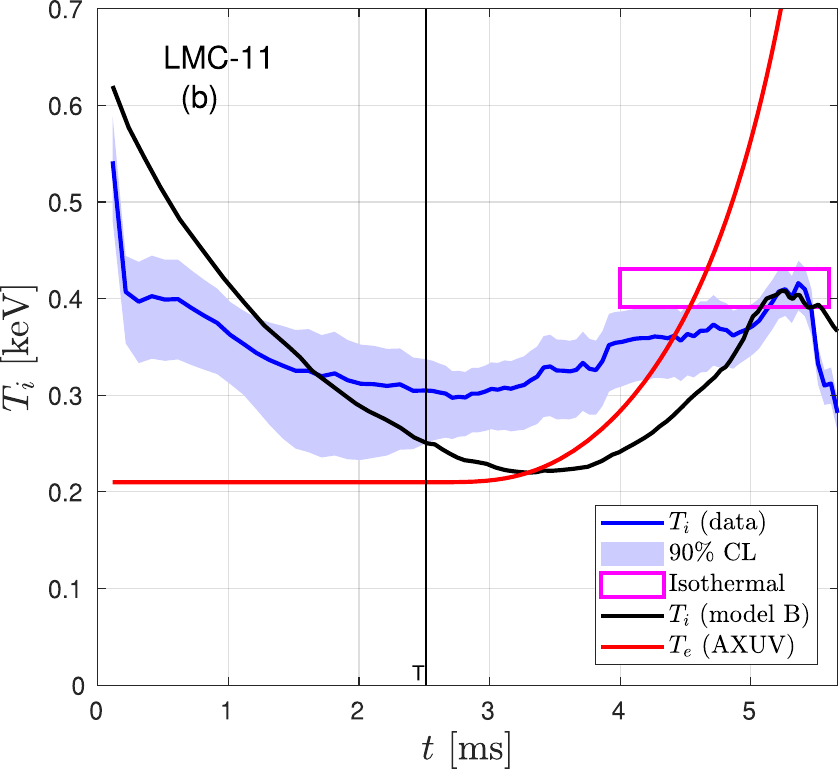}
    \caption{Ion temperature from neutron yield data (blue curve and confidence band) for shot LMC-9 (upper panel) and LMC-11 (lower panel).  The magenta box indicates the $T_i$ from the isothermal analysis of scintillator data (Table~\ref{tab:isothermal-Ti}, max Li case). Also shown are model results for $T_i$ (black curve) using $T_e$ from experiment (red curve) in the simple zero-dimensional Model B, given by equation (\ref{eq:model-2}), that separately accounts for e-i collisions, ion heat transport ($\hat\chi_i=4\mathrm{\,m^2/s}$), and compressional heating.  Electron temperature is based on the AXUV data during compression. Compression starts at time indicated by vertical line.  In LMC-9 an {\it ad~hoc} early-time $T_e(t)$ is imposed to illustrate the possible initial fast cooling of ions by e-i collisions.  In LMC-11 a constant early-time $T_e$ is assumed.
    }
    \label{fig:lmc-9-thermal-model}
\end{figure}

We next apply a more detailed (but still 0D) model that separately accounts for electron-ion energy exchange, which we call ``Model B'':
\begin{equation}
    \dot T_i=-\frac23
    \frac{\dot V}{V} T_i - \hat\chi_i k^2 T_i + \bar\nu_\epsilon^{i|e}(T_e-T_i)
    \label{eq:model-2}
\end{equation}
Fig.~\ref{fig:lmc-9-thermal-model} compares the neutron $T_i$ data from LMC-9 and LMC-11 to this model.
For the rate-of-compression factor $\dot V/V$ we again use the volume $V(t)$ of a small flux surface with constant toroidal flux.
The \mbox{e-i} collisional term involves rate $\bar\nu_\epsilon^{i|e}$, defined in Appendix~\ref{app:ISM}, depending on $n_e$ and $T_e$.  For $n_e(t)$
we use the pre-compression core density from interferometry, scaling it during compression inversely as the volume $V(t)$.  For $T_e$ we specify $T_e(t)=(238\mathrm{\,eV})C_R^{0.971}$ for LMC-9 and $T_e(t)=(210\mathrm{\,eV})C_R^{1.53}$ for LMC-11, on the basis of AXUV data (Fig.~\ref{fig:axuv-Te-vs-CR}).  Early in the discharge, the decay of neutron yield allows us, in principal, to estimate a lower bound for $T_e$ by attributing the decay to ion cooling via e-i collisions.  In the case of LMC-9 we use an {\em ad hoc} $T_e(t)$ during the pre-compression phase to illustrate, in Fig.~\ref{fig:lmc-9-thermal-model}, a possible scenario in which $\hat\chi_i$ is constant and the initial ion cooling is dominated by e-i collisions.  The model curves are calculated using for the effective thermal diffusivity the value $\hat\chi_i\simeq 4\mathrm{\,m^2/s}$. The initial $T_i$ is chosen  so the curve passes through the compressed $T_i$ determined by the isothermal analysis of Table~\ref{tab:isothermal-Ti}.  
For the pre-compression geometry, the chosen value of $\hat\chi_i$ corresponds to an ion energy confinement time $\tau_{E,i}\simeq 5\mathrm{\,ms}$.  Note that when comparing to more sophisticated models that properly account for density profile and flux surface geometry, $\tau_{E,i}$ is a more appropriate quantity than $\hat\chi_i$.

More sophisticated 1D modeling of the electron and ion temperature evolution using ISM-plasma is described in Sec.~\ref{sec:CompressHeatConfine}.  The levels of ion heat transport inferred from the experiment using ISM-plasma are similar to those appearing in the simple models described above, when compared on the basis of energy confinement time.

\subsection{Compressional heating and confinement}\label{sec:CompressHeatConfine}

Time-dependent reconstruction is performed using the ISM code described in Sec.~\ref{sec:time-dep-recon}.
The ISM reconstructions of LMC-9 and LMC-11 are constrained by magnetic field probes, interferometers, AXUV filter ratios, Thomson scattering (for LMC-11 only), and inferred ion temperature. In addition to providing direct magnetic information, the time evolution of the magnetic field probes yields information about electron temperature due to the resistive evolution of the plasma. The time-dependent approach of ISM reconstruction leverages this relationship, along with other time-dependencies like electron-ion heat exchange, leading to a higher degree of consistency across measurement domains.

Another benefit of the time-dependent reconstruction approach is that  translation of the compressing plasma across AXUV view cones reveals information about the gradient of electron temperature. A two-parameter $\hat\chi_e$ profile was therefore used in the ISM reconstruction to improve agreement with observed AXUV data,
\begin{align}
    \hat\chi_e(\bar\psi) = \hat\chi_{e}^\text{axis} + (\hat\chi_e^\text{LCFS}-\hat\chi_{e}^\text{axis}) \bar\psi^2, \label{eq:chi_e_prof}
\end{align}
where $\hat\chi_{e}^\text{axis}$ and $\hat\chi_{e}^\text{LCFS}$ are the effective electron thermal diffusivities at the magnetic axis and LCFS, respectively. 
Here $\bar\psi$ is, as usual, defined to be linear in $\psi$ and having values $0$ at the magnetic axis and $1$ at the LCFS.
The ion thermal diffusivity was assumed to be uniform, $\hat\chi_{i}^\text{axis}=\hat\chi_{i}^\text{LCFS}$, as there is insufficient data to constrain the ion temperature profile.

It is important to note that the $\hat\chi_i$ parameter in ISM simulations can have a rather different value than in the 0D simulations of Sec.~\ref{sec:simpleIons}.  This is because in ISM the density profile is taken into account, whereas in the 0D simulations the density profile is implicitly incorporated into $\hat\chi_i$.  Also, the ISM utilizes full flux surface geometry whereas the 0D models assume cylindrical flux surfaces. The relevant quantity for  comparison is instead the energy confinement time, $\tau_{E,i}$.

\begin{figure}
    \centering
    \includegraphics[width=\linewidth]{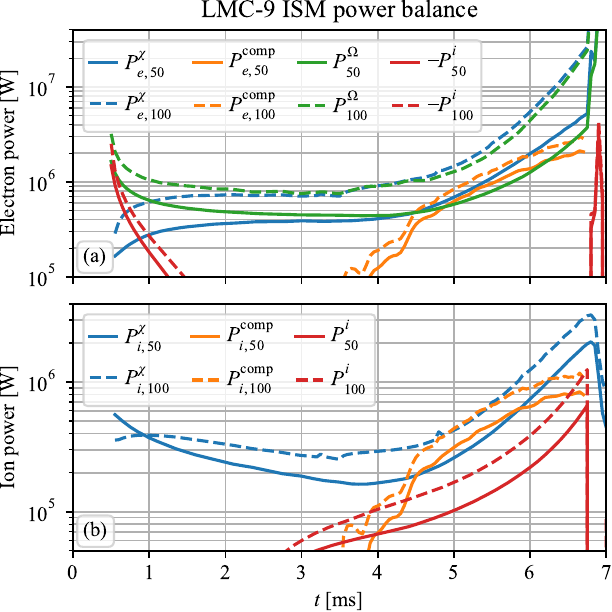}
    
    \medskip 
    \includegraphics[width=\linewidth]{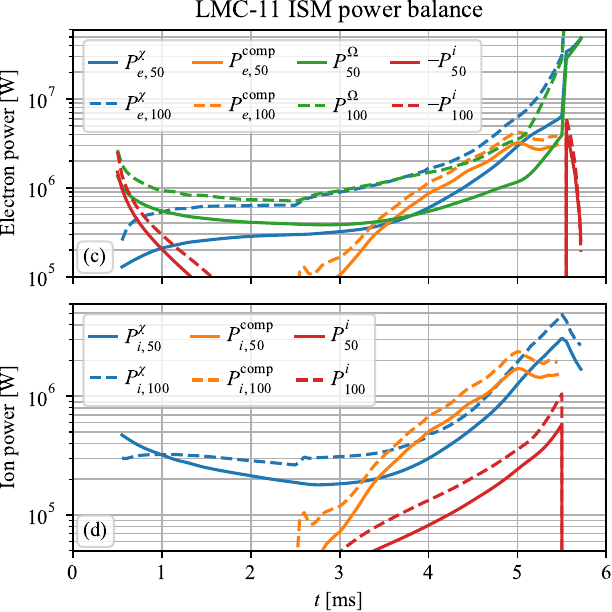}
    \caption{Power terms from the ISM reconstructions of (a, b) LMC-9 and (c, d) LMC-11, separated by (a, c) electron and (b, d) ion channels. Power due to thermal transport losses $P^\chi$ (blue), compressional heating $P^\text{comp}$ (orange), Ohmic heating $P^\Omega$ (green), and electron-ion exchange $P^i$ (red) are shown. To illustrate the variation of power terms across the core and edge, each term is evaluated within the 50\,\% poloidal flux surface (solid lines) as well as the LCFS (dashed lines), denoted by subscripts 50 and 100 respectively. Compression power data after temperature crashes is not reliable due to step changes in $\hat\chi_e$ and is therefore omitted. Compression begins at $t = 3.5$\,ms for LMC-9 and 2.5\,ms for LMC-11
    }
    \label{fig:ISM_power}
\end{figure}

\begin{figure}
    \centering
    \includegraphics[width=\linewidth]{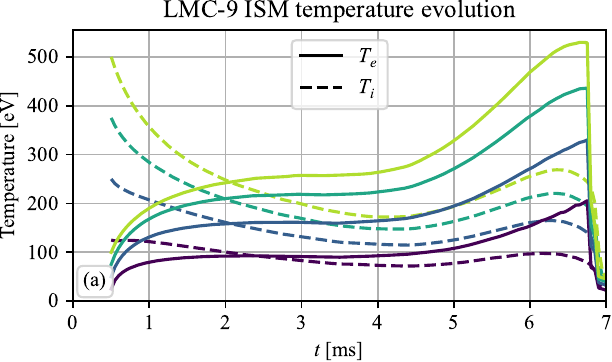}
    
    \medskip 
    \includegraphics[width=\linewidth]{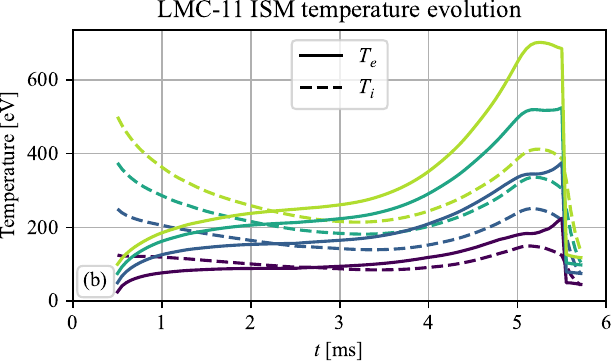}
    \caption{Temperature evolution from the ISM reconstructions of (a) LMC-9 and (b) LMC-11. Temperatures are shown for electrons (solid lines) and ions (dashed lines) at positions progressing from the magnetic axis (brightest color) to $\bar\psi=0.75$ (darkest color) in uniform increments. Both temperatures are fixed to 10\,eV at the LCFS. Compression begins at $t = 3.5$\,ms for LMC-9 and 2.5\,ms for LMC-11.}
    \label{fig:ISM_temps}
\end{figure}

\begin{figure}
    \centering
    \includegraphics[width=\linewidth]{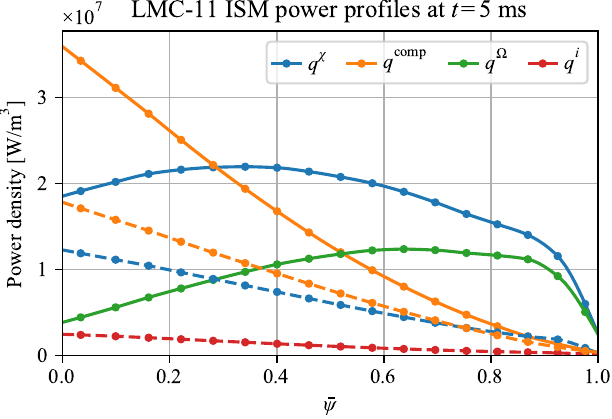}
    \caption{Power density profiles taken from the ISM reconstruction of LMC-11 at $t=5$\,ms. Profiles are shown for thermal transport losses $q^\chi$ (blue), compressional heating $q^\text{comp}$ (orange), Ohmic heating $q^\Omega$ (green), and electron-ion exchange $q^i$ (red). Electron and ion terms are denoted by solid and dashed lines, respectively. 
    }
    \label{fig:ISM_LMC11_power_profile}
\end{figure}

The best fit diffusivities and resulting thermal confinement times are summarized in Table~\ref{tab:ISM_params}. The evolution of the corresponding power balance terms and temperatures are shown in Fig.~\ref{fig:ISM_power} and Fig.~\ref{fig:ISM_temps}, respectively. The power balance of LMC-11 is expanded in terms of 1D power density profiles in Fig.~\ref{fig:ISM_LMC11_power_profile}, sampled at $t=5$\,ms.
These ISM results assume an effective ion charge $Z_\text{eff}=2$ and average ion charge $Z_\text{avg}=1.2$. 

\begin{table}
    \centering
    \caption{Summary of thermal confinement parameters for the ISM reconstructions of LMC-9 and LMC-11. Effective thermal diffusivities $\hat\chi$ are given in $\text{m}^2/\text{s}$. Thermal confinement times within the 50\,\% poloidal flux surface and the LCFS, $\tau_{E,50}$ and $\tau_{E,100}$ respectively, are taken at the start of compression and given in ms. For reconstruction purposes, these fitted values of $\hat\chi$ lump together all possible thermal transport processes, including radiation, and therefore do not represent the true diffusivity in the plasma.}
    \begin{tabular}{lrcccc}
        Shot & Channel & $\hat\chi^\text{axis}$ & $\hat\chi^\text{LCFS}$ & $\tau_{E,50}$ & $\tau_{E,100}$ \\
        \hline
        \rule{0pt}{2ex}%
        \multirow{3}*{LMC-9} & Electrons & 9.0 & 27 & 3.2 & 2.4 \\
        & Ions & 12 & 12 & 4.3 & 4.3 \\
        & Combined & -- & -- & 3.6 & 3.0 \\
        \hline
        \rule{0pt}{2ex}%
        \multirow{3}*{LMC-11} & Electrons & 6.5 & 24 & 4.3 & 2.6 \\
        & Ions & 10 & 10 & 5.0 & 5.3 \\
        & Combined & -- & -- & 4.6 & 3.5
    \end{tabular}
    \label{tab:ISM_params}
\end{table}

In order to match the final rapid turnover of the measured magnetic field, it is necessary to increase $\hat\chi_{e}$ by a large factor, resulting in a rapid decrease in temperature which accelerates the resistive decay of the plasma. For LMC-9 this change occurs at 6.75\,ms, a timing which is consistent with the MHD stability analysis discussed in Sec.~\ref{sec:stability}.

Before compression begins, the electron temperature profiles are approaching a steady state with transport losses being largely balanced by Ohmic heating. 
The onset of compression leads to a subsequent rise in electron temperature, due to compressional and Ohmic heating.  Ohmic heating increases during compression as the plasma current density grows due to flux compression.  Transport of heat becomes more effective as the plasma becomes smaller.

For MTF, the important power balance is that of the core plasma, taken here to be the central 50\,\% of poloidal flux and represented in Fig.~\ref{fig:ISM_power} by the solid curves.  The power balance of the plasma as a whole is represented by the dashed curves.  The reason the core plasma is relevant for MTF is that compressional heating, being proportional to pressure, predominantly occurs in the core, and it is there the fusion yield will be generated.  In addition, in the present experiments the Ohmic heating in the edge plasma is enhanced by the lag of shaft current, discussed in Sec.~\ref{sec:stability}, generating heat that transports quickly to the wall.
The 1D distribution of these power balance terms is exemplified by Fig.~\ref{fig:ISM_LMC11_power_profile}.

Focusing on the core plasma, the simulation shows that compressional heating is fully exceeding Ohmic heating at 5\,ms for LMC-9 and 3.5\,ms for LMC-11. In the case of LMC-11, compressional heating dominates Ohmic heating by a factor of 2 at 5\,ms.
Ion compressional heating exceeds ion losses in both cases, yielding peaks in ion temperature consistent with those in Table~\ref{tab:isothermal-Ti}.

Even for the power balance of the plasma as a whole, compressional heating becomes similar to Ohmic heating during compression in the case of LMC-11.

\subsection{Discussion}\label{sec:disc}

In these LM26 experiments, the energy confinement time is lower than observed in PI3 \cite{Pi3_tauE_NucFus_2025}.  The energy confinement time for the combined electron and ion energy is $\tau_E\simeq 3.5\mathrm{\,ms}$ from Table~\ref{tab:ISM_params}.
This value is below the range of values $\tau_E\simeq 4\text{--}12\mathrm{\,ms}$ observed in PI3 with fresh lithium coating.  A lower $\tau_E$ in LM26 is to be expected because it is a smaller plasma (scaling the LM26 energy confinement time to PI3 major radius as $R^2$ gives $\tau_{E,\mathrm{scaled}}\simeq 6\mathrm{\,ms}$, in the range of PI3 observed values, but at the low end).   We believe a more stringent comparison would be confounded by various effects that we now discuss.  The LM26 plasma is more elongated, and increasing elongation typically improves $\tau_E$, which would reduce the effect of the geometry change.  
The observed lower $\tau_E$ in LM26 may also be partly explained by impurities, which might be more prevalent due to the short period of time between machine assembly and the compression experiment, a few weeks, compared to the extended periods of time spent under vacuum in PI3 campaigns, many months.  
In addition, the $\tau_E$ in PI3 is from measurements made $5\mathrm{\,ms}$ after plasma formation, and the plasmas in LM26 are compressed earlier than that.
Finally, and possibly most importantly, the $\tau_E$ value for LM26 is an effective value, referenced to pre-compression geometry, describing transport for the plasma before and during compression, and thus is not expected to fully correspond to the non-compressing PI3 plasmas.

In the PCS experiments \cite{Howard_2025}, the ion heat transport during compression was similar to what we have observed in LM26, with $\hat\chi_i\approx 5\mathrm{\,m^2/s}$ for the effective thermal diffusivity in a 0D model, giving $\tau_{E,i}\simeq 122\text{\,\textmu s}$ when referred to pre-compression geometry.
The compression time was $\tau_C\simeq 140\text{\,\textmu s}$, giving $\tau_{E,i}/\tau_C\simeq 0.87$.
In LMC-11 the energy confinement time is significantly higher, $\tau_{E,i}\simeq 5.3\mathrm{\,ms}$ (Table \ref{tab:ISM_params}), and the compression time is $\tau_C=R_\mathrm{eq,0}/v=3.6\mathrm{\,ms}$, giving $\tau_{E,i}/\tau_C\simeq 1.5$.  Going from PCS to LM26 the ratio $\tau_{E,i}/\tau_C$ increases by a factor of 1.7, consistent with the modeled ion temperature increasing in LM26 and not in PCS.  The reason the $\tau_{E,i}/\tau_C$ ratio was not increased linearly with machine size is that the explosively driven compression velocity in PCS is much faster than the coil-driven compression velocity in LM26.
The electron energy confinement in PCS was worse than in LM26, in terms of $\hat\chi_e$. This can be attributed to the very high current density in PCS (Hugill number $H\sim 0.04$).  So the improvement of electron heating in LM26 compression has a large contribution from its lower current density ($H\sim 0.5$).

We note here that the PCS observation of enhanced ion energy confinement ($\tau_{E,i}\approx 1.6\mathrm{\,ms}$, $\hat\chi_i\approx 0.52\mathrm{\,m^2/s}$) in a time interval following plasma formation \cite{Howard_2025} has also been observed in the neutron data for PI3 \cite{radichIonTemperatureInference2026a} and in early shots in LM26.  This enhanced confinement behavior does not occur in the later LM26 compression shots considered in the detailed transport analysis.

The inferred transport level in the present experiments is similar to neoclassical ion heat diffusivities predicted for these plasma parameters. Comparable transport levels have also been observed in NSTX experiments with similar magnetic field ($B\approx 0.4\mathrm{\,T}$) and aspect ratio \cite{kayeScalingElectronIon2007}.  
Gyrokinetic simulations of the similar PI3 machine showed that ion heat transport can be predominantly driven by ITG turbulence \cite{kumarITGTurbulenceGyrokinetic2025}. In NSTX, ITG modes were largely suppressed due to strong $\mathbf{E} \times \mathbf{B}$ shear generated by neutral beam injection, resulting in ion thermal transport close to neoclassical levels. In contrast, PI3 (and LM26) has no significant external momentum injection source to stabilize ITG modes, and turbulent contributions to the transport may therefore still contribute significantly to observed ion heat losses.  ITG turbulence would also contribute to electron heat transport.  Distinguishing between neoclassical and turbulence-driven transport mechanism is beyond the scope of the present work.

Assuming the transport approximately follows gyro-Bohm scaling, $\chi_\mathrm{i,GB} \sim \rho_*\rho_iv_{\mathrm{th},i} \sim \rho_i^2v_{\mathrm{th},i}/a$, where $\rho_*=\rho_i/a$ is the normalized ion gyroradius, $\rho_i=\sqrt{m_iT_i}/eB$ is the thermal ion gyroradius, $a$ is the minor radius, and $v_{\mathrm{th},i}=\sqrt{T_i/m_i}$ is the ion thermal speed. In terms of the magnetic field $B$, gyro-Bohm scaling means the transport scales as $\chi_i\sim\rho_i^2 v_{\mathrm{th},i}/a \propto 1/B^2$. Assuming this scaling remains approximately valid, a 50\,\% increase in magnetic field strength $B$ would reduce the transport by more than a factor of two.  Experimentally, the energy confinement time in STs is observed to improve roughly linearly with $B$ \cite{Kurskiev_NF2022,kayeScalingElectronIon2007}. Even in this less optimistic scenario, increasing $B$ could be an effective way to improve the outcome from weak ion heating to strong compressional ion heating, in the LM26 machine. 

\section{Conclusions}\label{sec:Conclude}

In this study,  we presented a series of experimental results from the LM26 device at General Fusion and explore relevant analyses and modeling to draw conclusions about future directions for MTF research.  
The initial pre-compression state of the CHI-formed ST deuterium plasmas has sufficiently good confinement properties to remain in a nearly-axisymmetric magnetic field configuration until reaching radial compression factors of $C_R > 2.8$. 
Significant electron heating is accomplished by compression, due to the liner remaining sufficiently symmetric deep into compression, in addition to compression-enhanced Ohmic heating from increasing plasma current density.
As seen in Table~\ref{tab:AXUV-Te_summary}, the experiments achieve a three-fold increase in electron temperature in LMC-11, with the majority of this increase attributable to compressional heating (as shown in Sec.~\ref{sec:CompressHeatConfine}).

Plasma density also increases significantly during compression (Fig.~\ref{fig:IFsummary}, \ref{fig:lmc-8-inventories}) with trends that typically show a monotonically decaying particle inventory until the very end of compression. 
At that point, rapid additions to the electron population are best explained by ingress of higher $Z$ impurities from the wall that ionize and cool the edge plasma by radiation. 
These impurities are most likely 
vaporized $\mathrm{LiD}$ from the surface of the lithium liner and possibly trapped argon (Ar) released from voids within the liner. 
We conclude that the impurities are primarily in gaseous form because the LSE (laser scattering from ejecta) diagnostic used on LMC-10 detected no particulate ejecta where the liner first contacted the wall, nor was any indicated by spectroscopy data, AXUV error analysis, and analysis of the color fast camera data.

Magnetic stability is maintained quite well until a radial compression of $C_R \sim 1.8$ ($C_R=2.2$ in LMC-11), after which a weak $n=1$ mode begins to grow and is likely to cause a moderate increase in energy loss rate due to transport. 
The appearance of this mode tends to occur slightly before MHD instability is identified in reconstructed equilibria.  
However, even small differences in reconstructions can result in significant changes in the predicted timing of the instability. 
One possible cause of the mode is that liner buckling, as observed with the SLR and plasma fast cameras, grows sufficiently to begin coupling to the plasma edge and drive instability. 
Another possible cause is that as edge $q$ falls, spontaneous instability becomes more likely, where even small differences in reconstructions can result in significant changes in the predicted timing of the instability. In any case, this small amplitude rotating $n=1$ mode of the plasma poloidal magnetic field does not lead to disruption immediately, and the plasma continues to survive significant further compression.

This general observation of stability is further supported by suppression of AXUV crashes after the first 1 ms of liner compression, as evidenced by comparing compression shots to their preceding prototype shots. 
In the prototype shots, non-disruptive X-ray crashes with $n=0$ magnetic structure typically occur with an interval of 2--3 ms.  Each crash transiently increases the plasma current, and at least one would have fallen within the compression window. 
However, once rapid compression has begun, the next expected AXUV crash is suppressed, or at least delayed, due to plasma conditions shifting away from what would have caused a crash to happen. Stability analysis supports this conclusion as well, with ideal disruptive instabilities delayed until the end of compression.

The data collected by the scintillator array and the ion Doppler spectrometers are consistent with the interpretation that ion transport causes $T_i$ to fall below $T_e$ during the early slow phase of compression, followed by an ion reheating in the final faster phase of compression.  Our best-fitting models show that contributions from compressional heating and thermal transfer from the hotter electrons exceed overall ion cooling from transport.  This results in a net positive slope of $T_i(t)$ at deep compression on the best performing shots (see section \ref{sec:CompressHeatConfine}).
We believe the observed level of ion heating is due to ion heat transport being too high for the overall size of the plasma and the existing speed of compression. 
Transport modeling and experimental evidence from the literature (see section \ref{sec:disc}) indicate that ion transport could be sufficiently reduced to result in significant compressional heating of ions with the existing achieved liner velocities if the plasma magnetic field strength is increased by 50\,\%. 
Shaping can also have a significant impact on performance. Negative triangularity is known to have both a positive impact on stability and offer stronger compressional heating by enhancing the self-similarity of the equilibrium states during compression~\cite{Brennan2021}. These improvements will be implemented in the upcoming upgrade to the LM26 machine.

\section*{Acknowledgments}
The authors would like to thank Martin Cox, Tony Donne, Kurt Schoenberg, Brian Lloyd, and Ned Sauthoff for their insightful discussion and constructive comments. The authors also thank UKAEA for their support with Thomson scattering diagnostics.

\begin{appendices}
\section{Sources of error in AXUV \texorpdfstring{$T_e$}{Te} calculations and modeling}\label{App:A}
The AXUV diagnostic arrays use the Opto Diode model AXUV20HS1 silicon photodiode which has a nearly flat responsivity above 20 eV photon energies. 
The manufacturer provides an absolute calibration for responsivity ($R$) in $\mathrm{A/W}$ as a function of photon wavelength (5.2 to 1100 nm), which shows an average value of $ R= 0.261\,\mathrm{A/W}$ for photon energy $E$ in the range of $20\,\mathrm{eV} < E < 235\,\mathrm{eV}$, which we use for converting raw signal in amperes to total power incident on the diode in watts. These photodiodes continue to have similar responsivity up to $E \sim 10\,\mathrm{keV}$. 
However, the manufacturer does not specify an expected diode-to-diode variation in responsivity. Therefore, starting with the LMC-9 campaign, we have implemented our own process to determine a relative calibration correction within the set of 4 photodiodes of a single AXUV diagnostic module. 
Errors in overall responsivity of the whole diagnostic, including electronics, can then be compensated for, which is important since systematic errors in individual diode responsivities are magnified in the error of the ratio of those signals (5\,\% errors in diode $R$ can cause a maximum of 11\,\% error in their ratio). The calibration setup uses a 40 kV, 80 $\mathrm{\mu A}$ X-ray source, and relies on attenuation of soft X-ray components through 20 cm of air such that only a beam of hard X-rays illuminates the array of 4 photodiodes. This enables doing the calibration with the filters in place just before installation on LM26. As such it measures relative variation in the diode responsivity above $E >  5\,\mathrm{keV}$. We find that most of the diodes are within 3\,\% of each other, however some have been found to deviate from neighbors by as much as 25\,\%, which if left uncompensated for would lead to significant errors in $T_e$ estimation. As such, we have augmented the reported error bars for AXUV data before LMC-9 to represent this increase in measurement uncertainty.  In the future, this X-ray calibration setup could also better characterize the filter transmission curves throughout the VUV to soft X-ray range, which could reduce the present $\sim20\,\%$ uncertainty in $T_e$ associated with filter thickness. For the results reported here, we are relying on established calculated transmission functions \cite{Henke_1993} for our calculations. The effect of filter thickness uncertainty was directly calculated for LMC-11 results and retroactively applied as a minimum of $20\,\%$ uncertainty in $T_e$ for previous shots.  Reported error bars in Table~\ref{tab:AXUV-Te_summary} and Fig.~\ref{fig:axuv-Te-vs-CR} will exceed this minimum if there is a larger disagreement in $T_e$ values calculated from alternate ratios at the same position, in which case the error bars represent the spread in different calculated values.

Theoretical modeling from FLYCHK \cite{chung_flychk_2005} data was used to estimate the direction and magnitude of particular species of impurities that may impact the $T_e$ estimate derived from the X-ray diagnostics using the simple radiation model.  The simple radiation model is based on approximating the emission spectrum using the bremsstrahlung spectral shape $\sim \exp{(-E/T_e)}$, where $E$ is the photon energy.

Synthetic AXUV ratios for a given element are determined as follows: given ion fraction concentrations $c_k$ (including deuterium concentration to make $\sum_k c_k = 1$), we determine the ion dilution factor 
\begin{equation}
    f_i = \frac{n_{i}}{n_e} = \frac{c_i}{\sum_{j} c_{j}\bar Z_{j}} = \frac{c_i}{Z_\mathrm{avg}}
    \label{eq:ion_dilution_factor}
\end{equation}
noting that mean charge state $\bar Z_i$ may vary with temperature. Total power seen by an AXUV diode, assuming fixed density and temperature, is
\begin{equation}
    P_{k}(T_e) = C\sum_i f_i(T_e)n_e^2 \int_0^\infty\! T_k(E)p_{i}(E, T_e)\, dE
    \label{eq:impurity_mix_power}
\end{equation}
where $T_k(E)$ is the transmission curve for filter $k$, and $p_{i}(E, T_e)$ is the emission spectrum for element $i$ downloaded from FLYCHK, normalized by the $n_in_e$ used in the FLYCHK calculation. The prefactor $C$ disappears in the ratio $r_{ab}(T_e) = P_a(T_e)/P_b(T_e)$. The apparent $T_e$ is determined by inverting the derived ratio with the curve from the simple radiation model as used in our primary calculations: $T_{e,\mathrm{apparent}}(T_e) = r^{-1}_{ab,\mathrm{H}}\left(r_{ab,\mathrm{mix}}(T_e)\right)$, where $r_{ab,\mathrm{H}}(T_e)$ is the ratio curve for the simple radiation model and $r_{ab,\mathrm{mix}}(T_e)$ is the ratio curve calculated for the mixture.

\begin{figure}
    \centering
    \includegraphics[width=\linewidth]{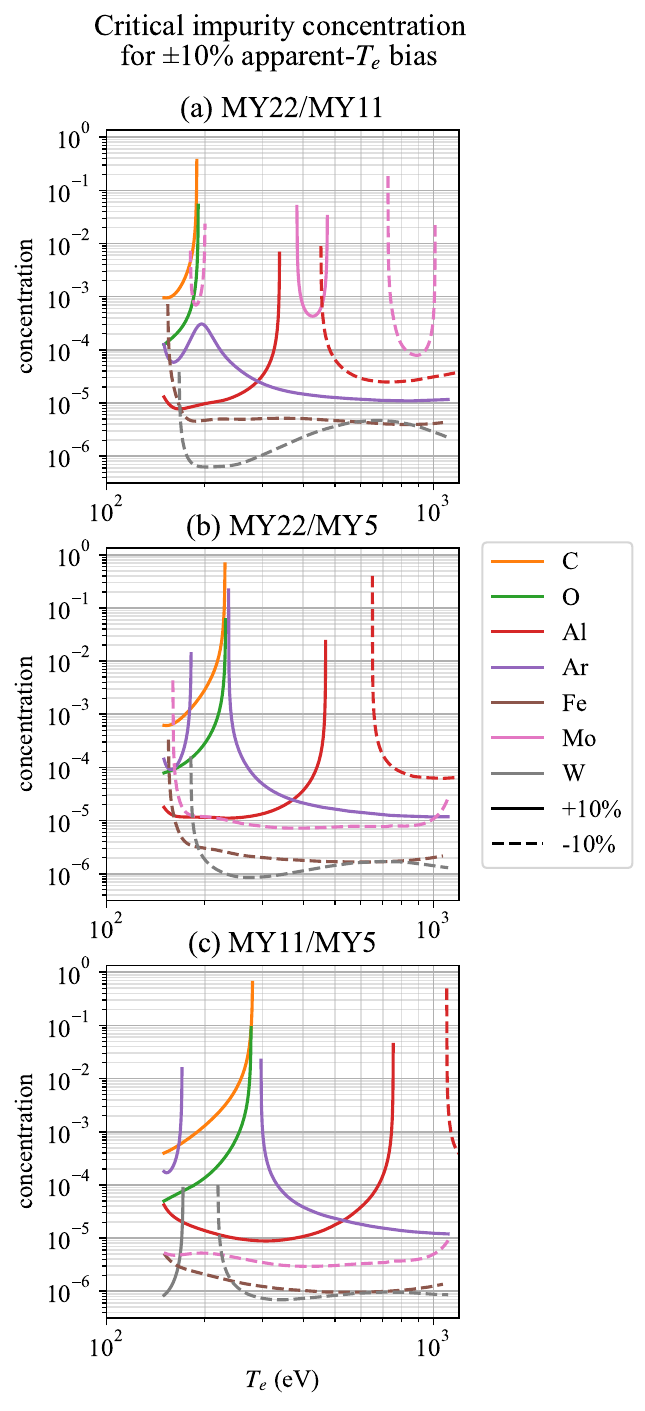}
    \caption{Concentrations of impurities in an otherwise pure hydrogen plasma at which there is a significant deviation of  10\,\% between the true temperature and the temperature estimated from AXUV ratios assuming the simple radiation model. The Li spectral shape is extremely similar to bremsstrahlung for the transmitted region, so there is no such concentration. For elements with a solid line, at concentrations above that line there is potential for the temperature obtained using the simple radiation model to overestimate the true temperature.}
    \label{fig:Te_impurity_concentration}
\end{figure}

\begin{figure}
    \centering
    \includegraphics[width=1\linewidth]{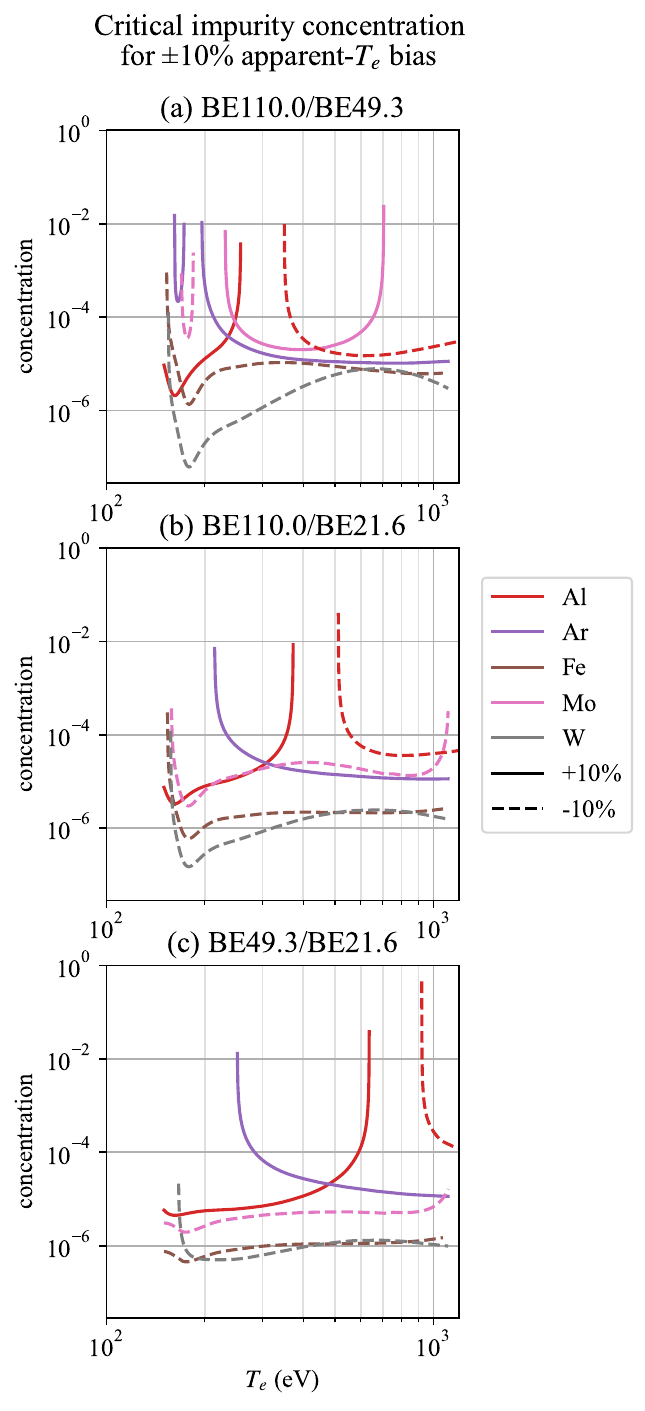}
    \caption{Concentrations of impurities that would give rise to a 10\,\% error in estimated $T_e$ value when using beryllium filters, repeating the analysis method shown in Fig.~\ref{fig:Te_impurity_concentration}. Carbon, oxygen and lithium were included in the analysis but do not appear in the graph because there is no concentration of these low-Z impurities that would give rise to a 10\,\% error in $T_e$ output, as can be seen in Fig.~\ref{fig:Be_fiter_Te_lookup}. }
    \label{fig:Be_filter_imp_conc}
\end{figure}

For hydrogen plasmas with a single impurity, the concentration at which the impurity causes $|T_{e,\mathrm{apparent}} - T_e|/T_e>0.1$ is shown in Fig.~\ref{fig:Te_impurity_concentration} for the Mylar filters and in Fig.~\ref{fig:Be_filter_imp_conc} for beryllium filters. Emissions from light elements such as lithium and oxygen have limited effect on measured ratios, especially at higher temperatures. Even small traces of iron and tungsten can cause the apparent temperature to be lower than the true temperature, with the effect more pronounced at higher temperatures.

\begin{figure}
    \centering
    \includegraphics[width=\linewidth]{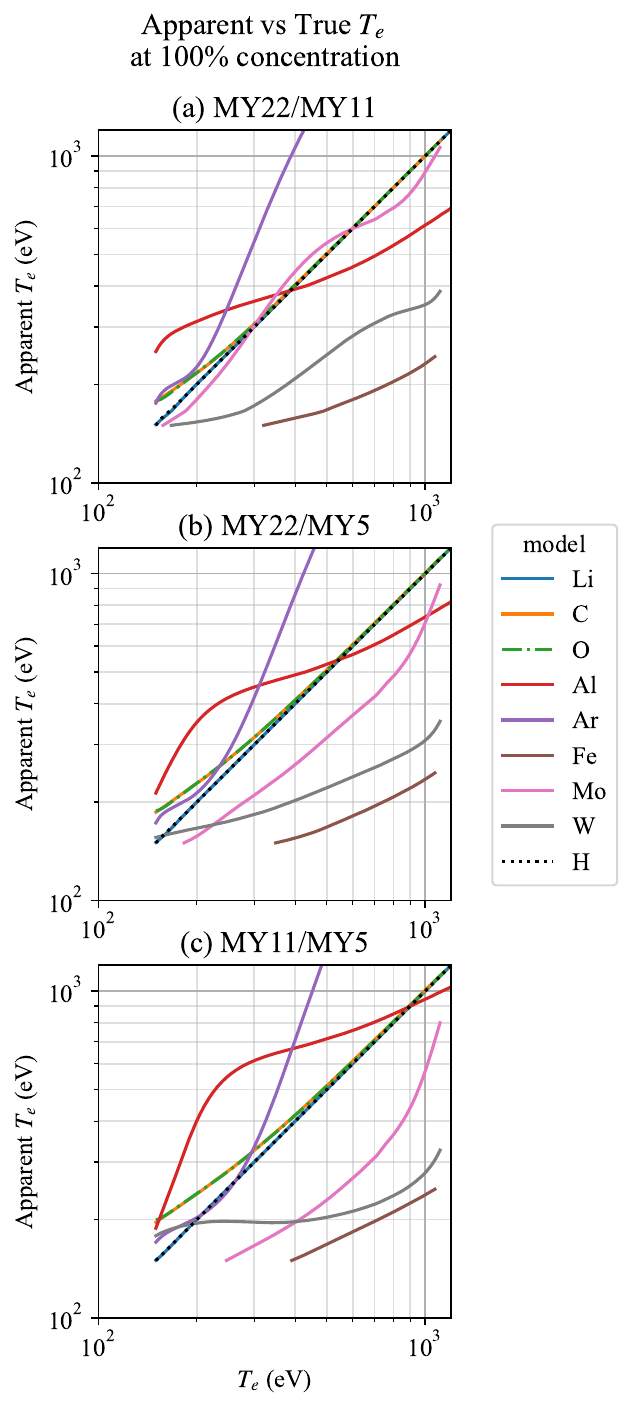}
    \caption{Electron temperature $T_e$ derived from applying a simple-radiation-model-based ratio-$T_e$ lookup curve to ratios forward-modeled from pure (100\,\% concentration) plasmas of various elements. This bounds the possible effect of each impurity species on AXUV $T_e$ calculations.}
    \label{fig:Te_impurity_max}
\end{figure}

\begin{figure}
    \centering
    \includegraphics[width=1\linewidth]{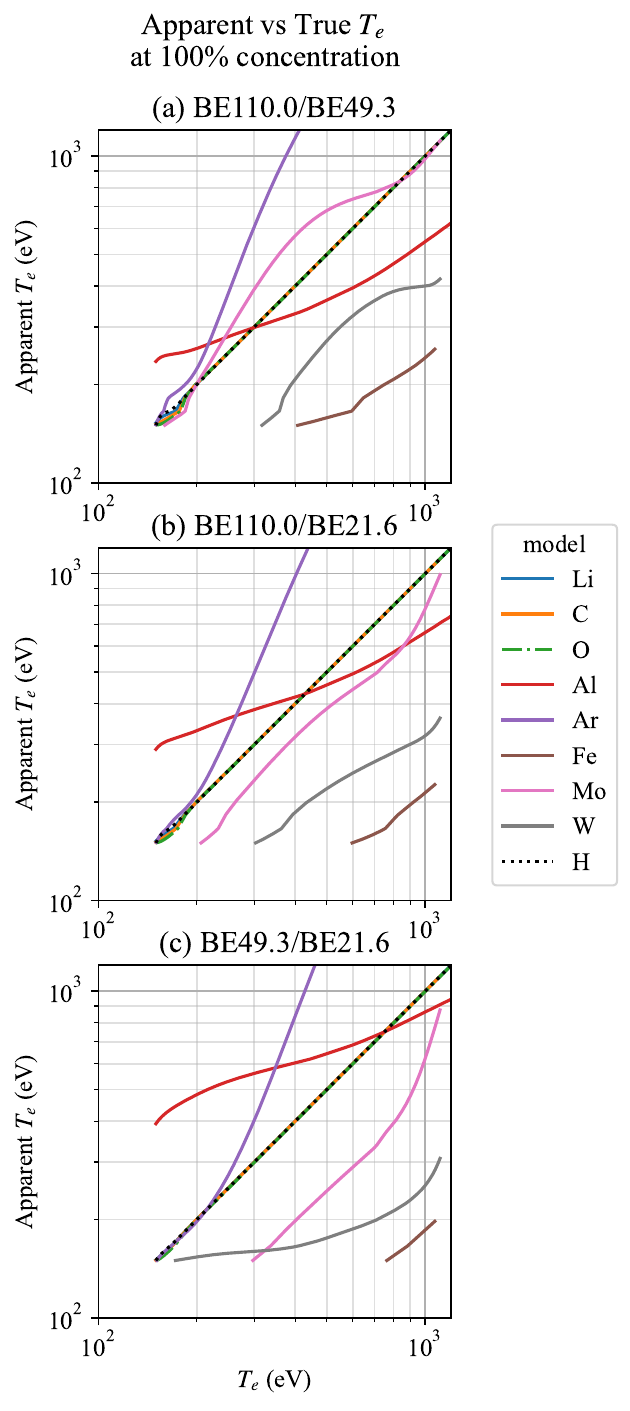}
    \caption{$T_e$ lookup curves similar to Fig.~\ref{fig:Te_impurity_max} but with beryllium filters, this shows a reduction in error for low-Z impurities when using beryllium filters rather than aluminized Mylar, however high-Z impurities have line radiation at higher photon energy which can cause either overestimates (for Ar and Al) or underestimates (for Fe, W, Mo) to a similar degree for beryllium filters as the Mylar filters. }
    \label{fig:Be_fiter_Te_lookup}
\end{figure}

However, at some temperatures, certain impurities skew the apparent temperature higher than the true temperature. Of these, aluminum and argon deserve particular consideration due to their prominence in machine construction and operations. RGA studies before compression shots estimate a background argon density of $10^{14} \mathrm{\,m^{-3}}$, which at our measured temperatures would have an effect within our existing error bars, but pockets of argon released from the liner as it deforms or aluminum scattered during liner touchdown have the potential to impact results. Figures \ref{fig:Te_impurity_max} and \ref{fig:Be_fiter_Te_lookup} show the worst-case impact of each species on the measured temperature, when using the Mylar or beryllium filter sets, which are qualitatively similar for both types of filter.

One remaining practical consideration that is worth discussing as a potential error source, and a factor in filter selection in an MTF-relevant application is the comparison of the visible-light rejection properties of aluminized Mylar and beryllium filters, as well as their mechanical properties. 
The vapor-deposited coating of aluminum on the Mylar substrate has the potential to develop small pinholes which can pass near-UV and visible light with wavelength $\lambda > 250\,\mathrm{nm}$. 
Al-layer pinholes can be present from the fabrication process, or can form during exposure to plasma. 
As a simple, preliminary measure to protect against introducing visible light contamination, a quality control step has been conducted prior to each installation of AXUV filter-diode assemblies on LM26 and just after disassembly after a compression shot, in which a high-brightness white LED flashlight is applied directly on the filtered aperture while a digital multimeter measures the forward-biased resistance across each AXUV diode. 
A new diode typically has a resistance value (when in complete darkness) of $R_d \approx 1.700 \, \mathrm{M\Omega}$, with measurement resolution of $1\, \mathrm{k\Omega}$,  using the multimeter applying a forward polarity test voltage of $V_{\text{test}} = +0.5\, \mathrm{VDC}$. 
Diode-to-diode variability for $R_d$ is around $1\,\%$ but has a repeatable measurement value for each diode. 
Because photovoltaic current is in the negative direction, absorption of light causes the measured forward-biased current to decrease, this reduction of current at $I(V_{\text{test}})$ is shown on the multimeter as an increase in measured resistance. 
For an AXUV diode without any filter, the measured resistance will increase to $R_d \approx 4 \, \mathrm{M\Omega}$ when the flashlight is applied. 
This corresponds to a photovoltaic current of $I_{\text{pv}} = 0.169\, \mathrm{\mu A}$ when compared to the nominal dark resistance. 
A filtered photodiode assembly is considered to fail this light-leak flashlight test if the measured resistance increases by $1\, \mathrm{k\Omega}$ or more, in which case the filter is removed and replaced with a new filter of the same thickness. The photovoltaic current required to cause this null upper bound for filter acceptance is 
\begin{equation*}
    I_{\text{null}} = (0.5/1.7 - 0.5/1.701)\times10^{-6} \mathrm{A} = 1.73\times10^{-10} \mathrm{A}
\end{equation*}
 For a filter that passes this test, the fraction of incident visible light transmitted by the filter will be 
 \begin{equation*}
 f_{\text{vis}} < I_{\text{null}}/I_{\text{pv}} = 0.00102,
\end{equation*}
When considering the overall effect of visible light leakage on $T_e$ measurement error, it is important to note that visible light makes up only a small fraction of total EM emissions from the plasma. 
The upper limit of the fraction of near-visible emissions $(1\,\mathrm{eV} < E < 5\,\mathrm{eV})$, emerges from the optimistic case of a pure H plasma, which at $T_e = 150\, \mathrm{eV}$ was calculated to be 2.35\,\% of total light from the plasma. This visible fraction decreases to 0.75\,\% for pure H at $T_e = 510\, \mathrm{eV}$, which is the upper limit of our FLYCHK scan.  
The final fractional error also needs to take into account the fact that the higher energy photons are also being filtered. This error can be defined as
\begin{equation}
    {\epsilon_{\text{vis}}}(k) = \frac{f_{\text{vis}}P_{\text{vis}} }{\left(f_{\text{vis}}P_{\text{vis}} + P_k(T_e)\right)}
\end{equation}
using (\ref{eq:impurity_mix_power}) for an individual diode signal collecting light through an Al-Mylar filter that has passed this visible-light leak test. The maximum estimate for possible error for this term comes from the $150\, \mathrm{eV}$ pure-H extreme case: 
\begin{align*}
    \epsilon_{\text{vis}}(\text{MY 5} \mathrm{\mu m}) &< 1.23\,\%\\
    \epsilon_{\text{vis}}(\text{MY 11} \mathrm{\mu m}) &< 12.5\,\%\\
    \epsilon_{\text{vis}}(\text{MY 22} \mathrm{\mu m}) &< 69.6\,\%.
\end{align*}

As an example of more realistic impurity composition, (1\,\% Li, 0.1\,\% O) can be added to the H simulation and the visible light fraction drops to  1.0\,\% at $T_e = 150\, \mathrm{eV}$ falling further to 0.5\,\% at $510\, \mathrm{eV}$.  For this example case, the visible light fractional errors would have upper bounds of:
\begin{align*}
    \epsilon^*_{\text{vis}}(\text{MY 5} \mathrm{\mu m}) &< 0.18\,\%\\
    \epsilon^*_{\text{vis}}(\text{MY 11} \mathrm{\mu m}) &< 1.14\,\%\\
    \epsilon^*_{\text{vis}}(\text{MY 22} \mathrm{\mu m}) &< 9.1\,\%.
\end{align*}

For any particular plasma composition this is a systematic effect that will preferentially increase ratios involving the thicker filters, and so will generally cause a larger over-estimate of $T_e$ with those ratios. The effect of this error, however,  will also decrease as true $T_e$ increases.  Additionally, it is also possible that during plasma operations, we could have $f_\text{vis}$ increase above the initial null level on any particular filter, thick or thin,  due to formation of new pinholes. 
 
Such cases can be detected within the redundant array of measurements at similar radius, and indicated retroactively by failing the flashlight leak test during post-compression disassembly. 
Due to the fact that high energy photons are only emitted from a smaller region near the core of the plasma, visible light contamination is easily noticed in the raw signals by the presence of significant ($>10\,\%$) signals during early compression when equivalent neighboring signals are below the minimum detection floor, or when ratios of thick/thin filtered signals exceed unity. 
All identified cases of measurable visible-light leakage have been excluded from consideration in AXUV $T_e$ estimates in this study. 

An advantage of beryllium filters is their robust ability to completely block all near-visible light, eliminating the possibility of pinhole formation during plasma exposure. 
This was recognized as a way to produce more reliable measurements, requiring less data-quality control effort and so beryllium filters were installed in one AXUV package on the LMC-10 campaign to compare to the Al-Mylar filter performance and also to confirm that the more brittle beryllium foils did not fracture from the intense mechanical shock from the high-speed liner impact. The first test was successful and so the use of beryllium filters was expanded to three AXUV modules on LMC-11.   
Overall, we find that the beryllium and properly-curated aluminized Mylar filters give similar results on the LM26 system. Both give similar $T_e$ values, which are further validated with the Thomson scattering measurements obtained on LMC-11. Both types of filters are intrinsically susceptible to plasma-gradient effects if not properly arranged, as well to the effects of possible high-Z impurities, these being the dominant error contributions to either AXUV measurement method. 


\section{Thomson scattering \texorpdfstring{$T_e$}{Te} calculation and error analysis}
\label{app:thomson}

Electron temperature $T_e$ from Thomson scattering is calculated by finding the best fit between noise-weighted polychromator channel measurements and synthetic signals from a forward model. The required scaling of the signal can also provide the relative change in electron density $n_e$ between laser pulses. 

A closed form estimate of $T_e$ uncertainty due to photon shot noise and intrinsic detector electrical noise is calculated as described in \cite{Maslov2012}. The background shot noise and intrinsic electrical noise is the standard deviation of a channel's trace before the laser pulse. 

Thomson scattering shot noise is conservatively estimated from the variation of the channel 1 stray light pulse height in a series of 20 pulses. For an average signal of 10 mV, the variation was 2 mV and at least half of this variation could be accounted for by the background and intrinsic noise measurement before the pulse. Therefore, we take the Thomson shot noise to be 10\,\% for a 10 mV signal. Behaving as a Poisson process, the relative Thomson shot noise then scales as $1/\sqrt{N}$ where $N$ is the number of Thomson photons.

The scattering angle is a significant source of error. Sensitivity analysis shows that at a 10~degree scattering angle, for every 0.5 degrees in error there is a 10\,\% error in $T_e$. The scattering angle was estimated from a 3D model of the machine taking into account potential deviations between the model and reality, and the alignment uncertainty of the laser and the view vectors. Four reference points were defined to calculate the scattering angle. These point are listed in Table~\ref{tab:scattering_angle_calc}. The cavity wall spot reference point is created by backfilling the fiber with white light and projecting the light through the aligned collection lens onto the formation cavity wall inside the vacuum. 
The resulting spot can then be imaged with a camera through a nearby window and compared to the 3D model to determine the coordinates of the spot using convenient port features on the cavity wall. Solving for the angle and propagating the error gives $9.9 \pm 0.5 \deg$. This contributes a $10\,\%$ uncertainty in $T_e$ which is added in quadrature to the uncertainty generated from the $T_e$ fit.

\begin{table}
    \centering
    \caption{Reference point coordinates and uncertainties used to define the Thomson measurement scattering angle. The $z$ coordinate is along the axis of the machine, with the equator at $z=2367\mathrm{\,mm}$.}
    \label{tab:scattering_angle_calc}
    \begin{tabular}{lccc}
        Reference point & $x$ [mm] & $y$ [mm] & $z$ [mm]  \\ 
        \hline
        \rule{0pt}{2ex}%
        View port & $108\pm2$ & $63\pm2$ & $3499\pm2$ \\
        Cavity wall spot & $283\pm10$ & $164\pm10$ & $1743\pm2$ \\
        Beam exit port & $266\pm4$ & $154\pm4$ & $3615\pm2$ \\
        Beam entry port & $0\pm8$ & $0\pm8$ & $-1640\pm20$  \\
    \end{tabular}
\end{table}

\section{Description of PsiBC}
\label{app:psibc}

Here we provide a description of the method used in the PsiBC code to reconstruct the soaked flux boundary condition from Mirnov data.  A finite difference method is used to solve both the GS equation and the flux diffusion equation. A Cartesian grid in $(R,Z)$ is used, with a typical resolution $\Delta R=\Delta Z=5\mathrm{\,mm}$.  Points on the grid are identified as being in metal or plasma (vacuum) regions based on the machine geometry and shape of the liner, coming from the COMSOL liner reconstruction, at the time $t$ under consideration.  In the vacuum region PsiBC solves the GS equation for the poloidal flux field $\psi(R,Z)$, namely
\begin{equation}
\Delta^*\psi = -FF'_\psi
\label{eq:gs}
\end{equation}
Pressure is neglected.
The poloidal current $F(\psi)=RB_\varphi(R,Z)$ is a flux function given in terms of the profile function $f$ (described below) by
\begin{equation}
    F(\psi) = F_0 + \lambda f(\bar\psi)
\end{equation}
Here $\bar\psi$ is, as usual, defined to be linear in $\psi$ and having values $0$ at the magnetic axis and $1$ at the LCFS.  The shaft current $F_0(t)$ is set to the value measured by toroidal-field Mirnov coils.
The derivative of $F$ with respect to $\psi$ is
\begin{equation}
    F'_\psi = \lambda \bar\psi'_\psi f'(\bar\psi)
\end{equation}
where $\bar\psi'_\psi\equiv d\bar\psi/d\psi$.
The scaling parameter, $\lambda$, with units of inverse length, is adjusted during Picard iteration to achieve the solution $\psi(R,Z)$ satisfying the GS equation with specified boundary conditions and desired plasma current.
The one-parameter shape function used in PsiBC is
\begin{equation}
    f'(\bar\psi)=-1 - a\bar\psi + (1+a)\bar\psi^2
\end{equation}
We have edge value $f'(1)=0$ to avoid discontinuity in $J_\varphi$ at the edge of the plasma that would interfere with stable solution of the GS equation.   Positive values of the shape parameter, $a$, correspond to a hollow plasma current profile, negative values to peaked.  In order to keep the toroidal current density $J_\varphi>0$ in the plasma we need to restrict $a$ to satisfy $a>-2$.
The integral of the shape function, with edge value $f(1)=0$, is
\begin{equation}
    f(\bar\psi)=1-\bar\psi-\frac{a}{2} \left(\bar\psi^2-1\right)+\frac{1+a}{3}\left(\bar\psi^3-1\right)
\end{equation}
The boundary condition $\psi=\psi_0+\delta\psi$ for the GS equation uses $\psi_0$ from COMSOL and $\delta\psi$ from the previous time step of PsiBC.   A marching algorithm identifies the LCFS at each Picard iteration, so that $\bar\psi(R,Z)$ can be calculated within the LCFS.  During the iteration the scale factor $\lambda$ and the solution $\delta\psi$ are rescaled to match the Mirnov data of the equatorial probe, while $a$ is held constant.  This makes the GS solution a function of one variable, $a$.  This function is wrapped in a 1D optimizer that varies $a$ to fit the Mirnov probes along the shaft and (early in compression) on the endplates.  Mirnov probes on the endplates are progressively excluded from the fit as the liner moves inward.   After $a$ is determined, the next step is to update the flux soak, which is done by solving the magnetic diffusion equation for poloidal flux,
\begin{equation}
\frac{\partial(\delta\psi)}{\partial t} = \frac{\eta}{\mu_0} 
    \Delta^*(\delta\psi)
\end{equation}
inside the metal (machine components and liner), with $\eta$ the resistivity,
and, simultaneously, solving
\begin{equation}
    \Delta^*\psi = -\mu_0 R J_\varphi
\end{equation}
outside the metal, with $J_\varphi(R,Z)$ the plasma current density from the preceding GS solve.  This system of equations is solved by a series of implicit backward Euler time steps.  
After this is done, the soaked magnetic flux is advected by interpolating $\delta\psi$ from the Cartesian grid to the Lagrangian mesh of the liner, moving the liner to its new position, and interpolating back onto the Cartesian grid.
By repeating this cycle of GS solve, flux soak update, and advection, PsiBC produces a time series of plasma parameters during the discharge, including $\psi$ boundary condition, informed by Mirnov data. 

Reconstruction of the density profile from PsiBC output is performed as a post-processing step.
The method used is an unconstrained linear fit to a set of polynomial shapes chosen based on the number of interferometers, both unconstrained sparse polynomials in $\bar\psi$ and sparse polynomials constrained to be low density at $\bar\psi=1$, and taking the best fit that satisfies the positivity constraint ($n_e(\bar\psi)\ge0$).

\section{Description of ISM}
\label{app:ISM}
This section describes the method used for time-dependent simulations in the Integrated System Model (ISM)~\cite{khalzov_ism_2021, khalzov_ism-plasma_2023, Khalzov_2024}. The ISM-plasma model, being a 1.5D plasma transport code, is divided into an equilibrium solver and transport model. The former is implemented as a direct equilibrium $q$-solver wherein the safety factor and entropy profiles are used to determine the plasma equilibrium. These profiles evolve as determined by 1D flux surface averaged transport processes. The poloidal flux along the boundary of the plasma domain is calculated self-consistently by solving the diffusion and advection of magnetic flux within the liner and center conductor. For reconstructions, the external toroidal field function $F_0(t)$ is prescribed from an average of toroidal field probes located on the center conductor.

The evolution of thermal energy in ISM-plasma is determined by a balance between thermal transport, Ohmic heating, and compressional heating. 
Heat flow between flux surfaces for species $s$ is given by power $P_s$, calculated using
\begin{equation}
    P_s = -\frac32 n_s(\psi) \hat\chi_s \frac{dT_s}{d\psi} \int\nabla\psi\cdot d\textbf{S}.
\end{equation}
Here $s=e,i$ indicate electron and ion transport, respectively, and $\psi$ is the poloidal flux (but can be any flux surface label).  In general, $\hat\chi_s$ can be a function of $\psi$.

The collisional energy transfer $Q_i$ between electrons and ions is given by:
\begin{equation}
    Q_i=\frac{3m_e}{m_i}n_e\nu_e(T_e-T_i)
\label{eq:e-i}
\end{equation}
with electron-ion collision frequency $\nu_e$ given by
\begin{equation}
    \nu_e=\frac{\sqrt{2}}{12 \pi^{3/2}}\frac{n_iZ^2e^4 \ln\Lambda}{\varepsilon_0^2 m_e^{1/2}T_e^{3/2}}
\end{equation}
where $\ln \Lambda$ is the Coulomb logarithm.
This results in a thermalization rate, $\bar\nu_\epsilon^{i|e}\equiv \dot T_i/(T_e-T_i)$, given by 
\begin{equation}
    \bar\nu_\epsilon^{i|e}=\frac{2m_en_e}{m_in_i}\nu_e.
\end{equation}

The electron and ion pressure profiles are evolved according to the following equations
\begin{align}
    \frac{3/2}{(V'_\rho)^{5/3}}\frac{\partial}{\partial t}\!\left[p_e(V'_\rho)^{5/3}\right]
    +\frac{1}{V'_\rho}
    \frac{\partial P_e}{\partial \rho}
    &=
    Q_\Omega - Q_i,
    \label{eq:evo-pe-alt}
\\
    \frac{3/2}{(V'_\rho)^{5/3}}\frac{\partial}{\partial t}\!\left[p_i(V'_\rho)^{5/3}\right]
    +\frac{1}{V'_\rho}
    \frac{\partial P_i}{\partial \rho}
    &=Q_i.
        \label{eq:evo-pi-alt}
\end{align}
Here $p_s=n_s T_s$ is the pressure of species $s$.  Densities are related by $n_e=Z_\mathrm{avg}n_i$, where the average ion charge $Z_\mathrm{avg}$ is a parameter in the code.
The coordinate $\rho$ labels the flux surfaces and $V'_\rho\,d\rho$ is the volume between surface $\rho$ and surface $\rho+d\rho$.  In ISM-plasma, $\rho$ is determined by the enclosed mass (and vice versa). Since particles are conserved, this choice of flux label means density is determined simply as the ratio of the fixed mass to the varying volume between flux surfaces. The first term on the left-hand sides of (\ref{eq:evo-pe-alt}) and (\ref{eq:evo-pi-alt}), the pressure time derivative term, is written in a form suitable for a Lagrangian treatment of compressional heating, keeping $pV^{5/3}$ constant in the ideal adiabatic case.  The second term represents the divergence of the heat flux.  On the right-hand side of (\ref{eq:evo-pe-alt}), $Q_\Omega= \langle \eta J^2\rangle$ is the volumetric flux surface average of the electron heating by Ohmic dissipation of magnetic energy, calculated using resistivity $\eta=\eta_\|$ with $\eta_\|(T_e)$ the Spitzer parallel resistivity with enhancement factors for impurities and neoclassical effects \cite{Sauter_1999}. ISM results presented in the present paper assume an effective ion charge $Z_\text{eff}=2$ and average ion charge $Z_\text{avg}=1.2$.

The resistive diffusion and dissipation of toroidal and poloidal fluxes in the plasma evolve according to
\begin{align}
    \frac{\partial\Phi}{\partial t} &= \frac{\eta}{2\pi\mu_0} \int\!\frac{1}{R^2}\nabla F\cdot d\mathbf{S}, \label{eq:evo-phi}\\
    \frac{\partial\psi}{\partial t} &= -\frac{\eta}{\mu_0} \left(  FF'_{\psi} + \frac{\mu_0}{2\pi} V'_L p'_{\psi} \right), \label{eq:evo-psi}
\end{align}
where $\Phi$ is the enclosed toroidal flux and $V'_L$ denotes the derivative of enclosed volume with respect to inductance $L$ with
\begin{align}
    L = \int\! \frac{1}{R}\,dR\,dZ.
\end{align}

Equations (\ref{eq:evo-pe-alt}) to (\ref{eq:evo-psi}) are discretized into $N$ flux surfaces, resulting in a system of ordinary differential equations which are advanced in time using a second order Runge--Kutta scheme. The plasma equilibrium is re-solved between each time advance substep using the updated profiles, boundary flux, and boundary geometry.  The results are sufficiently accurate for subsequent interpretive modeling of the processes governing the plasma confinement, including MHD stability and compressional heating power analyses. 

\end{appendices}

\clearpage

\bibliographystyle{apsrev4-2}
\bibliography{refs}

\end{document}